\documentclass[aps,prd,12pt,twocolumn,superscriptaddress,showpacs,showkeys,nofootinbib,reprint,longbibliography]{revtex4-1}
\usepackage{url}
\usepackage[utf8]{inputenc}
\usepackage[T1]{fontenc}
\usepackage[allcolors=blue,colorlinks]{hyperref}
\usepackage{mathtools,amsmath,amssymb,amsfonts,dsfont,mathrsfs,amsthm,latexsym}

\usepackage{bm}
\usepackage{graphicx}
\usepackage{centernot}
\usepackage{xcolor}
\usepackage{comment}
\usepackage{siunitx}
\usepackage{array}
\newcolumntype{C}[1]{>{\centering\let\newline\\\arraybackslash\hspace{0pt}}m{#1}}
\usepackage{tikz}
\usepackage{pgfplots}
\pgfplotsset{width=7cm}
\usepackage{braket}
\usepackage{xparse}
\usetikzlibrary{
  arrows.meta,
  shapes,
  spy,
  decorations,
  decorations.pathmorphing,
  decorations.markings,
  calc,
  shadows.blur,
  shadings}

\tikzset{
  Gr/.style={circle,draw,minimum size=8mm},
  Ga/.style={circle,draw,minimum size=8mm,fill=black!10,delta angle=40},
}

\newtheorem{Def}{Definition}[section]

\DeclareDocumentCommand{\Ag}{ s o }{ \IfBooleanTF{#1}
    { \IfValueTF{#2}{ \bm{\mathcal{A}}_{(#2)} }{ \bm{\mathcal{A}} } }
    { \IfValueTF{#2}{    {\mathcal{A}}_{(#2)} }{    {\mathcal{A}} } } }
\DeclareDocumentCommand{\Af}{ s o }{ \IfBooleanTF{#1}
    { \IfValueTF{#2}{ \boldsymbol{A}_{(#2)} }{ \boldsymbol{A} } }
    { \IfValueTF{#2}{            {A}_{(#2)} }{            {A} } } }

\DeclareMathOperator{\End}{End}

\DeclareDocumentCommand{\Fg}{ s o }{ \IfBooleanTF{#1}
    { \IfValueTF{#2}{ \bm{\mathcal{F}}_{(#2)} }{ \bm{\mathcal{F}} } }
    { \IfValueTF{#2}{    {\mathcal{F}}_{(#2)} }{    {\mathcal{F}} } } }
\DeclareDocumentCommand{\Ff}{ s o }{ \IfBooleanTF{#1}
    { \IfValueTF{#2}{ \boldsymbol{F}_{(#2)} }{ \boldsymbol{F} } }
    { \IfValueTF{#2}{            {F}_{(#2)} }{            {F} } } }

\newcommand{\ga}{\gamma}

\newcommand{\Lag}{\mathscr{L}}

\newcommand{\N}{\ensuremath{\mathscr{N}}}

\newcommand{\Op}{\mathcal{O}}

\DeclareDocumentCommand{\PB}{ O{m} O{q} O{p} m m }{ \frac{ \partial #4 }{\partial {#2}^{#1} } \frac{ \partial #5 }{\partial {#3}_{#1} } - \frac{ \partial #4 }{\partial {#3}_{#1} } \frac{ \partial #5 }{\partial {#2}^{#1} } }

\newcommand{\R}{\mathbb{R}}

\DeclareDocumentCommand\Te{o o O{\,} m }{{T}_{#3}{}^{#1}_{#2}(#4)}

\newcommand{\W}{\ensuremath{\mathscr{W}}}

\DeclareDocumentCommand{\BM}{ s }{ \IfBooleanTF{#1} {\hat{\bm{M}}}{\bm{M}} }
\DeclareDocumentCommand{\BN}{ s }{ \IfBooleanTF{#1} {\hat{\bm{N}}}{\bm{N}} }
\DeclareDocumentCommand{\BP}{ s }{ \IfBooleanTF{#1} {\hat{\bm{P}}}{\bm{P}} }
\DeclareDocumentCommand{\BQ}{ s }{ \IfBooleanTF{#1} {\hat{\bm{Q}}}{\bm{Q}} }
\DeclareDocumentCommand{\BR}{ s }{ \IfBooleanTF{#1} {\hat{\bm{R}}}{\bm{R}} }
\DeclareDocumentCommand{\BS}{ s }{ \IfBooleanTF{#1} {\hat{\bm{S}}}{\bm{S}} }
\DeclareDocumentCommand{\BU}{ s }{ \IfBooleanTF{#1} {\hat{\bm{U}}}{\bm{U}} }
\DeclareDocumentCommand{\BV}{ s }{ \IfBooleanTF{#1} {\hat{\bm{V}}}{\bm{V}} }

\NewDocumentCommand\MyAc{ m }{#1}
\DeclareDocumentCommand{\vif}{ t. t, t- s s m }{
  \RenewDocumentCommand\MyAc{ m }{##1}
  \IfBooleanT{#1}{\RenewDocumentCommand\MyAc{ m }{ \mathring{##1} } }
  \IfBooleanT{#2}{\RenewDocumentCommand\MyAc{ m }{ \tilde{##1} } }
  \IfBooleanT{#3}{\RenewDocumentCommand\MyAc{ m }{ \bar{##1} } }
  \IfBooleanTF{#4}
  { \IfBooleanTF{#5} { \hat{\MyAc{\boldsymbol{e}}}^{\hat{#6}} }{ \hat{\MyAc{\boldsymbol{e}}}^{{#6}} } }
  { \MyAc{\boldsymbol{e}}^{{#6}} } }
\DeclareDocumentCommand{\vi}{ t. t, t- s s m m}{
  \RenewDocumentCommand\MyAc{ m }{##1}
  \IfBooleanT{#1}{\RenewDocumentCommand\MyAc{ m }{ \mathring{##1} } }
  \IfBooleanT{#2}{\RenewDocumentCommand\MyAc{ m }{ \tilde{##1} } }
  \IfBooleanT{#3}{\RenewDocumentCommand\MyAc{ m }{ \bar{##1} } }
  \IfBooleanTF{#4}
  { \IfBooleanTF{#5} { \hat{\MyAc{e}}^{\hat{#6}}_{\hat{#7}} }{ \hat{\MyAc{e}}^{#6}_{{#7}} } }
  { \MyAc{e}^{{#6}}_{{#7}} } }

\DeclareDocumentCommand{\bt}{ t. t, t- s s m m m }{
  \RenewDocumentCommand\MyAc{ m }{##1}
  \IfBooleanT{#1}{\RenewDocumentCommand\MyAc{ m }{ \mathring{##1} } }
  \IfBooleanT{#2}{\RenewDocumentCommand\MyAc{ m }{ \tilde{##1} } }
  \IfBooleanT{#3}{\RenewDocumentCommand\MyAc{ m }{ \bar{##1} } }
  \IfBooleanTF{#4}
  { \IfBooleanTF{#5} { \hat{\MyAc{\mathcal{B}}}_{{#6}}{}^{\hat{#7}}{}_{\hat{#8}} }{ \hat{\MyAc{\mathcal{B}}}_{{#6}}{}^{{#7}}{}_{{#8}} } }
  { \MyAc{\mathcal{B}}_{{#6}}{}^{{#7}}{}_{{#8}} } }

\DeclareDocumentCommand{\ct}{ t. t, t- s s m m m }{
  \RenewDocumentCommand\MyAc{ m }{##1}
  \IfBooleanT{#1}{\RenewDocumentCommand\MyAc{ m }{ \mathring{##1} } }
  \IfBooleanT{#2}{\RenewDocumentCommand\MyAc{ m }{ \tilde{##1} } }
  \IfBooleanT{#3}{\RenewDocumentCommand\MyAc{ m }{ \bar{##1} } }
  \IfBooleanTF{#4}
  { \IfBooleanTF{#5} { \hat{\MyAc{\Gamma}}_{{#6}}{}^{\hat{#7}}{}_{\hat{#8}} }{ \hat{\MyAc{\Gamma}}_{{#6}}{}^{{#7}}{}_{{#8}} } }
  { \MyAc{\Gamma}_{{#6}}{}^{{#7}}{}_{{#8}} } }
\DeclareDocumentCommand{\spif}{ t. t, t- s s m m }{
  \RenewDocumentCommand\MyAc{ m }{##1}
  \IfBooleanT{#1}{\RenewDocumentCommand\MyAc{ m }{ \mathring{##1} } }
  \IfBooleanT{#2}{\RenewDocumentCommand\MyAc{ m }{ \tilde{##1} } }
  \IfBooleanT{#3}{\RenewDocumentCommand\MyAc{ m }{ \bar{##1} } }
  \IfBooleanTF{#4}
  { \IfBooleanTF{#5} { \hat{\MyAc{\boldsymbol{\omega}}}^{\hat{#6}}{}_{\hat{#7}} }{ \hat{\MyAc{\boldsymbol{\omega}}}^{{#6}}{}_{{#7}} } }
  { \MyAc{\boldsymbol{\omega}}^{{#6}}{}_{{#7}} } }
\DeclareDocumentCommand{\spi}{ t. t, t- s s m m m }{
  \RenewDocumentCommand\MyAc{ m }{##1}
  \IfBooleanT{#1}{\RenewDocumentCommand\MyAc{ m }{ \mathring{##1} } }
  \IfBooleanT{#2}{\RenewDocumentCommand\MyAc{ m }{ \tilde{##1} } }
  \IfBooleanT{#3}{\RenewDocumentCommand\MyAc{ m }{ \bar{##1} } }
  \IfBooleanTF{#4}
  { \IfBooleanTF{#5} { \hat{\MyAc{{\omega}}}_{\hat{#6}}{}^{\hat{#7}}{}_{\hat{#8}} }{ \hat{\MyAc{{\omega}}}_{{#6}}{}^{{#7}}{}_{{#8}} } }
  { \MyAc{{\omega}}_{{#6}}{}^{{#7}}{}_{{#8}} } }

\DeclareDocumentCommand{\rif}{ t. t, t- s s m m }{
  \RenewDocumentCommand\MyAc{ m }{##1}
  \IfBooleanT{#1}{\RenewDocumentCommand\MyAc{ m }{ \mathring{##1} } }
  \IfBooleanT{#2}{\RenewDocumentCommand\MyAc{ m }{ \tilde{##1} } }
  \IfBooleanT{#3}{\RenewDocumentCommand\MyAc{ m }{ \bar{##1} } }
  \IfBooleanTF{#4}
  { \IfBooleanTF{#5} { \hat{\MyAc{\bm{\mathcal{R}}}}{}^{\hat{#6}}{}_{\hat{#7}} }{ \hat{\MyAc{\bm{\mathcal{R}}}}{}^{{#6}}{}_{{#7}} } }
  { \MyAc{\bm{\mathcal{R}}}{}^{{#6}}{}_{{#7}} } }
\DeclareDocumentCommand{\ri}{ t. t, t- s s m m m }{
  \RenewDocumentCommand\MyAc{ m }{##1}
  \IfBooleanT{#1}{\RenewDocumentCommand\MyAc{ m }{ \mathring{##1} } }
  \IfBooleanT{#2}{\RenewDocumentCommand\MyAc{ m }{ \tilde{##1} } }
  \IfBooleanT{#3}{\RenewDocumentCommand\MyAc{ m }{ \bar{##1} } }
  \IfBooleanTF{#4}
  { \IfBooleanTF{#5} { \hat{\MyAc{\mathcal{R}}}_{{#6}}{}^{\hat{#7}}{}_{\hat{#8}} }{ \hat{\MyAc{\mathcal{R}}}_{{#6}}{}^{{#7}}{}_{{#8}} } }
  { \MyAc{\mathcal{R}}_{{#6}}{}^{{#7}}{}_{{#8}} } }

\DeclareDocumentCommand{\kf}{ t. t, t- s s m m }{
  \RenewDocumentCommand\MyAc{ m }{##1}
  \IfBooleanT{#1}{\RenewDocumentCommand\MyAc{ m }{ \mathring{##1} } }
  \IfBooleanT{#2}{\RenewDocumentCommand\MyAc{ m }{ \tilde{##1} } }
  \IfBooleanT{#3}{\RenewDocumentCommand\MyAc{ m }{ \bar{##1} } }
  \IfBooleanTF{#4}
  { \IfBooleanTF{#5} { \hat{\MyAc{\bm{\mathcal{K}}}}^{\hat{#6}}{}_{\hat{#7}} }{ \hat{\MyAc{\bm{\mathcal{K}}}}^{{#6}}{}_{{#7}} } }
  { \MyAc{\bm{\mathcal{K}}}^{{#6}}{}_{{#7}} } }
\DeclareDocumentCommand{\ko}{ t. t, t- s s m m m }{
  \RenewDocumentCommand\MyAc{ m }{##1}
  \IfBooleanT{#1}{\RenewDocumentCommand\MyAc{ m }{ \mathring{##1} } }
  \IfBooleanT{#2}{\RenewDocumentCommand\MyAc{ m }{ \tilde{##1} } }
  \IfBooleanT{#3}{\RenewDocumentCommand\MyAc{ m }{ \bar{##1} } }
  \IfBooleanTF{#4}
  { \IfBooleanTF{#5} { \hat{\MyAc{\mathcal{K}}}_{\hat{#6}}{}^{\hat{#7}}{}_{\hat{#8}} }{ \hat{\MyAc{\mathcal{K}}}_{{#6}}{}^{{#7}}{}_{{#8}} } }
  { \MyAc{\mathcal{K}}_{{#6}}{}^{{#7}}{}_{{#8}} } }

\DeclareDocumentCommand{\tf}{ t. t, t- s s m }{
  \RenewDocumentCommand\MyAc{ m }{##1}
  \IfBooleanT{#1}{\RenewDocumentCommand\MyAc{ m }{ \mathring{##1} } }
  \IfBooleanT{#2}{\RenewDocumentCommand\MyAc{ m }{ \tilde{##1} } }
  \IfBooleanT{#3}{\RenewDocumentCommand\MyAc{ m }{ \bar{##1} } }
  \IfBooleanTF{#4}
  { \IfBooleanTF{#5} { \hat{\MyAc{\bm{\mathcal{T}}}}^{\hat{#6}} }{ \hat{\MyAc{\bm{\mathcal{T}}}}^{{#6}} } }
  { \MyAc{\bm{\mathcal{T}}}^{{#6}} } }
\DeclareDocumentCommand{\tt}{ t. t, t- s s m m m }{
  \RenewDocumentCommand\MyAc{ m }{##1}
  \IfBooleanT{#1}{\RenewDocumentCommand\MyAc{ m }{ \mathring{##1} } }
  \IfBooleanT{#2}{\RenewDocumentCommand\MyAc{ m }{ \tilde{##1} } }
  \IfBooleanT{#3}{\RenewDocumentCommand\MyAc{ m }{ \bar{##1} } }
  \IfBooleanTF{#4}
  { \IfBooleanTF{#5} { \hat{\MyAc{\mathcal{T}}}_{\hat{#6}}{}^{\hat{#7}}{}_{\hat{#8}} }{ \hat{\MyAc{\mathcal{T}}}_{{#6}}{}^{{#7}}{}_{{#8}} } }
  { \MyAc{\mathcal{T}}_{{#6}}{}^{{#7}}{}_{{#8}} } }
\newcommand{\tor}{\mathcal{T}}

\NewDocumentCommand{\MyLe}{}{}
\DeclareDocumentCommand{\PG}{ s s O{\Pi} m m m }{
  \RenewDocumentCommand\MyLe{}{\Gamma}
  \IfBooleanT{#1}{\RenewDocumentCommand{\MyLe}{}{ \bt{ }{ }{ } }}
  \IfBooleanT{#2}{\RenewDocumentCommand{\MyLe}{}{\Ag}}
  { {#3}_{\MyLe}{}^{#4}{}_{#5}{}^{#6} }
}

\newcommand{\Riem}{\operatorname{Riem}}
\newcommand{\Ric}{\operatorname{Ric}}

\newcommand{\beq}{\begin{equation}}
\newcommand{\eeq}{\end{equation}}
\newcommand{\ber}{\begin{eqnarray}}
\newcommand{\eer}{\end{eqnarray}}

\NewDocumentCommand{\tak}{ s m m}{
  \IfBooleanTF{#1}{ \big( {#2} \big) \big[ {#3} \big] }
              { \big( {#2} \big] \big[ {#3} \big) }
}

\newcommand{\dn}[2]{{\mathrm{d}}^{#1}\!{#2}\;}

\newcommand*{\de}[1]{\mathop{\mathrm{d}#1}\nolimits}
\newcommand{\der}[2]{\frac{\de{#1}}{\de{#2}}}

\newcommand\UTFSM{Departamento de F\'isica, Universidad T\'{e}cnica Federico Santa Mar\'\i a\\ Casilla 110-V, Valpara\'iso, Chile}
\newcommand\UTFSMmat{Departamento de Matem\'aticas, Universidad T\'{e}cnica Federico Santa Mar\'\i a\\ Casilla 110-V, Valpara\'iso, Chile}
\newcommand\CCTVal{Centro Cient\'ifico Tecnol\'ogico de Valpara\'iso\\ Casilla 110-V, Valpara\'\i so, Chile}

\date{\today}
\title{}
\hypersetup{
 pdfauthor={Oscar Castillo-Felisola},
 pdftitle={},
 pdfkeywords={},
 pdfsubject={},
 pdfcreator={Emacs 26.1 (Org mode 9.2.5)}, 
 pdflang={English}}
\begin{document}

\title{Emergent metric and geodesic analysis in cosmological solutions of (torsion-free) Polynomial Affine Gravity}

\author{Oscar Castillo-Felisola}
\email{o.castillo.felisola@gmail.com}
\affiliation{\UTFSM}
\affiliation{\CCTVal}

\author{Jos\'e Perdiguero}
\affiliation{\UTFSM}

\author{Oscar Orellana}
\affiliation{\UTFSMmat}

\author{Alfonso R. Zerwekh}
\affiliation{\UTFSM}
\affiliation{\CCTVal}

\date{\today}

\begin{abstract}
Starting from an affinely connected space, we consider a model of
gravity whose fundamental field is the connection. We build up the
action using as sole premise the invariance under diffeomorphisms, and
study the consequences of a cosmological ansatz for the affine
connection in the torsion-free sector. Although the model is built
without requiring a metric, we show that the nondegenerated Ricci
curvature of the affine connection can be interpreted as an
\emph{emergent} metric on the manifold. We show that there exists a
parametrization in which the \((r,\varphi)\)-restriction of the
geodesics coincides with that of the Friedman--Robertson--Walker
model.  Additionally, for connections with nondegenerated Ricci we are
able to distinguish between space-, time- and null-like self-parallel
curves, providing a way to differentiate \emph{trajectories} of
massive and massless particles.
\end{abstract}


\maketitle

\section{Introduction}
\label{sec:intro}
General Relativity was proposed by A.~Einstein as an attempt to
compatibilise the gravitational interactions with the postulates of
special relativity
\cite{einstein15_grund_allgem_relat_und_anwen,einstein15_zur_allgem_relat,einstein15_erklar_perih_merkur_aus_allgem_relat,einstein15_feldg_gravit,einstein16_grund_allgem_relat}. The
ground-breaking idea behind the proposal was the interpretation of the
gravitational interaction as the effect of properties of the
spacetime, represented by a nontrivial geometry. The spacetime is
modelled by a Riemannian manifold, whose geometric properties are
determined by the metric tensor, and therefore it is the natural field
describing the dynamics of the spacetime. The properties of the
\emph{matter} distribution are encoded in the energy-momentum
tensor. Einstein's field equations are the extrema of the
Einstein--Hilbert action when varied with respect to the metric.

General Relativity has been tested extensively with magnificent
agreement with the experimental data, as one can appreciate in the
excellent review Ref. \cite{will14_confr_between_geren_relat}. The most
recent triumph of the theory was the direct measurement of
gravitational waves by the LIGO-Virgo collaborations
\cite{abbott16_gw151,abbott17_gravit_waves_gamma}.

Although General Relativity is, by far, the most successful theory of
gravitational interactions, there is an increasing interest in
alternative models of gravity, particularly driven for the lack of a
complete framework of quantum
gravity~\cite{dewitt67_quant_theor_gravit_I,dewitt67_quant_theor_gravit_II,deser74_one_loop_diver_quant_einst_maxwel_field,deser74_nonren_quant_dirac_einst_system,hooft74_one_loop_diver_theor,ashtekar86_new_variab_class_quant_gravit,ashtekar87_new_hamil_formul_gener_relat},
and the necessity of hypothesising a dark sector that accounts for
approximately 96\% of the energy content of the
Universe~\cite{zwicky37_masses_nebul_clust_nebul,rubin70_rotat_androm_nebul_from_spect,sofue01_rotat_curves_spiral,riess98_obser_eviden_from_super_accel,perlmutter99_measur_oemeg_lambd_from_high_redsh_super}.

The existence of these problems is a signal of \emph{new physics}, and
their solutions require either including new fields or changing the
gravitational theory. The latter suggests that Einstein's theory is an
effective theory of gravity and, therefore, one may consider
alternative models. Among the generalisations one encounters for
example: the Einstein--Cartan theory, which extends General Relativity
by allowing a non-symmetric connection, but considers the same
action~\cite{cartan22_sur_une_de_la_notion,cartan23_sur_les_connex_affin_et,cartan24_sur_les_connex_affin_et,cartan25_sur_les_connex_affin_et};
models with extra dimensions, firstly proposed by T.~Kaluza and
O.~Klein~\cite{kaluza21_probl_unity_physic,klein26_quant_theor_five_dimen_theor}; Lovelock
models, which are build under the same premises than General
Relativity, but in any dimension~\cite{lovelock71_einst_tensor_its};
the metric-affine models, in which the conditions of metricity and
vanishing torsion are generally dropped~\cite{hehl95_metric_affin_gauge_theor_gravit}; the
Lovelock--Cartan model, which are the extension of Lovelock models
with a non-symmetric connection~\cite{mardones91_lovel}; and many
others.

Inspired by the fact that fundamental interactions (other than
gravity) are described by gauge theories whose dynamical field is a
connection, seems reasonable to search for a model of gravity
described solely by a connection. The first affine model of gravity
was proposed by Sir~A.~Eddington, who considered an action
defined by the square root of the Ricci tensor~\cite{eddington23}
(See also Ref.~\cite{schroedinger50_space}). Moderns attempts to
describe gravity as a theory for the affine connection have been
proposed in
Refs.~\cite{kijowski78_new_variat_princ_gener_relat,krasnov06_renor_non_metric_quant_gravit,krasnov07_non_metric_gravit,krasnov08_non_metric_gravit_I,krasnov08_non_metric_gravit_II,krasnov11_pure_connec_action_princ_gener_relat,poplawski07_unified_purel_affin_theor_gravit_elect,poplawski07_nonsy_purel_affin,poplawski14_affin_theor}.

Recently, a novel model has been proposed, dubbed Polynomial Affine
Gravity, which is built out with polynomial terms of the irreducible
components of the
connection~\cite{castillo-felisola15_polyn_model_purel_affin_gravit,castillo-felisola18_einst_gravit_from_polyn_affin_model,castillo-felisola18_beyond_einstein,castillo-felisola18_cosmol},
assuming invariance under the group of diffeomorphims and no explicit
use of a metric, i.e. even when the spacetime is metric this field
plays no role in the mediation of gravitational interactions.\footnote{In General Relativity the metric plays a double role: it is the
instrument that serves to measure distances, and also it is the field
that mediates gravity. The idea behind metric-affine models is that
those roles are played by different fields, but both fields are
dynamic. In our construction, we manage to build a model of gravity
without the need of an instrument that measures distances.}
The action of Polynomial Affine Gravity has very interesting features:
(i) It is power-counting renormalisable;\footnote{We highlight that this is a necessary but not sufficient
condition for the model to be renormalisable.} (ii) No other term can
be added, i.e. it is not possible to add counter-terms; (iii) All the
couplings are dimensionless, suggesting that the model is a conformal
theory at tree level; (iv) The torsion-free sector is a consistent
truncation compatible with General Relativity, i.e. it passes the
classical test of gravity; and (v) The structure of the model yields
no three-point graviton vertices, which might allow to overcome the
no-go theorems found in
Refs. \cite{mcgady14_higher_spin_massl_s_dimen,camanho16_causal_const_correc_to_gravit}.\footnote{Regarding the aforementioned no-go theorem, a similar
phenomenon happens in the case of a massive spin-1 field coupled to a
non-Abelian gauge field. A necessary condition for making the theory
consistent with perturbative unitarity is the absence of a three-point
vetex for the massive field
\cite{zerwekh13_quant_chrom_massiv_vector_field_adjoin_repres}.}

There are several conceptual subtleties one has to rethink about when
working with an affine model. The most recurrent question is: How do
we measure distances if the model lacks a metric? In this paper we aim
to broaden current knowledge on those subtleties. With this in mind,
in Sec. \ref{sec:PAG_model} we briefly review the polynomial affine model
of gravity, and re-formulate the model in terms of geometrical objects
with simpler interpretation. In addition, we argue that the limit of
zero torsion is well defined, i.e. it is a consistent \emph{truncation} of
the model, and show that the field equations on this sector are a
known generalisation of the Einstein field equations. Then, in
Sec. \ref{sec:cosm_sol} we proceed---by restricting ourselves to the
torsion-free sector---to find cosmological solutions of the field
equations \cite{castillo-felisola18_beyond_einstein}, that extend the
results reported in Ref. \cite{castillo-felisola18_cosmol}. In the
absence of a metric there is no concept of geodesic, however the
concept of self-parallel curve is still valid. Assuming that the
trajectories of \emph{free falling} test particles are self-parallel
curves, we analyse them in Sec.\ref{sec:self-parallel} and show that, there
exist a parametrisation in which the \((r,\varphi)\)-restriction of
the equations is nothing but the expected from the
Friedman--Robertson--Walker model. At this point, the obstruction is
that without a metric it is not possible to differentiate between
trajectories of massive and massless particles. In
Sec. \ref{sec:ricci_metric} we show that under certain conditions the Ricci
tensor is a well-behaved (emergent) metric,\footnote{We prefer to call this metric ``emergent'', because it is a
derived instead of a fundamental geometrical object.} allowing to define \emph{space-like},
\emph{time-like} or \emph{light-like} vectors, and providing the arena for an
affine definition of Einstein manifolds. Then, in Sec. \ref{sec:redshift}
we mention how some basic cosmological quantities are defined in terms
of the parametric functions of the connection. In Sec. \ref{sec:discussion}
we conclude with a discussion of the results. For the sake of
completeness we include some appendixes. In Appendix \ref{app:build_PAG} we
review the dubbed \emph{dimensional analysis} that allowed us to build the
action. In Appendix \ref{app:field_equations} we show the explicit
contribution of each term of the action to the field equations.

\section{The model of Polynomial Affine Gravity}
\label{sec:PAG_model}
The polynomial affine gravity is an alternative theory of gravitation,
whose sole fundamental field is the affine connection,
\(\ct*{}{}{}\). Notice that without the use of a metric, one
calculates the curvature and the Ricci tensors, but not the curvature
scalar. Therefore, it is not possible to write an equivalent action to
the Einstein--Hilbert action.

Our goal is to build up the most general action which is invariant
under the group of diffeomorphims. Firstly, one could try to use the
connection as a whole, however this yields no interesting models,
since the obtained terms are topological invariants,\footnote{All other possible terms are related to these, up to boundary
terms.}
\begin{equation}
  S[\ct*{ }{ }{ }] = \int \Big( a_1 \ri*{\mu\nu}{\alpha}{\beta} \ri*{\lambda\rho}{\beta}{\alpha} + a_2 \ri*{\mu\nu}{\alpha}{\alpha} \ri*{\lambda\rho}{\beta}{\beta} \Big) \de{V}^{\mu\nu\lambda\rho},
\end{equation}
which are the four-dimensional Pontryagin density and the product of
(generalised) two-dimensional Pontryagin densities. In the above
equation we have introduced the \emph{natural} volume form, defined as the
wedge product of the coordinates, i.e.
\(\de{V}^{\alpha\beta\gamma\delta} = J(x) \de{x}^\alpha \wedge
\de{x}^\beta \wedge \de{x}^\gamma \wedge \de{x}^\delta\) for an
arbitrary nonvanishing---within a chart---function \(J(x)\).

The irreducible components of the connection, used to
build the action, are defined as\footnote{In a model with a metric, the connection can always be
decomposed into the Levi-Cività component and the distorsion. The
upper index of the later can be lowered with the metric, and since the
distorsion is a tensor, it decomposes according with the Young
projection, i.e. it has a completely symmetric and antisymmetric parts
plus a component with mixed symmetries. A detailed analysis of the
irreducible components of the connection in the metric case can be
found in
Ref. \cite{iosifidis19_metric_affin_gravit_cosmol_torsion,iosifidis19_exact_solvab_connec_metric_affin_gravit}. The
central difference with our approach is that without the use of a
metric, only the lower indices can be decomposed through Young
projection, as shown in Eq. \eqref{eq:conn_decomp}.}
\begin{equation}
  \begin{aligned}
    \ct*{\mu}{\lambda}{\nu} 
    & = \ct*{(\mu}{\lambda}{\nu)} + \ct*{[\mu}{\lambda}{\nu]} \\
    & = \ct{\mu}{\lambda}{\nu} + \epsilon_{\mu\nu\sigma\kappa}\tor^{\lambda,\sigma\kappa} + \Ag_{[\mu}\delta^\lambda_{\nu]}
    \\
    & = \ct{\mu}{\lambda}{\nu} + \bt{\mu}{\lambda}{\nu} + \delta^\lambda_{[\mu} \Ag_{\nu]},
  \end{aligned}
  \label{eq:conn_decomp}
\end{equation}
where we have first separated the symmetric and antisymmetric part in
the lower indices (the latter is nothing but the torsion of the
connection), in the second line we have renamed the symmetric part of
the affine connection \(\ct*{(\mu}{\lambda}{\nu)} \equiv
\ct{\mu}{\lambda}{\nu}\) and provided a reparametrisation of the
torsion in terms of its trace (\(\Ag_\mu\)) and a \emph{dual} of a
Curtright-like tensor,\footnote{This was the parametrisation of the decomposition utilised in
Refs. \cite{castillo-felisola15_polyn_model_purel_affin_gravit,castillo-felisola18_einst_gravit_from_polyn_affin_model,castillo-felisola18_beyond_einstein}.} while in the third line the \(\bt{}{}{}\)
field is just the traceless part of the torsion. All these elements
transform as tensors under diffeomorphism, except for the symmetric
part of the affine connection, \(\ct{\mu}{\lambda}{\nu}\), which
consequently must be included in the action solely through the
covariant derivative.\footnote{Notice that under similar requirements but in odd dimensions,
there are Chern--Simons terms, in which the connection enters
explicitly in the action. See for example the three-dimensional
construction in
Ref. \cite{castillo-felisola15_polyn_model_purel_affin_gravit}.}

In order to build the most general action using the irreducible
fields, \(\ct{\mu}{\lambda}{\nu}\), \(\bt{\mu}{\lambda}{\nu}\) and
\(\Ag_\mu\), the strategy to write down the action is to define the
most general scalar density, where the dynamics is given by the
covariant derivative with respect to the symmetric part of the
connection.

Using the second parametrisation in Eq. \eqref{eq:conn_decomp}, the most
general action is
\begin{widetext}
  \begin{equation}
    \label{eq:new-action}
    \begin{split}
      S & = \int \de{V}^{\alpha \beta \gamma \delta} \bigg[ B_1
      \ri{\mu\nu}{\mu}{\rho} \bt{\alpha}{\nu}{\beta}
      \bt{\gamma}{\rho}{\delta} + B_2 \ri{\alpha\beta}{\mu}{\rho}
      \bt{\gamma}{\nu}{\delta} \bt{\mu}{\rho}{\nu} + B_3
      \ri{\mu\nu}{\mu}{\alpha} \bt{\beta}{\nu}{\ga} \Ag_\delta + B_4
      \ri{\alpha\beta}{\sigma}{\rho} \bt{\ga}{\rho}{\delta} \Ag_\sigma
      \\
      & \quad + B_5 \ri{\alpha\beta}{\rho}{\rho}
      \bt{\ga}{\sigma}{\delta} \Ag_\sigma + C_1
      \ri{\mu\alpha}{\mu}{\nu} \nabla_\beta \bt{\ga}{\nu}{\delta} +
      C_2 \ri{\alpha\beta}{\rho}{\rho} \nabla_\sigma
      \bt{\ga}{\sigma}{\delta} + D_1 \bt{\nu}{\mu}{\lambda}
      \bt{\mu}{\nu}{\alpha} \nabla_\beta \bt{\ga}{\lambda}{\delta}
      \\
      & \quad + D_2 \bt{\alpha}{\mu}{\beta} \bt{\mu}{\lambda}{\nu}
      \nabla_\lambda \bt{\ga}{\nu}{\delta} + D_3 \bt{\alpha}{\mu}{\nu}
      \bt{\beta}{\lambda}{\ga} \nabla_\lambda \bt{\mu}{\nu}{\delta} +
      D_4 \bt{\alpha}{\lambda}{\beta} \bt{\ga}{\sigma}{\delta}
      \nabla_\lambda \Ag_\sigma + D_5 \bt{\alpha}{\lambda}{\beta}
      \Ag_\sigma \nabla_\lambda \bt{\ga}{\sigma}{\delta}
      \\
      &\quad + D_6 \bt{\alpha}{\lambda}{\beta} \Ag_\ga \nabla_\lambda
      \Ag_\delta + D_7 \bt{\alpha}{\lambda}{\beta} \Ag_\lambda
      \nabla_\ga \Ag_\delta + E_1 \nabla_\rho \bt{\alpha}{\rho}{\beta}
      \nabla_\sigma \bt{\ga}{\sigma}{\delta} + E_2 \nabla_\rho
      \bt{\alpha}{\rho}{\beta} \nabla_\ga \Ag_{\delta}
      \\
      &\quad + F_1 \bt{\alpha}{\mu}{\beta} \bt{\ga}{\sigma}{\delta}
      \bt{\mu}{\lambda}{\rho} \bt{\sigma}{\rho}{\lambda} + F_2
      \bt{\alpha}{\mu}{\beta} \bt{\ga}{\nu}{\lambda}
      \bt{\delta}{\lambda}{\rho} \bt{\mu}{\rho}{\nu} + F_3
      \bt{\nu}{\mu}{\lambda} \bt{\mu}{\nu}{\alpha}
      \bt{\beta}{\lambda}{\ga} \Ag_\delta + F_4
      \bt{\alpha}{\mu}{\beta} \bt{\ga}{\nu}{\delta} \Ag_\mu \Ag_\nu
      \bigg].
    \end{split}
  \end{equation}
\end{widetext}
where the covariant derivation and the curvature are defined with
respect to the symmetric connection, i.e. \(\nabla = \nabla^\Gamma\)
and \(\ri{}{}{}=\ri{}{\Gamma}{}\). The action is defined up to
boundary and topological terms. Although the dropped terms are
relevant when studying global aspects of the model, they do not
contribute to the equations of motion. In order to write down the
action we used a variation of the \emph{dimensional analysis} method
introduced in
Ref. \cite{castillo-felisola18_einst_gravit_from_polyn_affin_model}. Details
of the dimensional analysis are shown in
Appendix \ref{app:build_PAG}.

Interestingly, all of the coupling constant are dimensionless, which
from the view point of Quantum Field Theory is desirable if one is
interested in trying to quantise the model. In addition, as shown by
the dimensional analysis in
Ref. \cite{castillo-felisola18_einst_gravit_from_polyn_affin_model},
there is a finite number of possible terms contributing to the action
(once those ignored in Eq. \eqref{eq:new-action} are included), which we
interpret as a \emph{rigidity} of the model, given than in the hypothetical
scenario of quantisation of Polynomial Affine Gravity all the
counter-terms should have the form of terms already present in the
original action.

\subsection{Field equations}
\label{sec:org866fbcb}

In what follows, we will obtain the field equations of the model using
the standard variational principle. It is well-known that to ensure
the well-posedness of the variational problem no second-class
constraints should be present, or higher derivative terms might appear
in the field equations. In General Relativity, one needs to add the
Gibbons--Hawking--York term to solve this problem. Although the
absence of second-class constraints in Polynomial Affine Gravity has
not been proven yet, the structure of the action in
Eq. \eqref{eq:new-action} suggests that the variational problem is
well-posed.\footnote{Analysis of affine analogues to the Gibbons--Hawking--York term
can be found in
Refs. \cite{parattu16_bound_term_gravit_action_with_null_bound,krishnan17_robin_gravit,krishnan17_neuman_bound_term_gravit,lehner16_gravit_action_with_null_bound,hopfmueller17_gravit_degrees_freed_null_surfac,jubb17_bound_corner_terms_action_gener_relat}.}

Under the assumption that no Gibbons--Hawking--York term is needed,
and since the action contains up to first derivatives of the fields,
the field equations are obtained through the Euler-Lagrange equations,
\begin{equation}
  \begin{aligned}
    \partial_\mu \left(\frac{\partial \Lag}{\partial\left(\partial_\mu{\ct{\nu}{\lambda}{\rho}}\right)}\right)
    - \frac{\partial \Lag}{\partial {\ct{\nu}{\lambda}{\rho}}} & = 0,
    \\
    \partial_\mu \left(\frac{\partial \Lag}{\partial\left(\partial_\mu{\bt{\nu}{\lambda}{\rho}}\right)}\right)
    - \frac{\partial \Lag}{\partial {\bt{\nu}{\lambda}{\rho}}} & = 0,
    \\
    \partial_\mu \left(\frac{\partial \Lag}{\partial\left(\partial_\mu{ \Ag_{\nu}}\right)}\right) -
    \frac{\partial \Lag}{\partial { \Ag_{\nu}}} & = 0.
  \end{aligned}
  \label{euler_lagrange} 
\end{equation}
A simple way of dealing with the field equations was introduced by
Kijowski in Ref. \cite{kijowski78_new_variat_princ_gener_relat}, and
we will reviewed in what follows.\footnote{Our notation is inspired by that of the cited work, but it
might differ in both symbols and signs. We advise to be careful if
you would like to compare the results.}

In the following, we present an alternative form of writing the
Euler--Lagrange equations presented above, in a way that the
calculations are easier to follow. Nonetheless, the explicit
calculations are given in Appendix \ref{app:field_equations}.

\subsubsection{Field equations for the symmetric connection}
\label{sec:field_eq_G}
The canonically conjugated momenta of the connection are defined by
\begin{equation}
  \PG{\mu\nu}{\lambda}{\rho} = \frac{ \partial \Lag }{
    \partial \, (\partial_\mu \ct{\nu}{\lambda}{\rho}) } \equiv \frac{ \partial \Lag }{
    \partial \,  \ct{\mu\nu}{\lambda}{\rho} },
\end{equation}
and since the derivative of the connection appears only in the
curvature tensor, it follows that
\begin{equation}
  \label{eq:def-PG}
  \PG{\mu\nu}{\lambda}{\rho} = \frac{\partial \Lag}{\partial
    \ri{\alpha\beta}{\gamma}{\delta}}
  \frac{\partial \ri{\alpha\beta}{\ga}{\delta}}{\partial
    \ct{\mu\nu}{\lambda}{\rho}} \equiv \PG[z]{\alpha\beta}{\ga}{\delta}
  \frac{\partial \ri{\alpha\beta}{\ga}{\delta}}{\partial
    \ct{\mu\nu}{\lambda}{\rho}}.
\end{equation}
The last term in Eq. \eqref{eq:def-PG} can be calculated explicitly from
the definition of the curvature tensor,
\begin{equation}
  \label{eq:der-R-wrt-pG}
  \frac{\partial \ri{\alpha\beta}{\ga}{\delta}}{\partial
    \ct{\mu\nu}{\lambda}{\rho}} = 4 \, \delta ^\ga_\lambda
  \delta^\mu_{[\alpha} \delta^{(\nu}_{\beta]} \delta^{\rho)}_\delta,
\end{equation}
from which it follows that
\begin{equation}
  \label{eq:PG-from-z}
  \PG{\mu\nu}{\lambda}{\rho} = 2 \PG[z]{[\mu\nu]}{\lambda}{\rho} + 2 \PG[z]{[\mu\rho]}{\lambda}{\nu}.
\end{equation}
The last equation implies that the canonical momenta satisfy the
Jacobi--Bianchi identity,
\begin{equation}
  \PG{[\mu\nu}{\lambda}{\rho]} = 0.
\end{equation}
On the other hand, the second term in the Euler--Lagrange equations
for \(\Gamma\) yields
\begin{equation}
  \frac{\partial \Lag}{\partial \ct{\nu}{\lambda}{\rho}} =
  \frac{\partial \Lag}{\partial \ri{\alpha\beta}{\ga}{\delta}}
  \frac{\partial \ri{\alpha\beta}{\ga}{\delta}}{\partial \ct{\nu}{\lambda}{\rho}}.
\end{equation}
Once again the last term can be calculated from the definition of
curvature
\begin{equation}
  \frac{\partial \ri{\alpha\beta}{\ga}{\delta}}{\partial
    \ct{\nu}{\lambda}{\rho}} = 4 \Big[  \delta^\ga_\lambda
  \delta^{(\nu}_{[\alpha} \ct{\beta]}{\rho)}{\delta} + \delta^\rho_\delta
  \delta^{(\nu}_{[\beta} \ct{\alpha]}{\ga)}{\lambda} \Big],
\end{equation}
and then,
\begin{align}
  \frac{ \partial \Lag }{ \partial \ct{\nu}{\lambda}{\rho} }
  & =
    2 \, \Big[
    \PG[z]{[\nu\beta]}{\lambda}{\delta} \ct{\beta}{\rho}{\delta}
    + \PG[z]{[\rho\beta]}{\lambda}{\delta} \ct{\beta}{\nu}{\delta}
    \notag
  \\
  & \quad
    + \PG[z]{[\beta\nu]}{\ga}{\rho} \ct{\beta}{\ga}{\lambda}
    + \PG[z]{[\beta\rho]}{\ga}{\nu} \ct{\beta}{\ga}{\lambda}
    \Big] \notag
  \\
  & =
    \frac{1}{2} \PG{\nu\beta}{\lambda}{\delta} \ct{\beta}{\rho}{\delta}
    + \frac{1}{2} \PG{\rho\beta}{\lambda}{\delta}
    \ct{\beta}{\nu}{\delta}
    + \PG{\beta\nu}{\ga}{\rho} \ct{\beta}{\ga}{\lambda} \notag
  \\
  & =
    - \PG{\mu\nu}{\lambda}{\delta} \ct{\mu}{\rho}{\delta}
    - \PG{\mu\rho}{\lambda}{\delta}
    \ct{\mu}{\nu}{\delta}
    + \PG{\mu\nu}{\ga}{\rho} \ct{\mu}{\ga}{\lambda}
    .
    \label{eq:der-L-wrt-G}
\end{align}
Therefore, the field equations for the symmetric part of the
connection are
\begin{equation}
  \label{eq:EOM-G}
  \nabla_\mu \PG{\mu\nu}{\lambda}{\rho} = \frac{\partial^* \Lag}{\partial \ct{\nu}{\lambda}{\rho} }.
\end{equation}
In this last equation, we have used the fact that the canonical
momenta are densities, thus there are two term (which seems to be
missing above) that cancel themselves.\footnote{The covariant derivative of this tensor density is given by the
expression,
\begin{align*}
  \nabla_\sigma \PG{\mu\nu}{\lambda}{\rho}
  & = \partial_\sigma \PG{\mu\nu}{\lambda}{\rho}
    + \ct{\sigma}{\mu}{\tau} \PG{\tau\nu}{\lambda}{\rho}
    + \ct{\sigma}{\nu}{\tau} \PG{\mu\tau}{\lambda}{\rho}
  \\
  &
    - \ct{\sigma}{\tau}{\lambda} \PG{\mu\nu}{\tau}{\rho}
    + \ct{\sigma}{\rho}{\tau} \PG{\mu\nu}{\lambda}{\tau}
    - \ct{\sigma}{\tau}{\tau} \PG{\mu\nu}{\lambda}{\rho},
\end{align*}
then, when one contract the \(\sigma\) and \(\mu\) indices, the second
and sixth terms in the right-hand side cancel each other.} The asterisk on the
right-hand side of Eq. \eqref{eq:EOM-G} denotes the partial derivative
with respect to the connection that is not contained in the curvature
tensor.

Notice that there are seven term in which the symmetric part of the
connection enters through the curvature tensor, while it enters
through the covariant derivative of the tensors \(\bt{}{}{}\) and \(\Ag\) in
eleven terms. Noteworthily, there are only two terms in which the
symmetric part of the connection enters in both ways, these are the
terms in the action with coupling constants \(C_1\) and
\(C_2\). Furthermore, the terms in the action with couplings \(C_1\) and
\(C_2\) are the only which are linear in either \(\bt{}{}{}\) and \(\Ag\)
fields. Hence, these are the terms which could possibly contribute
to the field equations in the sector of vanishing torsion, i.e.,
\(\Ag \to 0\) and \(\bt{}{}{} \to 0\).

\subsubsection{Field equations for the \(\bt{}{}{}\) field}
\label{sec:field_eq_B}
Using the relations
\begin{equation}
  \frac{ \partial \nabla_\alpha \bt{\beta}{\gamma}{\delta} }{ \partial
    ( \partial_\mu \bt{\nu}{\lambda}{\rho} ) }
  =
  2 \delta^\mu_\alpha \delta^\gamma_\lambda \delta^{[\nu}_\beta \delta^{\rho]}_\delta
\end{equation}
and
\begin{equation}
  \frac{ \partial \nabla_\alpha \bt{\beta}{\ga}{\delta} }{ \partial
    \bt{\nu}{\lambda}{\rho} }
  =
  - 2 \ct{\alpha}{[\nu}{\beta} \delta^{\rho]}_\delta
  \delta^\ga_\lambda
  - 2 \ct{\alpha}{[\rho}{\delta} \delta^{\nu]}_\beta
  \delta^\ga_\lambda
  + 2 \ct{\alpha}{\ga}{\lambda} \delta^{[\nu}_\beta \delta^{\rho]}_\delta
\end{equation}
it is straightforward to show that the equations of motion for the
\(\bt{}{}{}\) field are
\begin{equation}
  \label{eq:EOM-B}
  \nabla_\mu \PG*{\mu\nu}{\lambda}{\rho} = \frac{\partial \Lag}{\partial \bt{\nu}{\lambda}{\rho} }.
\end{equation}

\subsubsection{Field equations for the \(\Ag\) field}
\label{sec:field_eq_A}
The field equations for the \(\Ag\) field are simpler to calculate, but
a rigorous  approach as in the previous sections, yields the field
equations,
\begin{equation}
  \label{eq:EOM-A}
  \nabla_\mu \PG**{\mu\nu}{}{} = \frac{\partial \Lag}{\partial \Ag_{\nu} }.
\end{equation}

\subsection{The torsion-less limit}
\label{sec:torsion_less}
An important result, obtained in
Ref. \cite{castillo-felisola18_einst_gravit_from_polyn_affin_model}, is
that within the torsion-less sector of the conection, the field
equations admits all vacuum solutions of Einstein's gravity, as
solutions of the polynomial affine gravity.

The sector of vanishing torsion is equivalent to the limit
\(\Ag \to 0\) and \(\bt{}{}{} \to 0\). Clearly, such limit cannot be taken at
the level of the action, but at the equation of motions. In the
Appendix \ref{app:field_equations}, the explicit field equations are
shown from Eq. \eqref{eq:first_feq} to Eq. \eqref{eq:last_feq}, and it
can be checked that the mentioned limit is well-defined. In the
torsion-free sector, the only nontrivial field equations are
Eqs. \eqref{eq:C1} and \eqref{eq:C2}, i.e.
\begin{equation}
  \label{eq:nontrivial_feq}
  \nabla_\mu \left( \ri{\sigma \alpha}{\sigma}{\lambda}
    \de{V}^{\mu \nu \rho \alpha} + C \ri{\alpha \beta}{\sigma}{\sigma}
    \delta^{\mu}_{\lambda} \de{V}^{\nu \rho \alpha \beta} \right) = 0,
\end{equation}
where \(C\) is the ratio of the original parameters of the model, \(C
= \tfrac{C_2}{C_1}\). The second term in Eq. \eqref{eq:nontrivial_feq}
is proportional to the trace of the curvature 2-form, which vanishes
in General Relativity. Also, the volume form is not necessarily
compatible with the connection. However, if one restricts oneself to
connections that preserve a volume form, such connections are dubbed
\emph{equiaffine} connections \cite{nomizu94_affin}, the trace of the
curvature 2-form is ensured and the field equations simplify further
to
\begin{equation}
  \label{eq:simple_eom}
  \nabla_{[\mu} \ri{\nu] \lambda}{}{} = 0, 
\end{equation}
which can be written as
\begin{equation}
  \label{eq:harm_curv}
  \nabla_\rho \ri{\mu\nu}{\rho}{\lambda} = 0,
\end{equation}
after using the second Bianchi
identity. Equations \eqref{eq:simple_eom} and \eqref{eq:harm_curv} are
part of a set of well-known generalisations of Einstein's field
equations (see for example Chap. 16 of Ref. \cite{besse07_einst}).

Particularly, Eq. \eqref{eq:harm_curv} can be obtained as the field
equation for the connection of a gravitational Yang--Mills theory,
\begin{equation}
  \label{eq:SKY_action}
  S_{\textsc{sky}} = \int \dn{4}{x} \sqrt{g} \ri{\mu\nu}{\lambda}{\rho} \ri{}{\mu\nu\rho}{\lambda}.
\end{equation}
The above model is known as Stephenson--Kilmister--Yang (or SKY for
short)
\cite{stephenson58_quadr_lagran_gener_relat,kilmister61_use_alg_struct_phys,yang74_integ_formal_gauge_field},
but its structure requires the inclusion of the metric in order to
build the action. Therefore, besides Eq. \eqref{eq:harm_curv} there is a
field equation for the metric, and it spoils desirable features of the
Stephenson--Kilmister--Yang model
\cite{pavelle75_unphy_solut_yangs_gravit_field_equat,thompson75_geomet_degen_solut_kilmis_yang_equat}.\footnote{Notice for example that in terms of the connection,
Eq. \eqref{eq:simple_eom} is a set of second order differential
equations, while if we would interprete them as equations for the
metric, become a set of third order differential equations.}

The solutions to Eq. \eqref{eq:simple_eom} are classified in three
categories: (i) Ricci-flat connections, \(R_{\mu\nu} = 0\); (ii)
connections with parallel Ricci, \(\nabla_\lambda R_{\mu\nu} = 0\);
and (iii) connections with harmonic curvature, \(\nabla_\rho
\ri{\mu\nu}{\rho}{\lambda} = 0\). Interestingly, among the possible
solutions of Eq. \eqref{eq:simple_eom} one encounters the vacuum
solutions to the Einstein field equations, dubbed Einstein
manifolds. A key difference in Polynomial Affine Gravity is that
unlike General Relativity the cosmological constant appears as an
integration constant. The same feature occurs in other generalisations
of General Relativity, e.g. unimodular gravity
\cite{einstein19_spiel_gravit_im_auf_ber}.

\section{Cosmological solutions on the torsion-free sector}
\label{sec:cosm_sol}
In order to solve the equations \eqref{eq:simple_eom} one
proceeds---just as in General Relativity---by giving an ansatz
compatible with the symmetries of the problem. Using the Lie
derivative, we have found the most general torsion-free connection
compatible with the cosmological
principle \cite{castillo-felisola18_beyond_einstein}. Since we shall
restrict ourselves to the torsion-free sector, the nonvanishing
coefficients of the connection are
\begin{equation}
  \label{eq:cosm_conn}
  \begin{aligned}
    \ct{t}{t}{t} & = f(t),
    &
    \ct{i}{t}{j} & = g\left(t\right)S_{ij},
    \\
    \ct{t}{i}{j} & = h\left(t\right)\delta^{i}_{j},
    &
    \ct{i}{j}{k} & = \gamma_{i}{}^{j}{}_{k},
  \end{aligned}
\end{equation} 
where \(f\), \(g\) and \(h\) are functions of time, while \(S_{ij}\) and
\(\gamma_{i}{}^{j}{}_{k}\) are the three-dimensional rank two
symmetric tensor and connection compatible with isotropy and
homogeneity, defined by
\begin{align}
  S_{ij} =
  \begin{pmatrix} 
    \frac{1}{1 - \kappa r^2} & 0 & 0 \\
    0 & r^2 & 0 \\
    0 & 0 & r^2 \sin^2\theta 
  \end{pmatrix},
\end{align}
and
\begin{equation}
  \begin{aligned}
    \gamma_{r}{}^{r}{}_{r} & = \frac{\kappa r}{1 - \kappa r^2},
    &
    \gamma_{\theta}{}^{r}{}_{\theta} & = \kappa r^3 - r,
    \\
    \gamma_{\varphi}{}^{r}{}_{\varphi} & = \left(\kappa r^3 - r\right)\sin
    ^2\theta,
    &
    \gamma_{r}{}^{\theta}{}_{\theta} & = \frac{1}{r},
    \\
    \gamma_{\varphi}{}^{\theta}{}_{\varphi} & = - \cos \theta \sin \theta,
    &
    \gamma_{r}{}^{\varphi}{}_{\varphi} & =\frac{1}{r},
    \\
    \gamma_{\theta}{}^{\varphi}{}_{\varphi} & = \frac{\cos \theta}{\sin
      \theta}.
    & &
  \end{aligned}
\end{equation}

With the connection above, one can calculate the curvature,
\begin{equation}
  \begin{aligned}
    \ri{t i}{t}{j} = - \ri{i t}{t}{j} & = \left( \dot{g} + (f - h) g
                                        \right) S_{ij}
    \\
    \ri{m t }{n}{t} = - \ri{t m}{n}{t} & = - \delta_m^n \left( \dot{h} +
                                         h^2 - f h \right)
    \\
    \ri{m i}{n}{j} = - \ri{i m}{n}{j} & = \left( g h - \kappa\right)
                                        \delta_m^n S_{ij},
  \end{aligned}
\label{eq:Riem}
\end{equation}
and the Ricci tensor,
\begin{equation}
  \begin{aligned}
    \ri{t t}{}{} & = - 3 \left( \dot{h} + h^2 - f h \right)
    \\
    \ri{ij}{}{} & = \left(\dot{g} + (f + h) g + 2 \kappa \right) S_{ij}.
  \end{aligned}
  \label{eq:Ric}
\end{equation}

The covariant derivative of the Ricci yields
\begin{widetext}
  \begin{equation}
    \begin{aligned}
      \nabla_t \ri{t t}{}{} & =
      \partial_t \ri{t t}{}{} - 2 \ct{t}{t}{t} \ri{t t}{}{}
      =
      - 3 \left[ \ddot{h} + 2 h \dot{h} - 3
        f \dot{h} - \dot{f} h - 2 f h (f - h)  \right],
      \\
      \nabla_t \ri{i j}{}{} & =
      \partial \ri{i j}{}{} - 2 \ct{t}{k}{i}  \ri{k j}{}{}
      =
      \left[ \ddot{g} + \dot{f} g + \dot{h}
        g + (f - h) \dot{g} - 2 (f + h) g h - 4 \kappa h \right] S_{ij},
      \\
      \nabla_i \ri{t j}{}{} & =
      - \ct{i}{k}{t} \ri{k j}{}{} - \ct{i}{t}{j} \ri{t t}{}{}
      =
      \left[ 3 g \dot{h} - \dot{g} h - 4 f
      g  + 2 g h^2 - 2 \kappa h \right] S_{i j}.
    \end{aligned}
    \label{eq:DRic}
  \end{equation}
  Finally, the harmonic curvature expression has a single independent
  component,
  \begin{equation}
    \begin{aligned}
      \nabla_{[t} \ri{i] j}{}{} & =
      \partial_t \ri{i j}{}{} - \ct{t}{k}{i} \ri{k j}{}{} +
      \ct{i}{t}{j} \ri{t t}{}{} 
      \\
      & = \left( \ddot{g} + \dot{f} g + f
        \dot{g} - 2 g \dot{h} + 2 f g h - 4 g h ^2 - 2 \kappa h
      \right) S_{ij}.
    \end{aligned}
    \label{eq:HRiem}
  \end{equation}
\end{widetext}

In the remain of this section we will solve the
Eq. \eqref{eq:simple_eom}. Firstly, notice that the Levi-Cività
connection from Friedman--Robertson--Walker models is obtained from
Eq. \eqref{eq:cosm_conn} by setting \(f = 0\), \(g = a \dot{a}\) and
\(h = \frac{\dot{a}}{a}\), implying that all (vacuum) cosmological
models in General Relativity are in the space of solutions of
Polynomial Affine Gravity. Moreover, it was shown in
Ref. \cite{castillo-felisola18_cosmol}, that within this space of
solutions, suitable deviations from the vacuum
Friedman--Robertson--Walker connection mimic the behaviour of matter
content, even though Eq. \eqref{eq:simple_eom} aim to describe
geometric properties of the manifold exclusively. Secondly, our
classification of solutions---into Ricci-flat, parallel Ricci and
harmonic curvature---is hierarchic, in the sense that once a condition
is satisfied, the remaining are satisfied as well. Therefore, in order
to find a proper solution of the parallel Ricci equations, we have to
ensure that the connection is not Ricci-flat; and in order to find a
proper solution of the harmonic curvature equations, neither the
Ricci-flat or parallel Ricci condition should be satisfied.

\subsection{Cosmological solutions with vanishing Ricci}
\label{sec:cosm_ricci_flat}
A first kind of solutions can be found by solving the system of
equations determined by vanishing Ricci. From Eq. \eqref{eq:Ric} the
differential equations to solve are 
\begin{align}
  \dot{h} - (f - h) h & = 0,
  \label{eq:cosm_rf_1}
  \\
  \dot{g} + (f + h) g + 2 \kappa & = 0. 
  \label{eq:cosm_rf_2}
\end{align}
Since \(f\) is not a dynamical function, the system can be solved in
terms of \(f\) (see Ref. \cite{castillo-felisola18_cosmol}),
\begin{align}
  \label{eq:aff_cosm_fr_sol_h}
  h(t) & = \frac{\exp \left( F(t) \right)}{C_h + \int \de{t} \,\exp(F)},
  \\
  \label{eq:aff_cosm_fr_sol_g}
  g(t) & = \exp( - \Sigma(t) ) \left( C_g - 2 \kappa \left( \int \de{t} \exp( \Sigma(t) )
  \right) \right),
\end{align}
where \(F = \int \de{t} \, f\) and \(\Sigma(t) = \int \de{t} \, \left(
f(t) + h(t) \right)\) are integrals of the defining functions, while
\(C_h\) and \(C_g\) are integration constants.

\subsubsection{Friedman--Robertson--Walker-like models}
\label{sec:FRW-like-models}
In particular, Friedman--Robertson--Walker-like models are obtained by
setting \(f = 0\), and besides the trivial
solution---\(h = g = \kappa = 0\)---, yielding a parametric family of
nontrivial solutions,
\begin{equation}
  \begin{aligned}
    g(t) &= \frac{1}{t + C_h} \left( C_g - \kappa (t + C_h)^2 \right) ,
    &
    h(t) &= \frac{1}{t + C_h}.
  \end{aligned}
  \label{eq:sol_cosm_very_simple}
\end{equation}
Unlike in General Relativity---whose sole Ricci-flat cosmological
solution is a flat manifold---, the above solution is not flat in
general, since its curvature tensor has nonvanishing components
\begin{equation}
  \begin{aligned}
    \ri{ti}{t}{j} = - \ri{it}{t}{j} & = - 2 C_g h^2 S_{ij},
    \\
    \ri{mi}{n}{j} = - \ri{i m}{n}{j} & = C_g h^2 S_{ij} \delta_m^n.
  \end{aligned}
\end{equation}

\subsubsection{Case \(h = f\)}
\label{sec:sol_h_eq_f}
In these particular subspace, the solutions to Eqs. \eqref{eq:cosm_rf_1}
and \eqref{eq:cosm_rf_2} are given by
\begin{align}
  f   & = C_h
  &
    h & = C_h
  &
    g & = C_g \exp( - 2 C_h t ) - \frac{\kappa}{C_h}.
\end{align}

\subsubsection{Case \(h = - f\)}
\label{sec:sol_h_eq_mf}
Equations \eqref{eq:cosm_rf_1} and \eqref{eq:cosm_rf_2} are solved by
\begin{align}
  f   & =  - \frac{1}{2t + C_h}
  &
    h & =  \frac{1}{2t + C_h}
  &
    g & = C_g - 2 \kappa t.
\end{align}

\subsubsection{Case \(h = 0\) and a given \(f\)}
\label{sec:sol_h0_f}
In this case, equation \eqref{eq:cosm_rf_1} becomes an identity, and
Eq. \eqref{eq:cosm_rf_2} is solved by
\begin{equation}
  g(t) = \exp( - F(t) ) \left( C_g - 2 \kappa \left( \int \de{t} \exp( F(t) )
  \right) \right).
\end{equation}

\subsubsection{Case \(g = 0\) and a given \(f\)}
\label{sec:sol_g0_f}
In this case, equation \eqref{eq:cosm_rf_2} requires \(\kappa = 0\), and
\(h\) can still be solved for a given function \(f\) as
\begin{equation}
  h(t) = \frac{\exp \left( F(t) \right)}{C_h + \int \de{t} \,\exp(F)}.
\end{equation}

\subsection{Cosmological solutions with parallel Ricci}
\label{sec:cosm_parallel_ricci}
In this section we solve the Eqs. \eqref{eq:DRic}, under the condition
that the Ricci tensor is nonzero, i.e. \(\nabla_\lambda
\ri{\mu\nu}{}{} = 0\) but \(\ri{\mu\nu}{}{} \neq 0\). The strategy to
solve these equations is to propose an ansatz for the Ricci tensor,
and solve for \(f\), \(g\) and \(h\) accordingly. Nonetheless, a
broaden ansatz is useless, thus we focus in the two simpler cases.

\subsubsection{Parallel time-independent Ricci}
\label{sec:parallel_ricci_w_cte_ric}
A time-independent Ricci has the form,
\begin{align}
  \label{eq:constant_ricci}
  \ri{tt}{}{} & = R_1,
  &
    \ri{ij}{}{} & = R_2 S_{ij},
\end{align}
with constant \(R_1\) and \(R_2\). For this proposal of the Ricci
tensor the Eqs. \eqref{eq:DRic} yield the constraints
\begin{align}
  \nabla_t \ri{t t}{}{} & = 0 & \Rightarrow & \, f = 0 \lor R_1 = 0, 
  \\
  \nabla_t \ri{i j}{}{} & = 0 & \Rightarrow & \, h = 0 \lor R_2 = 0,
  \\
  \nabla_i \ri{t j}{}{} & = 0 & \Rightarrow & \, h = 0 \lor (g = 0
                                                       \land \kappa = 0).
\end{align}
Notice that \(h = 0\) implies \(R_1 = 0\), and \(g = 0 \land \kappa =
0\) implies \(R_2 = 0\). Therefore, there is no solution of the
Eqs. \eqref{eq:DRic} for a time-independent nondegenerated Ricci.

Given the above conditions, there are solely two solutions with
degenerated nonvanishing Ricci. Firstly, for vanishing \(h\),
one have
\begin{equation}
  \label{eq:parallel_cte_ricci_h0}
  \begin{aligned}
    h & = 0,
    \\
    f & = \text{undetermined function},
    \\
    g & = \exp(-F) \left( C_g + (R_2 - 2 \kappa) \int \de{t} \, \exp(F)  \right).
  \end{aligned}
\end{equation}
Similarly, for \(g = 0 \land \kappa = 0 \land f = 0\), the solution is
given by
\begin{equation}
  \label{eq:parallel_cte_ricci_g0}
  \begin{aligned}
    h & = \sqrt{\frac{R_1}{3}} \tanh \left( \sqrt{\frac{R_1}{3}} (t - t_*)  \right),
    \\
    f & = 0,
    \\
    g & = 0. 
  \end{aligned}
\end{equation}

\subsubsection{Parallel Ricci with a ``scale factor''}
\label{sec:parallel_ricci_w_sf}
We now consider an ansatz for the Ricci tensor with the form of a
Friedman--Robertson--Walker metric, given that this has the required
symmetries, i.e.,
\begin{align}
  \label{eq:scale_factor_Ricci}
  \ri{t t}{}{} & = - R_1,
  &
    \ri{i j}{}{} & = A(t) S_{i j}.
\end{align}

With the above ansatz, the parallel Ricci equations require,
\begin{equation}
  \begin{aligned}
    \nabla_t \ri{t t}{}{} & = 2 f R_1
    &
    & \Rightarrow \, f = 0 \lor R_1 = 0,
    \\
    \nabla_t \ri{i j}{}{} & = S_{i j} (\dot{A} - 2 h A)
    &
    & \Rightarrow \, A = C_A \exp(2 H),
    \\
    \nabla_i \ri{t j}{}{} & = - S_{i j} (h A - R_1 g)
    &
    & \Rightarrow \, g = \tfrac{h A}{R_1}; \, (R_1 \! \neq \! 0).
  \end{aligned}
  \label{eq:parallel_ricci_w_sf}
\end{equation}

Unlike the previous case, Eqs. \eqref{eq:parallel_ricci_w_sf}, accept a
nondegenerated solution given by
\begin{equation}
  \label{eq:sol_parallel_ricci_w_sf}
  \begin{aligned}
    f & = 0,
    \\
    h & = \sqrt{\frac{R_1}{3}} \tanh \left( \sqrt{\frac{R_1}{3} (t - t_*)} \right),
    \\
    A & = C_A \cosh^2 \left( \sqrt{\frac{R_1}{3} (t - t_*)} \right),
    \\
    g & = \frac{1}{\sqrt{12 R_1}} \sinh \left( \sqrt{\frac{4 R_1}{3} (t - t_*)} \right).
  \end{aligned}
\end{equation}
Notice that degenerated solutions, with \(R_1 = 0\), require either
vanishing \(C_A\) or a constant \(A\) for \(h = 0\), which are part of
previously considered cases.

\subsection{Cosmological solutions with harmonic curvature}
\label{sec:cosm_harm_curv}
\subsubsection{Harmonic curvature with time-independent Ricci}
\label{sec:harm_curv_const_ricci}
For a time-independent Ricci tensor, see Eq. \eqref{eq:constant_ricci},
the harmonic curvature condition, Eq. \eqref{eq:HRiem}, becomes
\begin{equation}
  \nabla_{[t} \ri{i]j}{}{} = - (g R_1 + h R_2) S_{ij},
  \label{eq:harm_curv_const_ricci_cond}
\end{equation}
i.e. the harmonic curvature requires
\begin{equation}
  g = - \frac{R_2}{R_1} h.
  \label{eq:harm_curv_const_ricci_g}
\end{equation}

Therefore, the consistency equations for the Ricci are rewritten,
after using Eq. \eqref{eq:harm_curv_const_ricci_g}, as
\begin{align}
  \dot{h} + h^2 & = \beta_1,
  &
    \beta_1 & = \frac{R_1}{3 R_2} (3 \kappa - R_2),
  \\
  f & = \frac{\beta_2}{h},
  &
    \beta_2 & = \frac{R_1}{3 R_2} (3 \kappa - 2 R_2).
\end{align}
Notice that the general solution is parameterised by the function
\(h\), and for vanishing \(\beta_2\) the function \(f\) is zero,
however this solution is degenerated for vanishing \(\kappa\).

We assume that the product of the constants \(R_1\) and \(R_2\) is
positive, i.e. \(R_1 R_2 > 0\). Then, the solution are 
\begin{equation}
  \begin{aligned}
    h & =
    \begin{cases}
      \omega \tanh \left( \omega (t - t_*) \right) & \phantom{-}
      \omega^2 = \beta_1 > 0,
      \\
      - \omega \tan \left( \omega (t - t_*) \right) & - \omega^2 =
      \beta_1 < 0,
      \\
      \frac{1}{t - C_h} & \phantom{- \omega^2 = \,} \beta_1 = 0,
    \end{cases}
    \\
    g & = - \frac{R_2}{R_1} h,
    \\
    f & = \frac{\beta_2}{h}.
  \end{aligned}
  \label{eq:harm_curv_const_ricci_sol}
\end{equation}

\subsubsection{Harmonic curvature from Ricci with a ``scale factor''}
\label{sec:harm_curv_w_sf_ricci}
Finally, the harmonic curvature condition for a Ricci with a \emph{scale
factor}, obtained from \eqref{eq:parallel_ricci_w_sf}, is
\begin{equation}
  \nabla_{[t} \ri{i]j}{}{} = (\dot{A} - g R_1 - h A) S_{ij},
  \label{eq:harm_curv_w_sf_cond}
\end{equation}
whose solution for \(g\) is
\begin{equation}
  g = \frac{\dot{A} - h A}{R_1}.
  \label{eq:harm_curv_w_sf_g}
\end{equation}

Substituting Eq. \eqref{eq:harm_curv_w_sf_g} into the consistency
equations for the Ricci, we obtain
\begin{align}
  \dot{h} + h^2 - f h & = \frac{R_1}{3},
  \label{eq:harm_curv_w_sf_cons1}
  \\
  \ddot{A} + f \dot{A} - \left( 2 f h + \frac{4 R_1}{3}  \right) A
                      & = - 2 \kappa R_1.
  \label{eq:harm_curv_w_sf_cons2}
\end{align}

Equation \eqref{eq:harm_curv_w_sf_cons1} can be solved for a constant
function \(f = C_f\), therefore, we shall restrict ourselves to that
case. In this particular case, Eq. \eqref{eq:harm_curv_w_sf_cons1} can
be rewritten as
\begin{align}
  \dot{h} + \left( h  - \frac{C_f}{2} \right)^2 & = \beta_3
  &
  \beta_3 & = \frac{1}{12} (3 C_f^2 + 4 R_1),
  \label{eq:harm_curv_w_sf_cons1a}
\end{align}
and its solutions are 
\begin{equation}
  h =
  \begin{cases}
    \omega \tanh ( \omega (t - t_*) ) + \frac{C_f}{2} & ; \phantom{-}
    \omega^2 = \beta_3 > 0,
    \\
    - \omega \tan ( \omega (t - t_*) ) + \frac{C_f}{2} & ; -
    \omega^2 = \beta_3 < 0,
    \\
    \frac{1}{t - t_*} + \frac{C_f}{2} & ; \phantom{- \omega^2 = } \,
    \beta_3 = 0,
    \\
    \pm \omega + \frac{C_f}{2} & ; \phantom{-}
    \omega^2 = \beta_3 > 0.
  \end{cases}
  \label{eq:harm_curv_w_sf_h_sol}
\end{equation}
Worth to highlight, the last case is a constant solution for \(h\).

For the simplest solutions of \(h\)---a constant function---the second
consistency condition, Eq. \eqref{eq:harm_curv_w_sf_cons2}, can be
integrated. Define the constant
\begin{equation*}
  \alpha = 2 C_f C_h + \frac{4 R_1}{3},
\end{equation*}
where \(h = C_h = \pm \omega + C_f/2\) is the constant determined from
Eq. \eqref{eq:harm_curv_w_sf_h_sol} above. Therefore, the exact solutions for the
\emph{scale factor} is\label{simplest-A}
\begin{equation}
  A
  = \mathbb{A}_1 \exp \Big( (2 C_h - C_f) t \Big) + \mathbb{A}_2 \exp
  \Big( - 2 C_h t \Big) + \frac{2 \kappa R_1}{\alpha}.
  \label{eq:harm_curv_A_simplest}
\end{equation}
Additionally, second simplest solution to the
Eq. \eqref{eq:harm_curv_w_sf_cons2} is obtained for \(\beta_3 = 0\), in
whose case the \(R_1\) constant is determined by the value of \(C_f\),
and the function \(h = \frac{1}{t} + \frac{C_f}{2}\). The solution for
\(A\) is
\begin{widetext}
  \begin{equation}
    A = t (C_f t + 2) \Big[ \mathbb{A}_1 + \mathbb{A}_2 C_f
    \Gamma(0,C_f t) \Big] - \mathbb{A}_2 e^{- C_f t} (C_f t + 1) +
    \frac{3}{4} \kappa C_f^2 t^2,
    \label{eq:harm_curv_a_simplest2}
  \end{equation}
\end{widetext}
with \(\Gamma(0,C_f t)\) an incomplete gamma function defined by the
expression 
\begin{equation*}
  \Gamma(0,C_f t) = \int_{1}^{\infty} \de{x} \, \frac{e^{-x C_f t}}{x}.
\end{equation*}

The \emph{scale factor} can be obtained for the other choices of \(h\) when
one sets the \(C_f = 0\). In those cases, the solutions to
Eq. \eqref{eq:harm_curv_w_sf_cons2} is
\begin{equation}
  A = 
  \mathbb{A}_1 \exp \left( \sqrt{\tfrac{4 R_1}{3}} t \right)
  + \mathbb{A}_2 \exp \left( - \sqrt{\tfrac{4 R_1}{3}} t \right)
  + \frac{3 \kappa}{2},
  \label{eq:harm_curv_a_complicated}
\end{equation} 
where the behaviour of the exponential functions is managed by the
sign of the constant \(R_1\).

\section{The affine self-parallel curves}
\label{sec:self-parallel}
In General Relativity, given that the gravitational connection is the
one of Levi-Cività, the concepts of geodesic and self-parallel curve
are equivalent, however, for generic connections these concepts
differ. Geodesics play an important role in General Relativity, since
they represent the trajectories followed by free falling particles.

Although it is not possible to define geodesics in (purely) affine
models of gravity, we shall postulate that trajectories of free
falling particles are described by self-parallel curves
\begin{equation}
  \label{eq:self_parallel}
  \ddot{x}^{\mu} + \Gamma_{\nu}{}^{\mu}{}_{\lambda} \dot{x}^{\nu} \dot{x}^{\lambda} = 0,
\end{equation}
where now \(\ct{}{}{}\) is a generic connection, and the \emph{dot}
represents---unlike in the previous sections---derivation with respect
to the affine parameter of the curve, \(\tau\).

With the coefficients of the cosmological affine connection,
Eq. \eqref{eq:cosm_conn}, the self-parallel curves are given by
\begin{align*}
  \ddot{t} + f\dot{t}^2 + g S_{ij} \dot{x}^i \dot{x}^j & = 0,
  \\
  \ddot{x}^i + 2 h \dot{t} \dot{x}^i + \gamma_j{}^i{}_k \dot{x}^j \dot{x}^k & = 0.
\end{align*}
The term that mixes the time with the spacial coordinates can be
eliminated by a redefinition of the affine parameter as
\begin{equation*}
  \der{}{l} = C_T e^{2 H} \der{}{\tau} = C_T e^{2 \int \de{t} h} \der{}{\tau},
\end{equation*}
with \(C_T\) an arbitrary constant. In terms of the new parameter
(were now the dot derivative is with respect to the parameter \(l\))
the self-parallel equations become
\begin{equation}
  \begin{aligned}
    \ddot{t} + (f - 2 h) \dot{t}^2 + g S_{ij} \dot{x}^i \dot{x}^j & = 0,
  \\
  \ddot{x}^i + \gamma_j{}^i{}_k \dot{x}^j \dot{x}^k & = 0.
  \end{aligned}
  \label{eq:decoupled_self_parallel}
\end{equation}

The three-dimensional restriction of
Eqs. \eqref{eq:decoupled_self_parallel} are the same as the spacial
geodesic equations for a Friedman--Robertson--Walker model in General
Relativity \cite{garfinkle18_shape_orbit_flrw_spacet}.

The above shows that there exists a parametrisation of the
self-parallel curves from Polynomial Affine Gravity, in which the
restriction to the spatial coordinates coincides with the geodesic of
the Friedman--Robertson--Walker models on General Relativity.

For the sake of completeness, we remind to the readers that due to the
isotropy, one can set \(\theta = \tfrac{\pi}{2}\), and thus the
spatial part of Eq. \eqref{eq:decoupled_self_parallel} are 
\begin{equation}
  \begin{aligned}
    \ddot{r} + \frac{k r}{1 - kr^2} \dot{r}^2 + \left(kr^3 - r \right)
    \dot{\varphi}^2 & = 0
    \\
    \ddot{\varphi} + 2 \frac{\dot{r} \dot{\varphi}}{r} & = 0.
  \end{aligned}
  \label{eq:parallel_r_phi}
\end{equation}
Hence, in terms of the new affine parameter, the
geometry of the \(r(\varphi)\)-curves is determined by
\begin{equation}
  r(\varphi) = 
  \begin{cases}
    \sin(l) & \text{for } \kappa = 1
    \\
    l & \text{for } \kappa = 0
    \\
    \sinh(l) & \text{for } \kappa = -1
  \end{cases},
\end{equation}
for \(\dot{\varphi} = 0\), i.e. for radial self-parallels, and 
\begin{equation}
  r(\varphi) = \frac{1}{\sqrt{\kappa + B^2 \cos^2 (\varphi + \beta)}},
\end{equation}
for \(\dot{\varphi} \neq 0\).

Interestingly, in this context, although the equation that describes
the geometry of the orbits is the same, in Polynomial Affine Gravity
there is no way to differentiate among the orbits of massive and
massless particles, since the absence of a metric precludes the
classification of vectors into time-like, space-like or null-like.

\section{The Ricci tensor as a metric}
\label{sec:ricci_metric}
From the last section, we understood that---generically---the geometry
of the self-parallel curves (or geodesics) is solely determined by the
notion of parallelism, however, their relation with physical notions
(as trajectories, equivalence principle and principles of relativity)
requires the existence of a metric. Nonetheless, our proposal stands
on the idea that no metric is required to formulate the model.

A coruscating fact is that, unlike in General Relativity, a metric
field is not necessarily a fundamental geometric object. In what
follows, we highlight that under certain conditions the Ricci tensor
could play the role of a \emph{emergent} (or derived) metric. Therefore, we
shall consider the affine connection as the fundamental field of
Polynomial Affine Gravity, while the (non-degenerated) symmetric part
of the Ricci tensor serves as metric. Let us first remind the formal
definition of a metric.

\begin{Def}[Metric]
  \label{def:metric}
  Let \(M\) be an \(m\)-dimensional differential manifold. A
  pseudo-Riemannian metric on \(M\) is a \(\binom{0}{2}\)-tensor field
  \(g\) on \(M\) satisfying:
  \begin{enumerate}
  \item \(g\) is a symmetric tensor field, and 
  \item \(g\) is nondegenerated, i.e. \(\forall \, X \in TM\) the
    quantity \(g(X,Y) = 0\) if and only if \(TM \ni Y = 0\).
  \end{enumerate}
  A pair \((M,g)\) is called a pseudo-Riemannian manifold.
\end{Def}

Notice that the second condition in Def. \ref{def:metric} ensures that
as a map, \(g: TM \to T^*M\), the metric field has trivial kernel and
thus it is invertible.

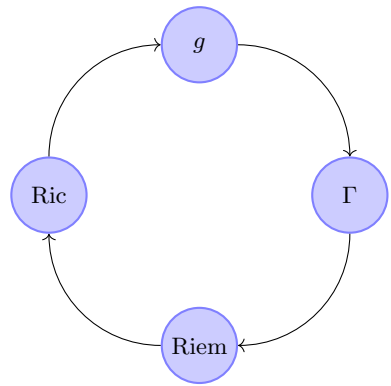
\begin{figure}[ht]
  \centering
  \begin{tikzpicture}
    [place/.style={circle,draw=blue!50,fill=blue!20,thick, inner
      sep=0pt,minimum size=10mm}, bend left=45]
    \node (g)  at ( 90:2cm) [place] {$g$};
    \node (G)  at (  0:2cm) [place] {$\Gamma$};
    \node (Rm) at (270:2cm) [place] {$\Riem$};
    \node (Ri) at (180:2cm) [place] {$\Ric$};
    \draw[->] (g)  to (G);
    \draw[->] (G)  to (Rm);
    \draw[->] (Rm) to (Ri);
    \draw[->] (Ri) to (g);
  \end{tikzpicture}
  \caption{Closed chain of derived quantities for (Riemannian)
    Einstein manifolds.\label{fig:riem_einst_manif}}
\end{figure}

The metric is the fundamental field in General Relativity, and all
other geometrical properties of the manifold---connection, curvature,
etc.---are derived from it,
\begin{equation}
  g \to \ct{}{}{} \to \Riem \to \Ric \to \ri{}{}{}.
\end{equation}
A peculiar, and particularly important type of manifolds are the
Einstein spaces, since the chain of derived quantities closes, see
Fig. \ref{fig:riem_einst_manif}.

The closure condition for Einstein manifolds is expressed as
\begin{equation}
  \ri{\mu \nu}{}{} = \Lambda g_{\mu\nu}.
  \label{eq:einstein-manifold}
\end{equation}
It is worth mentioning that in the vanishing Ricci case, despite the
Eq. \eqref{eq:einstein-manifold} is satisfied, one cannot say that the
chain really closes, since
\begin{equation*}
  g \to \ct{}{}{} \to \Riem \to \Ric = 0.
\end{equation*}

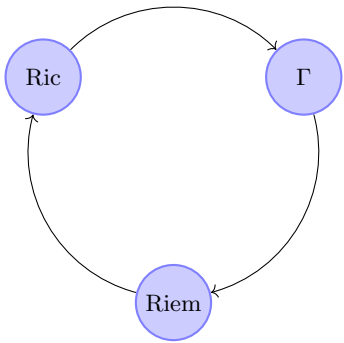
\begin{figure}[ht]
  \centering
  \begin{tikzpicture}
    [place/.style={circle,draw=blue!50,fill=blue!20,thick, inner
      sep=0pt,minimum size=10mm},bend left=45]
    \node  (G) at ( 30:2cm) [place] {$\Gamma$};
    \node (Rm) at (-90:2cm) [place] {$\Riem$};
    \node (Ri) at (150:2cm) [place] {$\Ric$};
    \draw[->]  (G) to (Rm);
    \draw[->] (Rm) to (Ri);
    \draw[->] (Ri) to  (G);
  \end{tikzpicture}
  \caption{Closed chain of derived quantities for affine
    Einstein manifolds.\label{fig:affine_einst_manif}}
\end{figure}

In our model, we start with an affinely connected manifold
\((M,\ct{}{}{})\), were the ansatz of the affine connection is
determined by the symmetries of the problem.\footnote{A noticeable difference with the approach in General
Relativity is that (in principle) a choice of the connection does not
determine the signature of the manifold.} Without the use of
a metric, the chain of derided products stops at the Ricci tensor,
\begin{equation}
  \ct{}{}{} \to \Riem \to \Ric.
\end{equation}
Nevertheless, the above chain could close if the Ricci tensor---as in
the case of Einstein manifolds---satisfies the conditions of a metric,
i.e. conditions in Def. \ref{def:metric}. Assuming such, the chain of
derived quantities closes, see Fig. \ref{fig:affine_einst_manif}, and
therefore it is possible to define an affine analogue of Einstein
manifolds.

Noticeable, along the developing of this work we restricted ourselves
to \emph{locally equiaffine} connections, which ensure the symmetry of the
Ricci tensor (see Proposition 3.1 in Ref. \citep{nomizu94_affin}). This
is precisely the requirement to fulfil the first condition in
Def. \ref{def:metric}. The second condition the Ricci tensor should
satisfy in order to be a good metric field is not to posses a zero
eigenvalue. 

Given the classification of solutions of the field equations
\eqref{eq:simple_eom}, one concludes that: (i) as in General Relativity,
the relation between the Ricci tensor and the metric interpretation is
lost for Ricci-flat connections; (ii) in the case of parallel
Ricci---if nondegenerated---it is a \emph{good} metric field, which
additionally satisfies the ``metricity'' condition; (iii) in the case
of proper solutions of the harmonic curvature, the Ricci tensor accept
interpretation of a metric, but it is not compatible with the
connection, i.e. does not satisfy the metricity condition.

Since we started from an affinely connected space, \((M, \ct{}{}{})\),
and a derived geometric object, the Ricci tensor, introduces a metric
structure, we call it an \emph{emergent} metric.

It can be shown with ease, given a locally equiaffine
connection whose Ricci tensor is parallel and nondegenerated, that 
\begin{equation}
  \ct{\mu}{\lambda}{\nu} = \frac{1}{2} \ri{}{\lambda \sigma}{}
  \left( \partial_\mu \ri{\sigma \nu}{}{}
    + \partial_\nu \ri{\mu \sigma}{}{}
    - \partial_\sigma \ri{\mu \nu}{}{} \right).
  \label{eq:conn_in_terms_ricci}
\end{equation}
Consequently, in the analysed case the geometry is naturally
Riemannian, and we can relate the Ricci tensor to a \emph{canonical}
covariantly constant \(\binom{0}{2}\)-tensor, \(g\), by
\begin{equation}
  \ri{\mu\nu}{}{} = \Lambda g_{\mu\nu}, \text{ with } \Lambda \in
  \R^*,
  \label{eq:einstein_manifold}
\end{equation}
i.e. the space is an Einstein manifold. An additional comment is that
given the relation in Eq. \eqref{eq:einstein_manifold} it follows that
\begin{equation*}
  \ct{}{}{} (\ri{}{}{}) = \ct{}{}{} (g).
\end{equation*}

The fact that the Ricci tensor---or its symmetric part---could play the
role of a metric was (somehow) anticipated by Schrödinger
\cite{schroedinger50_space}.

\section{Cosmological quantities}
\label{sec:redshift}
In the previous section we argue that even when in principle the space
does not posses a metric structure, under certain conditions a metric
structure emerges through the symmetric component of the Ricci tensor.

A consequence of the emergence of a metric is that one can now
distinguish between null-like and time-like self-parallel curves,
which could be interpreted as the trajectories of free-falling
particles---as mentioned in Sec. \ref{sec:self-parallel}. In particular,
it is possible to define a null-like self-parallel curve by the
equations
\begin{equation}
  \begin{aligned}
    \ddot{x}^\mu + \ct{\lambda}{\mu}{\rho} \dot{x}^\lambda \dot{x}^\rho
    & =
    0,
    \\
    \ri{\mu\nu}{}{} \dot{x}^\mu \dot{x}^\nu
    & =
    0.
  \end{aligned}
  \label{eq:null-self-parallel}
\end{equation}
The special case of a light ray coming in the radial direction yields
\begin{equation}
  \de{t} = \pm \sqrt{\frac{\ri{r r}{}{}}{\ri{tt}{}{}}} \frac{\de{r}}{\sqrt{1 - \kappa r^2}},
  \label{eq:scale-factor-cond}
\end{equation}
which by renaming 
\begin{equation}
  a(t) = \sqrt{\frac{\ri{r r}{}{}}{\ri{tt}{}{}}},
  \label{eq:our_scale_factor}
\end{equation}
is the same equation that serves to define the cosmological redshift,
i.e.
\begin{equation}
  1 + z = \frac{a(t_0)}{a(t_1)} = \sqrt{\frac{\ri{r r}{}{}(t_0)
      \ri{tt}{}{}(t_1)}{\ri{tt}{}{}(t_0) \ri{r r}{}{}(t_1)}}. 
  \label{eq:our_redshift}
\end{equation}

Not surprisingly, after the emergence of the metric, the physical
properties of the space are determined by a single function, i.e. the
scale factor. If we focus on the nondegenerated Ricci tensors found in
Sec. \ref{sec:cosm_sol}, the standard scale factor is defined as
\begin{equation*}
  a(t) = \sqrt{A(t)},
\end{equation*}
where \(A(t)\) is the function defining the Ricci tensor in
Eq. \eqref{eq:scale_factor_Ricci}.

\section{Conclusions and remarks}
\label{sec:discussion}
In this paper we have extended the set of known solutions to the
parallel Ricci and harmonic curvature equations, when these are
defined in terms of a connection instead of a metric.\footnote{For solutions of the mentioned equations in terms of the
metric, we refer the readers to Chapter 16 of
Ref. \cite{besse07_einst}. In addition, a detailed exposition of the SKY
model of gravity can be found in Chap. 7 of
Ref. \cite{mielke17_geomet_gauge_field}.} We also
analysed the equation of self-parallel curves and noticed that,
without the aid of a metric, it is not possible to distinguishing
between the (yet hypothetical) trajectories followed by massive or
massless free-falling particles. Then, we show that the Ricci derived
from the connection would, under certain conditions, be a good
metric. In these cases, it is possible to surpass the mentioned
limitation, and distinguish trajectories followed by massive and
massless free-falling particles. Moreover, the emergent metric allows
us to make contact with the standard cosmological quantities such as
the redshift, scale factor, Hubble and des-acceleration parameters,
etc.

We first would like to highlight the fact that a very important step
toward the simplification of the Polynomial Affine model of Gravity
was due to the change of the field decomposition of the connection,
Eq. \eqref{eq:conn_decomp}, which alleviate both the geometric
interpretation of the irreducible components of the connection and the
process of finding the complete field equations (See Appendix
\ref{app:field_equations}). An advantage of having the complete field
equations at hand, is that readers can convince themselves that the
torsion-free sector is a consistent truncation of the model.

Despite the field equations of Polynomial Affine Gravity are
well-defined in the torsion-free sector, the limit \(\Ag \to 0\) and
\(\bt{}{}{} \to 0\) is meaningless at the action level because all the
terms are at least linear in either \(\Ag\) or \(\bt{}{}{}\). Hence,
the Feynman rules for the model lack vertices with only
\emph{gravitons}.\footnote{Here we call gravitons to the spin-2 field within the
symmetric part of the connection, \(\ct{\mu}{\lambda}{\nu}\). This
feature suggests that the Polynomial Affine model of Gravity could
bypass the no-go theorems found in
Refs. \cite{mcgady14_higher_spin_massl_s_dimen,camanho16_causal_const_correc_to_gravit}.} Furthermore, the effective action from which one
can derive the field equations \eqref{eq:simple_eom} contains vertices
with three and four gravitons.

In the formulation of the Polynomial Affine model of Gravity there is
not something such as a cosmological constant. Nevertheless, the
integration constant included in the process of solving
Eq. \eqref{eq:simple_eom}, plays the role of cosmological constant. This
changes the paradigm on the \emph{cosmological problem}, similarly as in
Unimodular Gravity models. The most relevant feature of Unimodular
Gravity is that vacuum fluctuations of the energy-momentum tensor do
not gravitate \cite{weinberg89_cosmol_const_probl}, removing the
discrepancy between the observed and estimated values of the vacuum
energy
\cite{ng91_unimod_theor_gravit_cosmol_const,smolin09_quant_unimod_gravit_cosmol_const_probl,ellis11_trace_free_einst_equat_as}.\footnote{Very recently, it has been pointed that from an affine point
of view, the nature of the cosmological constant is related with the
volume preserving property (instead of being related to the sectional
curvature) \cite{boskoff19_recov_cosmol_const_from_affin_geomet}.}

Although cosmological solutions, in the torsion free sector of the
Polynomial Affine model of Gravity, were found in
Ref. \cite{castillo-felisola18_cosmol}, in this work we were able of
developing further arguments that allows us to obtain explicit
solutions in cases that were previously unexplored. 

The field equations of General Relativity---without cosmological
constant---in vacuum are equivalent to Ricci-flat manifolds. In
general these solutions are curved spacetimes (i.e. Schwarzschild
space), however, once one ask for cosmological solutions the field
equations require the manifold to be flat. In the analysis of the
solutions of Polynomial Affine Gravity field equations, we notices
that they accept Ricci-flat cosmological solutions which are not
flat. Let us work out the interpretation of this situation.

Remember first that a connection, \(\nabla\), on a vector bundle
\(\pi:E \to M\) assigns to each vector field, \(X\), a map
\(\nabla_X\) from the space of sections \(C^\infty(E)\) to
itself. Therefore, for a given direction, the connection represents an
endomorphism on the space of sections,
i.e. \(C^\infty(\End(E))\). Similarly, the curvature of a linear
connection, \(\ri{}{}{}^\nabla\), on a vector bundle \(\pi:E \to M\)
is a 2-form on \(M\) with values in \(C^\infty(\End(E))\).\footnote{See Refs. \cite{besse07_einst,ivey03_cartan_begin,baez94_gauge}.} In
the gravitational case, the vector bundle is the tangent bundle,
\(TM\), and in particular when one considers pseudo-Riemannian
geometries the group structure of \(\End(TM)\) is a subgroup of the
orthogonal group, \(O(p,q)\).\footnote{We are using the standard notation that \(n = p + q\) is the
dimension of the pseudo-Riemannian manifold, \(p\) the number of
space-like coordinates and \(q\) the number of time-like dimensions.}

Now, from Eqs. \eqref{eq:Riem} and \eqref{eq:Ric}, it follows that the
Ricci-flat condition allows the components \(\ri{t i}{t}{j} = - \ri{i
t}{t}{j}\) and \(\ri{m i}{n}{j} = - \ri{i m}{n}{j}\) of the curvature
not to vanish. Thus, the group structure underling the endomorphisms
of the tangent bundle is the homogeneous Carroll group
\cite{levy-leblond65_une_nouvel_limit_non_relat}.\footnote{We thank Dr. Zurab Silagadze for help us with the
identification of this group.} The homogeneous
Carroll groups can be obtained from the Lorentz group through the
Inönü--Wigner contraction in the limit \(c \to 0\). 

Notice that the nonvanishing components of curvature tensor for the
Ricci-flat manifolds obtained in Sec. \ref{sec:cosm_ricci_flat} are
\begin{equation}
  \begin{aligned}
    \ri{t i}{t}{j} = - \ri{i t}{t}{j} & = - 2 S_{ij} \big( g h + \kappa \big),
    \\
    \ri{mi}{n}{j} = - \ri{i m}{n}{j} & = S_{ij} \, \delta_m^n \, \big( g h + \kappa \big).
  \end{aligned}
\end{equation}
Thus, for example the solution in Eq. \eqref{eq:sol_cosm_very_simple} is
flat---irrespective of the value of \(\kappa\)---if and only if
\(C_g\) vanishes. 

The family of solutions in Sec. \ref{sec:sol_h_eq_f} has vanishing
curvature for \(C_g = 0\), while the family from Sec. \ref{sec:sol_h_eq_mf}
requires \(C_g = - \kappa C_h\). Solutions in Sec. \ref{sec:sol_h0_f} and
\ref{sec:sol_g0_f} are flat solely for \(\kappa = 0\). 

The second class of solutions of the field equations are those
connections with parallel Ricci tensor, \(\nabla_\lambda
\ri{\mu\nu}{}{} = 0\). In order to solve the field equations we
required the Ricci tensor to be compatible with the cosmological
principle, i.e. to preserve the isotropic and homogeneity
symmetries. We analyse the time-independent and
Friedman--Robertson--Walker-like cases.

Noticeable, there are no connections with parallel, nondegenerated,
time-independent Ricci tensor. However, we found solutions with
degenerated Ricci [see Eqs. \eqref{eq:parallel_cte_ricci_h0} and
\eqref{eq:parallel_cte_ricci_g0}]. 

There are connections with parallel, nondegenerated
Friedman--Robertson--Walker-like Ricci tensor. In this cases the Ricci
represents an emergent metric, and thus the underling structure of the
manifold is Riemannian. From Eq. \eqref{eq:sol_parallel_ricci_w_sf} one
notices that depending on the sign of the constant \(R_1\) the
manifold is a sphere- or hyperbolic-like space.\footnote{If one demand the signature to be Lorentzian, the solutions
are de Sitter and Anti de Sitter. However, as mentioned previously the
signature of the space is not fixed from the affine structure.} It can be
checked that the solution in Eq. \eqref{eq:sol_parallel_ricci_w_sf}
satisfies---as the standard Friedman--Robertson--Walker solutions---:
\begin{align*}
  f & = 0, & a & = \sqrt{A},
  \\
  g & = a \dot{a}, & h & = \frac{\dot{a}}{a}.
\end{align*}
Such result, as expected from the discussion in
Sec. \ref{sec:ricci_metric} is not very interesting from the view point
of the Polynomial Affine model of Gravity.

The third class of solutions to the field equations, i.e. connections
with harmonic curvature, does not accept degenerated solutions for the
time-independent Ricci case, since either \(f\) or \(g\) diverge when
\(R_1 \lor R_2\) vanish. The solutions shown in
Eq. \eqref{eq:harm_curv_const_ricci_sol} assume that \(R_1 R_2 > 0\),
which ensures, since the Ricci tensor endows the manifold with a
metric structure, a Lorentzian signature of the metric. Worth
mentioning that it is possible to set \(\beta_2 = 0\) which implies
that \(f\) vanishes (similar to the behaviour of standard
Friedman--Robertson--Walker models).

Even more interesting are the solutions of the harmonic curvature
equations, whose Ricci contains a scale factor. We found
solution with nondegenerated Ricci tensor, which again endows the
manifold with a metric structure. Although this metric structure has a
similar form than the expected from the Friedman--Robertson--Walker
models, the cosmological scale factor, \(a = \sqrt{A}\), admits richer
behaviour in the cosmological evolution. Of course, one has to remind
that that these are (so far) vacuum solutions in Polynomial Affine
Gravity. 

\begin{figure}[htbp]
\centering
\includegraphics[width=.9\linewidth]{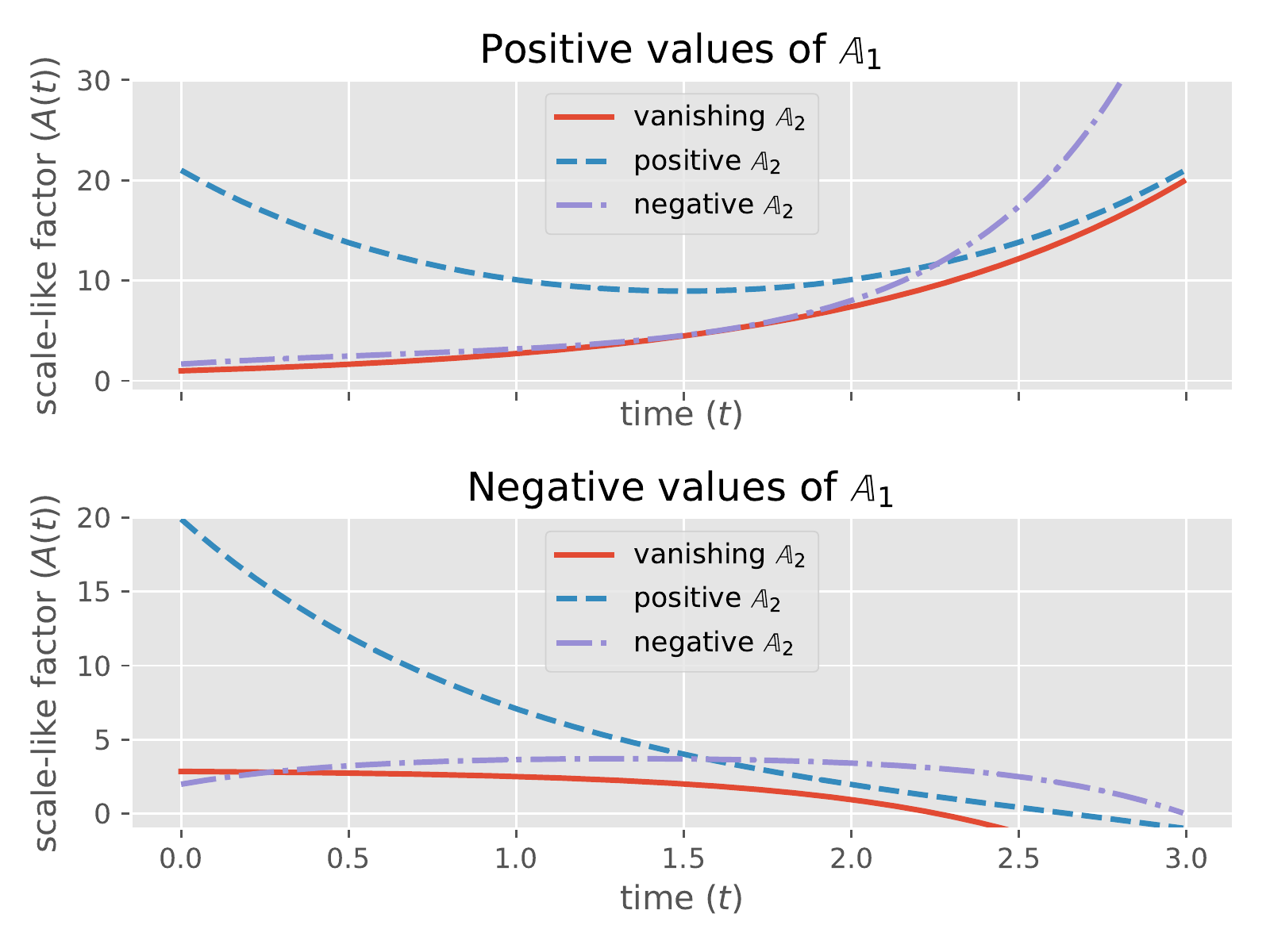}
\caption{\label{fig:plot_A}
Behaviour of the function \(A(t)\) for certain values of the parameters \((\mathbb{A}_1, \mathbb{A}_2, \kappa, C_h, C_f)\)}
\end{figure}

Notice for example that for constant \(f = C_f\) and \(h = C_h\),
Eq. \eqref{eq:harm_curv_w_sf_cons2} becomes the equation of motion of a
damped harmonic oscillator on which a constant force is exerted. This
case is the simplest case of a cosmological model in Polynomial Affine
Gravity with an emergent metric, with no equivalent in General
Relativity. These kind of solutions generalise the metric solutions to
the field equations \(\nabla_{[\mu} \ri{\nu] \lambda}{}{} = 0\)
reported in Ref. \cite{castillo-felisola18_cosmol}. The behaviour of the
\(A\) function for certain set of values of the parameters
\((\mathbb{A}_1, \mathbb{A}_2, \kappa, C_h, C_f)\) is shown in
Fig. \ref{fig:plot_A}.

\begin{figure}[htbp]
\centering
\includegraphics[width=.9\linewidth]{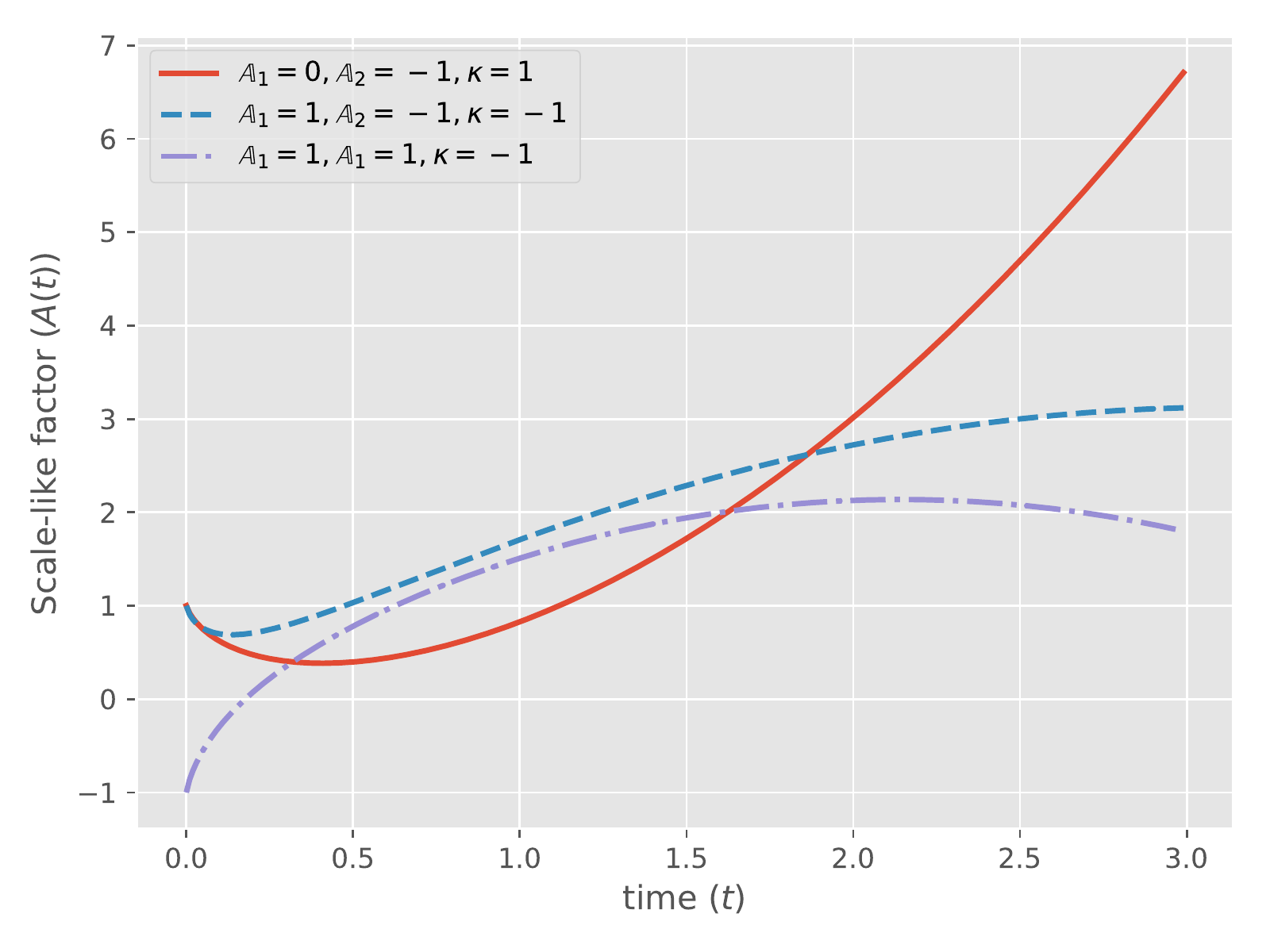}
\caption{\label{fig:plot_A2}
Behaviour of the function \(A(t)\) in Eq. \eqref{eq:harm_curv_a_simplest2}, for certain values of the parameters \((\mathbb{A}_1, \mathbb{A}_2, \kappa, C_f)\)}
\end{figure}

We were able to solve the field equations for the harmonic curvature
in the case \(\beta_3 = 0\), which introduces interesting models of
cosmologies through the appearance of the incomplete gamma function
within the scale factor [see Eq. \eqref{eq:harm_curv_a_simplest2}]. In
Fig. \ref{fig:plot_A2} the behaviour of the scale-like factor \(A(t)\)
is shown for a set of values of the parameters \((\mathbb{A}_1,
\mathbb{A}_2, \kappa, C_f)\). Notice the complex behaviour for the
particular case with dashed lines.

It is noticeable that connections with harmonic curvature could provide
an emergent metric (i.e. Ricci tensor) corresponding to the standard
metric on a Eucliean or Minkowskian space, even though their curvature
do not vanish. 

Further studies, which take the nature of the dark sector into
account, will need to be undertaken. Interesting proposals---in models
other than the Polynomial Affine model of Gravity---could be found in
Ref. \cite{josset17_dark_energ_from_violat_energ_conser,brensinger19_dark_energ_from_dynam_projec_connec,cervantes-cota16_const_purel_affin_theor_with_matter,barrow19_fried_like_univer_with_torsion,kilic19_diffeom_field_revis,roberts95_projec_connec_t,chothe19_cosmol_dark_sector_from_mimet,boskoff19_recov_cosmol_const_from_affin_geomet}.

\begin{acknowledgments}
We would like to thanks the following people for
their support and helpful discussions: Pietro Fr\'e, Aureliano
Skirzewski, Crist\'obal Corral, Claudio Dib, Iv\'an Schmidt,
Fernando Izaurrieta, Andr\'es Anabal\'on and Julio Oliva. We
gratefully acknowledge to Stefano Vignolo, without whose help this
work would never have been possible.  We are specially thankful to
the developers of the software SageMath~\cite{stein18_sage_mathem_softw_version} and
SageManifolds~\cite{gourgoulhon18_sagem_version,gourgoulhon15_tensor_calcul_with_open_sourc_softw,gourgoulhon18_symbol_tensor_calcul_manif} which
we have used extensively in our calculations.

The ``Centro Cient\'ifico y Tecnol\'ogico de Valpara\'iso''
\mbox{(CCTVal)} is funded by the Chilean Government through the
Centers of Excellence Base Financing Program of Comisi\'on Nacional
de Investigaci\'on Cient\'ifica y Tecnol\'ogica (CONICYT), by grant
number FS0821. The work of JP is funded by USM-PIIC grant
number 059/20128. AZ is funded by FONDECYT project 1160423. This
research benefited from a grant from the Universidad T\'ecnica Federico
Santa Mar\'ia \verb|PI_LI_19_02|.
\end{acknowledgments}

\appendix

\section{Building the simplified polynomial affine  action}
\label{app:build_PAG}
Our goal is to build the most general action using the irreducible
fields, \(\ct{\mu}{\lambda}{\nu}\), \(\bt{\mu}{\lambda}{\nu}\) and
\(\Ag_\mu\), where the dynamics is given by the covariant derivative
with respect to the symmetric part of the connection, i.e., \(\nabla
= \nabla^{\hat{\Gamma}}\). Now, the strategy to write down the action
is to define the most general scalar density. For that sake, inspired
by the procedure described in
Ref. \cite{castillo-felisola18_einst_gravit_from_polyn_affin_model}, we
introduced a two operators, \(\N\) and \(\W\), that count the number
of free indices and the weight density---respectively---of a given
term.

The action of the operators \N and \W on the irreducible components of
the connection,
\begin{equation}
  \begin{aligned}
    \N(\Ag) & = -1, & \N(\bt{ }{ }{ }) & = -1,
    \\
    \N(\nabla) & = -1, & \N(\de{V}) & = 4,
    \\
    \W(\Ag) & = 0, & \W(\bt{ }{ }{ }) & = 0,
    \\
    \W(\nabla) & = 0, & \W(\de{V}) & = 1.
  \end{aligned}
\end{equation}
As example of how the dimensional analysis works, we consider a
general term of the form \(\mathcal{O} = \Ag^m \bt{}{}{}^n \nabla^p
\de{V}^q\), the action of the \N and \W operators on the term yield
the equations
\begin{equation}
  \begin{aligned}
    \N(\Op) & = 4q -m-n-p ,
    \\
    \W(\Op) &= q.
  \end{aligned}
  \label{eq:term_weight_eqs}
\end{equation}
We are interested in the Lagrangian, i.e. a scalar density. Equations
\eqref{eq:term_weight_eqs} require \(q = 1\) and \(m+n+p = 4\). The
terms contributing to this construction are shown in Table \ref{tab:action_terms}.

\begin{table}[htbp]
\caption{\label{tab:action_terms}
Possible terms contributing to the action of Polynomial Affine Gravity}
\centering
\begin{tabular}{C{.2\linewidth}|C{.2\linewidth}|C{.2\linewidth}|C{.2\linewidth}}
\(m\) & \(n\) & \(p\) & Terms\\
\hline
4 & 0 & 0 & \(\Ag \Ag \Ag \Ag\)\\
3 & 1 & 0 & \(\Ag \Ag \Ag \bt{}{}{}\)\\
3 & 0 & 1 & \(\Ag \Ag \Ag \nabla\)\\
2 & 2 & 0 & \(\Ag \Ag \bt{}{}{} \bt{}{}{}\)\\
2 & 1 & 1 & \(\Ag \Ag \bt{}{}{} \nabla\)\\
2 & 0 & 2 & \(\Ag \Ag \nabla \nabla\)\\
1 & 3 & 0 & \(\Ag \bt{}{}{} \bt{}{}{} \bt{}{}{}\)\\
1 & 2 & 1 & \(\Ag \bt{}{}{} \bt{}{}{} \nabla\)\\
1 & 1 & 2 & \(\Ag \bt{}{}{} \nabla \nabla\)\\
1 & 0 & 3 & \(\Ag \nabla \nabla \nabla\)\\
0 & 4 & 0 & \(\bt{}{}{} \bt{}{}{} \bt{}{}{} \bt{}{}{}\)\\
0 & 3 & 1 & \(\bt{}{}{} \bt{}{}{} \bt{}{}{} \nabla\)\\
0 & 2 & 2 & \(\bt{}{}{} \bt{}{}{} \nabla \nabla\)\\
0 & 1 & 3 & \(\bt{}{}{} \nabla \nabla \nabla\)\\
0 & 0 & 4 & \(\nabla \nabla \nabla \nabla\)\\
\end{tabular}
\end{table}

From Table \ref{tab:action_terms} one can straightforwardly read terms
that vanish, e.g. the term with four \(\Ag\) does not contribute to
the action since its contraction with the volume element is
identically zero. Whenever two covariant derivatives are contracted
with the volume form they give a curvature tensor, and since the
curvature is defined for the symmetric component of the connection,
such curvature satisfy the torsion-free Bianchi identities, which
relate some of the several possible contractions of indices. An
additional argument that helps to drop contraction of indices is that
\(\bt{}{}{}\) is traceless. Finally, the terms contributing to the
action come from
\begin{align*}
  \Ag \Ag \bt{}{}{} \bt{}{}{} & \rightarrow F_4
  &
    \Ag \Ag \bt{}{}{} \nabla & \rightarrow D_6 , D_7
  \\
  \Ag \bt{}{}{} \bt{}{}{} \bt{}{}{} & \rightarrow F_3
  &
    \Ag \bt{}{}{} \bt{}{}{} \nabla & \rightarrow D_4 , D_5
  \\
  \Ag \bt{}{}{} \nabla \nabla & \rightarrow B_3 , B_4 , B_5 , E_2
  &
    \bt{}{}{} \bt{}{}{} \bt{}{}{} \bt{}{}{} & \rightarrow F_1 , F_2
  \\
  \bt{}{}{} \bt{}{}{} \bt{}{}{} \nabla & \rightarrow D_1 , D_2 , D_3
  &
    \bt{}{}{} \bt{}{}{} \nabla \nabla & \rightarrow B_1 ,  B_2 , E_1
  \\
  \bt{}{}{} \nabla \nabla \nabla & \rightarrow C_1, C_2.
  & &
\end{align*}

\section{Explicit calculation of the field equations}
\label{app:field_equations}
In order to obtain the field equations we proceed as proposed in
Ref. \cite{kijowski78_new_variat_princ_gener_relat},\footnote{Notice that the original method was proposed by Tulczyjew in
Refs. \cite{tulczyjew74_hamil_lagr_syst,tulczyjew75_symp_form_part_dyna,tulczyjew75_symp_form_field_dyna}
and developed further by Kijowski and Tulczyjew in
Ref. \cite{kijowski79}.}\label{footnote_a} where an auxiliary field is introduced to
ease the process. Additionally, we calculate the contribution of each
term in the action to the field equations separately, since this
allows us to obtain manageable expressions, and to check explicitly
the possible \emph{consistent truncations} of the model.

\subsection{Field equations for \(\Gamma_{\mu}{}^{\lambda}{}_{\nu}\)}
\label{app:field_eq_conn}
We shown in Sec. \ref{sec:field_eq_G} that the field equations for the
symmetric part of the connection are
\begin{equation*}
  \nabla_\mu \PG{\mu\nu}{\lambda}{\rho} = \frac{\partial^*
    \Lag}{\partial \ct{\nu}{\lambda}{\rho} }.
\end{equation*}
The asterisk on the right-hand side denotes the partial derivative
with respect to the connection that is not contained in the curvature
tensor.

\subsubsection*{Calculation of the \(\PG{\mu\nu}{\lambda}{\rho}\)}
\label{sec:momenta_conn}
We start calculating the \(\PG[z]{\mu\nu}{\lambda}{\rho}\) and
then use the Eq. \eqref{eq:PG-from-z}, to obtain
\begin{align*}
  & B_1: &
  & \Big[ 2
    \delta^{[\mu}_\lambda \bt{\alpha}{\nu]}{\beta}
    \bt{\ga}{\rho}{\delta} + 2 \delta^{[\mu}_\lambda
    \bt{\alpha}{\rho]}{\beta} \bt{\ga}{\nu}{\delta} \Big]
    \de{V}^{\alpha\beta\ga\delta}
  \\
  & B_2: &
  &
    2 \bt{\ga}{\sigma}{\delta} \bt{\sigma}{(\rho}{\lambda} \de{V}^{\nu)\mu\ga\delta}
  \\
  & B_3: &
  & 
    2 \delta^{[\mu}_\lambda \bt{\beta}{\nu]}{\ga} \Ag_\delta
    \de{V}^{\rho\beta\ga\delta} + 2 \delta^{[\mu}_\lambda \bt{\beta}{\rho]}{\ga} \Ag_\delta
    \de{V}^{\nu\beta\ga\delta}
  \\
  & B_4: &
  &
    -2 \bt{\ga}{(\rho}{\delta} \Ag_\lambda \de{V}^{\nu)\mu\ga\delta}
  \\
  & B_5: &
  &
    -2 \bt{\ga}{\sigma}{\delta} \Ag_\sigma \delta^{(\rho}_\lambda
    \de{V}^{\nu)\mu\ga\delta}
  \\
  & C_1: &
  &
    2 \nabla_\beta \bt{\ga}{\rho}{\delta} \delta^{[\mu}_\lambda
    \de{V}^{\nu]\beta\ga\delta} + 2 \nabla_\beta \bt{\ga}{\nu}{\delta}
    \delta^{[\mu}_\lambda \de{V}^{\rho]\beta\ga\delta}
  \\
  & C_2: &
  &
    - 2 \nabla_\sigma \bt{\ga}{\sigma}{\delta} \delta^{(\rho}_\lambda \de{V}^{\nu)\mu\ga\delta}
\end{align*}

\subsubsection*{Calculation of \(\frac{\partial^* \Lag}{\partial \ct{\nu}{\lambda}{\rho} }\)}
\label{sec:interaction_conn}
In Eq. \eqref{eq:new-action}, the connection appears explicitly (non in
the curvature tensor) in the covariant derivative. However, the
covariant derivative contains different terms depending on the field
it is acting on. The two different terms in which the connection
appears are
\begin{equation*}
  \nabla_\alpha \bt{\beta}{\ga}{\delta} = \partial_\alpha
  \bt{\beta}{\ga}{\delta} + \ct{\alpha}{\lambda}{\beta}
  \bt{\lambda}{\ga}{\delta} + \ct{\alpha}{\lambda}{\delta}
  \bt{\beta}{\ga}{\lambda} - \ct{\alpha}{\ga}{\lambda} \bt{\beta}{\lambda}{\delta}
\end{equation*}
and
\begin{equation*}
  \nabla_\alpha \Ag_\beta = \partial_\alpha \Ag_\beta -
  \ct{\alpha}{\lambda}{\beta} \Ag_\lambda.
\end{equation*}
Now their partial derivatives with respect to the connection are
\begin{equation*}
  \frac{\partial \nabla_\alpha \bt{\beta}{\ga}{\delta} }{ \partial
    \ct{\nu}{\lambda}{\rho} } = 2 \delta^{(\nu}_{\alpha}
  \delta^{\rho)}_{\beta} \bt{\lambda}{\ga}{\delta} + 2 \delta^{(\nu}_{\alpha}
  \delta^{\rho)}_{\delta} \bt{\beta}{\ga}{\lambda} - 2
  \delta^c_\lambda \delta^{(\nu}_\alpha \bt{\beta}{\rho)}{\delta}
\end{equation*}
and
\begin{equation*}
  \frac{ \nabla_\alpha \Ag_\beta}{\partial \ct{\nu}{\lambda}{\rho}} = -
  2 \delta^{(\nu}_\alpha \delta^{\rho)}_\beta \Ag_\lambda.
\end{equation*}
The terms coming from these derivatives are
\begin{align*}
  & C_1: &
  &
    2 \ri{\mu\lambda}{\mu}{\alpha} 
    \bt{\ga}{(\rho}{\delta} \de{V}^{\nu)\alpha\ga\delta}
  \\
  & C_2: &
  &
    2 \ri{\alpha\beta}{\sigma}{\sigma} \Big[
    2 \bt{\lambda}{(\nu}{\delta} \de{V}^{\rho)\alpha\beta\delta}
    - \delta^{(\nu}_\lambda \bt{\ga}{\rho)}{\delta}
    \de{V}^{\alpha\beta\ga\delta}
    \Big]
  \\
  & D_1: &
  &
    2 \bt{\tau}{\sigma}{\lambda} \bt{\sigma}{\tau}{\alpha}
    \bt{\ga}{(\rho}{\delta}
    \de{V}^{\nu)\alpha\ga\delta}
  \\
  & D_2: &
  &
    2 \bt{\alpha}{\sigma}{\beta} \bt{\sigma}{(\nu}{\tau} \Big[
    2 \bt{\lambda}{\tau}{\delta}
    \de{V}^{|\rho)\alpha\beta\delta} -
    \delta^{\tau}_\lambda \bt{\ga}{|\rho)}{\delta}
    \de{V}^{\alpha\beta\ga\delta}
    \Big]
  \\
  & D_3: &
  &
    2 \bt{\alpha}{\sigma}{\tau} \bt{\beta}{(\nu}{\ga} \Big[
    \delta^{\rho)}_\sigma \bt{\lambda}{\tau}{\delta}
    + \delta^{\rho)}_\delta \bt{\sigma}{\tau}{\lambda}
    - \delta^\tau_\lambda \bt{\sigma}{|\rho)}{\delta}
    \Big]
    \de{V}^{\alpha\beta\ga\delta}
  \\
  & D_4: &
  &
    - 2 \bt{\alpha}{\nu}{\beta} \bt{\ga}{\rho}{\delta} \Ag_\lambda
    \de{V}^{\alpha\beta\ga\delta}
  \\
  & D_5: &
  &
    2 \bt{\alpha}{(\nu}{\beta} \Ag_\sigma \Big[
    2 \bt{\lambda}{\sigma}{\delta} \de{V}^{|\rho)\alpha\beta\delta}
    - \delta^\sigma_\lambda \bt{\ga}{|\rho)}{\delta}
    \de{V}^{\alpha\beta\ga\delta}
    \Big]
  \\
  & D_6: &
  &
    2 \bt{\alpha}{(\nu}{\beta} \Ag_\ga \Ag_\lambda
    \de{V}^{\rho)\alpha\beta\ga}
  \\
  & D_7: &
  &
    0
  \\
  & E_1: &
  &
    4 \nabla_\sigma \bt{\alpha}{\sigma}{\beta} \Big[
    2 \bt{\lambda}{(\rho}{\delta} \de{V}^{\nu)\alpha\beta\delta}
    - \delta^{(\nu}_\lambda \bt{\ga}{\rho)}{\delta}
    \de{V}^{\alpha\beta\ga\delta}
    \Big]
  \\
  & E_2: &
  &
    2 \Fg_{\alpha\beta} \Big[
    2 \bt{\lambda}{(\rho}{\delta} \de{V}^{\nu)\alpha\beta\delta}
    - \delta^{(\nu}_\lambda \bt{\ga}{\rho)}{\delta}
    \de{V}^{\alpha\beta\ga\delta}
    \Big]
\end{align*}
Notice that in the last set of contributions, the one coming from the
term accompanied by the coupling \(D_7\) yields zero. This is because
the antisymmetrization of \(\nabla \Ag\) is nothing but the field strength
of the potential \(\Ag\) (or the curvature of an Abelian 1-form). The
strength does not depend on the symmetric connection.

\subsection*{Complete contribution}
\label{sec:orgb3c1904}
\begin{widetext}
  \begin{align}
    & B_1: &
    & \nabla_\mu \left(\Big[ 2
      \delta^{[\mu}_\lambda \bt{\alpha}{\nu]}{\beta}
      \bt{\ga}{\rho}{\delta} + 2 \delta^{[\mu}_\lambda
      \bt{\alpha}{\rho]}{\beta} \bt{\ga}{\nu}{\delta} \Big]
      \de{V}^{\alpha\beta\ga\delta} \right) = 0
      \label{eq:first_feq}
    \\
    & B_2: &
    &
      \nabla_\mu \left(2 \bt{\ga}{\sigma}{\delta}
      \bt{\sigma}{(\rho}{\lambda} \de{V}^{\nu)\mu\ga\delta} \right) = 0
    \\
    & B_3: &
    & 
      \nabla_\mu \left(2 \delta^{[\mu}_\lambda \bt{\beta}{\nu]}{\ga} \Ag_\delta
      \de{V}^{\rho\beta\ga\delta} + 2 \delta^{[\mu}_\lambda \bt{\beta}{\rho]}{\ga} \Ag_\delta
      \de{V}^{\nu\beta\ga\delta} \right) = 0
    \\
    & B_4: &
    &
      \nabla_\mu \left(-2 \bt{\ga}{(\rho}{\delta} \Ag_\lambda
      \de{V}^{\nu)\mu\ga\delta} \right) = 0
    \\
    & B_5: &
    &
      \nabla_\mu \left(-2 \bt{\ga}{\sigma}{\delta} \Ag_\sigma \delta^{(\rho}_\lambda
      \de{V}^{\nu)\mu\ga\delta} \right) = 0
    \\
    & C_1: &
    &
      \nabla_\mu \left(2 \nabla_\beta \bt{\ga}{\rho}{\delta} \delta^{[\mu}_\lambda
      \de{V}^{\nu]\beta\ga\delta} + 2 \nabla_\beta \bt{\ga}{\nu}{\delta}
      \delta^{[\mu}_\lambda \de{V}^{\rho]\beta\ga\delta} \right) =
      2 \ri{\mu\lambda}{\mu}{\alpha} 
      \bt{\ga}{(\rho}{\delta} \de{V}^{\nu)\alpha\ga\delta}
    \\
    & C_2: &
    &
      \nabla_\mu \left(- 2 \nabla_\sigma \bt{\ga}{\sigma}{\delta}
      \delta^{(\rho}_\lambda \de{V}^{\nu)\mu\ga\delta} \right) =
      2 \ri{\alpha\beta}{\sigma}{\sigma} \Big[
      2 \bt{\lambda}{(\nu}{\delta} \de{V}^{\rho)\alpha\beta\delta}
      - \delta^{(\nu}_\lambda \bt{\ga}{\rho)}{\delta}
      \de{V}^{\alpha\beta\ga\delta}
      \Big]
    \\
    & D_1: &
    &
      2 \bt{\tau}{\sigma}{\lambda} \bt{\sigma}{\tau}{\alpha}
      \bt{\ga}{(\rho}{\delta}
      \de{V}^{\nu)\alpha\ga\delta} = 0
    \\
    & D_2: &
    &
      2 \bt{\alpha}{\sigma}{\beta} \bt{\sigma}{(\nu}{\tau} \Big[
      2 \bt{\lambda}{\tau}{\delta}
      \de{V}^{|\rho)\alpha\beta\delta} -
      \delta^{\tau}_\lambda \bt{\ga}{|\rho)}{\delta}
      \de{V}^{\alpha\beta\ga\delta}
      \Big] = 0
    \\
    & D_3: &
    &
      2 \bt{\alpha}{\sigma}{\tau} \bt{\beta}{(\nu}{\ga} \Big[
      \delta^{\rho)}_\sigma \bt{\lambda}{\tau}{\delta}
      + \delta^{\rho)}_\delta \bt{\sigma}{\tau}{\lambda}
      - \delta^\tau_\lambda \bt{\sigma}{|\rho)}{\delta}
      \Big]
      \de{V}^{\alpha\beta\gamma\delta} = 0
    \\
    & D_4: &
    &
      - 2 \bt{\alpha}{\nu}{\beta} \bt{\gamma}{\rho}{\delta} \Ag_\lambda
      \de{V}^{\alpha\beta\gamma\delta} = 0
    \\
    & D_5: &
    &
      2 \bt{\alpha}{(\nu}{\beta} \Ag_\sigma \Big[
      2 \bt{\lambda}{\sigma}{\delta} \de{V}^{|\rho)\alpha\beta\delta}
      - \delta^\sigma_\lambda \bt{\gamma}{|\rho)}{\delta}
      \de{V}^{\alpha\beta\gamma\delta}
      \Big] = 0
    \\
    & D_6: &
    &
      2 \bt{\alpha}{(\nu}{\beta} \Ag_\gamma \Ag_\lambda
      \de{V}^{\rho)\alpha\beta\gamma} = 0
    \\
    & E_1: &
    &
      4 \nabla_\sigma \bt{\alpha}{\sigma}{\beta} \Big[
      2 \bt{\lambda}{(\rho}{\delta} \de{V}^{\nu)\alpha\beta\delta}
      - \delta^{(\nu}_\lambda \bt{\ga}{\rho)}{\delta}
      \de{V}^{\alpha\beta\ga\delta}
      \Big] = 0
    \\
    & E_2: &
    &
      2 \Fg_{\alpha\beta} \Big[
      2 \bt{\lambda}{(\rho}{\delta} \de{V}^{\nu)\alpha\beta\delta}
      - \delta^{(\nu}_\lambda \bt{\ga}{\rho)}{\delta}
      \de{V}^{\alpha\beta\ga\delta}
      \Big] = 0
  \end{align}
\end{widetext}
\subsection{Field equations for \(B_{\mu}{}^{\lambda}{}_{\nu}\)}
\label{app:field_eq_B}
In Sec. \ref{sec:field_eq_B} we show that the field equations for the
\(\bt{}{}{}\) field are
\begin{equation*}
  \nabla_\mu \PG*{\mu \nu}{\lambda}{\rho} = \frac{ \partial \Lag }{
    \partial \bt{\nu}{\lambda}{\rho} }.
\end{equation*}
We now calculate explicitly the contribution of each term in the
action to the above field equation.

\subsubsection*{Calculation of \(\PG*{\mu \nu}{\lambda}{\rho}\)}
\label{sec:momenta_B}
The explicit calculation of \(\PG*{\mu \nu}{\lambda}{\rho}\) yield
\begin{align*}
  & C_1 : &
  &
    -2 \ri{\sigma \alpha}{\sigma}{\lambda} \de{V}^{\mu \nu \rho \alpha}
  \\
  & C_2 : &
  &
    2 \ri{\alpha \beta}{\sigma}{\sigma} \delta^{\mu}_{\lambda} \de{V}^{\nu \rho \alpha \beta}
  \\
  & D_1 : &
  &
    -2 \bt{\sigma}{\theta}{\lambda} \bt{\theta}{\sigma}{\alpha} \de{V}^{\mu\nu\rho\alpha}
  \\
  & D_2 : &
  &
    2 \bt{\alpha}{\sigma}{\beta} \bt{\sigma}{\mu}{\lambda} \de{V}^{\nu
    \rho \alpha \beta}
  \\
  & D_3 : &
  &
    - 2 \bt{\alpha}{\nu}{\lambda} \bt{\beta}{\mu}{\gamma} \de{V}^{\rho
    \alpha \beta \gamma}
  \\
  & D_5 : &
  &
    2 \bt{\alpha}{\mu}{\beta} \Ag_\lambda \de{V}^{\nu\rho\alpha\beta}
  \\
  & E_1 : &
  &
    4 \delta ^{\mu}_{\lambda} \nabla_{\sigma} \bt{\alpha}{\sigma}{\beta}\de{V}^{\nu\rho\alpha\beta}
  \\
  & E_2 : &
  &
    2 \delta ^{\mu}_{\lambda} \Fg_{\alpha \beta} \de{V}^{\nu\rho \alpha \beta}
\end{align*}

\subsubsection*{Calculation of \(\frac{\partial \Lag}{\partial \bt{\nu}{\lambda}{\rho}}\)}
\label{sec:interactions_B}
The contributions to the right-hand side of the field equations for the
\(\bt{}{}{}\)-field are,
\begin{align*}
  & B_1 : &
  &
    4 \ri{\mu(\sigma}{\mu}{\lambda)} \bt{\gamma}{\sigma}{\delta} \de{V}^{\nu\rho\gamma\delta}
  \\
  & B_2 : &
  &
    2 \ri{\alpha \beta}{\mu}{\sigma} \bt{\mu}{\sigma}{\lambda} \de{V}^{\nu \rho \alpha \beta}
    + 2 \ri{\alpha \beta}{[\nu}{\lambda} \bt{\gamma}{\rho]}{\delta} \de{V}^{\alpha \beta \gamma \delta}
  \\
  & B_3 : &
  &
    2 \ri{\mu \lambda}{\mu}{\alpha} \Ag_\beta \de{V}^{\nu \rho \alpha \beta}
  \\
  & B_4 : &
  &
    2 \ri{\alpha\beta}{\sigma}{\lambda} \Ag_\sigma \de{V}^{\nu \rho \alpha \beta}
  \\
  & B_5 : &
  &
    2 \ri{\alpha\beta}{\tau}{\tau} \Ag_\lambda \de{V}^{\nu \rho \alpha \beta}
  \\
  & D_1 : &
  &
    2 \bt{\lambda}{[\nu}{\alpha} \nabla_\beta
    \bt{\gamma}{\rho]}{\delta} \de{V}^{\alpha \beta \gamma \delta}
    + 2 \bt{\lambda}{[\nu|}{\sigma} \nabla_\beta
    \bt{\gamma}{\sigma}{\delta} \de{V}^{|\rho] \beta \gamma \delta} 
  \\
  & D_2 : &
  &
    2 \bt{\lambda}{\mu}{\sigma} \nabla_\mu \bt{\alpha}{\sigma}{\beta}
    \de{V}^{\nu \rho \alpha \beta} + 2 \bt{\alpha}{[\nu}{\beta}
    \nabla_\lambda \bt{\gamma}{\rho]}{\delta} \de{V}^{ \alpha \beta \gamma \delta}
  \\
  & D_3 : &
  &
    2 \bt{\beta}{\mu}{\gamma}  \nabla_\mu \bt{\lambda}{[\rho}{\delta}
    \de{V}^{\nu] \beta \gamma \delta} + 2 \bt{\gamma}{\mu}{\sigma}
    \nabla_\lambda \bt{\mu}{\sigma}{\delta} \de{V}^{\nu \rho \gamma \delta}
  \\
  & D_4 : &
  &
    4 \bt{\alpha}{\sigma}{\beta} \nabla_{(\lambda} \Ag_{\sigma)}
    \de{V}^{\nu \rho \alpha \beta}
  \\
  & D_5 : &
  &
    2 \nabla_\lambda \bt{\alpha}{\sigma}{\beta} \Ag_\sigma \de{V}^{\nu\rho\alpha\beta}
  \\
  & D_6 : &
  &
    2 \Ag_{\gamma} \nabla_{\lambda} \Ag_{\delta} \de{V}^{\nu\rho\gamma\delta}
  \\
  & D_7 : &
  &
    2 \Ag_\lambda F_{\gamma \delta} \de{V}^{\nu\rho\gamma\delta}
  \\
  & F_1 : &
  &
    4 \bt{\alpha}{\mu}{\beta} \bt{\mu}{\sigma}{\tau} \bt{\lambda}{\tau}{\sigma} \de{V}^{\nu \rho \alpha \beta}
    + 4 \bt{\alpha}{\mu}{\beta} \bt{\gamma}{[\nu}{\delta}
    \bt{\mu}{\rho]}{\lambda} \de{V}^{\alpha \beta \gamma \delta}
  \\
  & F_2 : &
  &
    2 \bt{\alpha}{\mu}{\sigma} \bt{\beta}{\sigma}{\tau}
    \bt{\lambda}{\tau}{\mu} \de{V}^{\nu \rho \alpha \beta}
    + 2 \bt{\alpha}{\mu}{\beta} \bt{\mu}{\sigma}{\lambda} 
    \bt{\gamma}{[\rho}{\sigma} \de{V}^{\nu] \alpha \beta \gamma}
    \notag
  \\
  &       &
  &
    - 2 \bt{\alpha}{\mu}{\beta} \bt{\gamma}{\sigma}{\lambda}
    \bt{\mu}{[\rho}{\sigma} \de{V}^{\nu] \alpha \beta \gamma}
    + 2 \bt{\alpha}{[\nu}{\beta} \bt{\gamma}{\rho]}{\sigma} \bt{\delta}{\sigma}{\lambda} \de{V}^{\alpha\beta\gamma\delta}
  \\
  & F_3 : &
  &
    2 \bt{\lambda}{[\nu}{\alpha} \bt{\beta}{\rho ]}{\gamma} \Ag_\delta
    \de{V}^{\alpha \beta \gamma \delta}
    + 2 \bt{\alpha}{\sigma}{\beta} \Ag_\gamma \bt{\lambda}{[\nu}{\sigma}
    \de{V}^{\rho] \alpha \beta \gamma}
    \notag
  \\
  &       &
  &
    \quad + 2 \bt{\sigma}{\mu}{\lambda} \bt{\mu}{\sigma}{\alpha} \Ag_\beta \de{V}^{\nu \rho \alpha \beta}
  \\
  & F_4 : &
  &
    4 \bt{\alpha}{\mu}{\beta} \Ag_\mu \Ag_\lambda \de{V}^{\nu\rho\alpha\beta}
\end{align*}

\label{sec:orge4dd6bb}
\begin{widetext}
  \subsection*{Complete contribution}
  \begin{align}
    & B_1 : &
    &
      4 \ri{\mu(\sigma}{\mu}{\lambda)} \bt{\gamma}{\sigma}{\delta}
      \de{V}^{\nu\rho\gamma\delta} = 0
    \\
    & B_2 : &
    &
      2 \ri{\alpha \beta}{\mu}{\sigma} \bt{\mu}{\sigma}{\lambda} \de{V}^{\nu \rho \alpha \beta}
      + 2 \ri{\alpha \beta}{[\nu}{\lambda} \bt{\gamma}{\rho]}{\delta}
      \de{V}^{\alpha \beta \gamma \delta} = 0
    \\
    & B_3 : &
    &
      2 \ri{\mu \lambda}{\mu}{\alpha} \Ag_\beta \de{V}^{\nu \rho \alpha
      \beta} = 0
    \\
    & B_4 : &
    &
      2 \ri{\alpha\beta}{\sigma}{\lambda} \Ag_\sigma \de{V}^{\nu \rho
      \alpha \beta} = 0
    \\
    & B_5 : &
    &
      2 \ri{\alpha\beta}{\tau}{\tau} \Ag_\lambda \de{V}^{\nu \rho \alpha
      \beta} = 0
    \\
    & C_1 : &
    &
      \nabla_\mu \left( -2 \ri{\sigma \alpha}{\sigma}{\lambda}
      \de{V}^{\mu \nu \rho \alpha} \right) = 0
      \label{eq:C1}
    \\
    & C_2 : &
    &
      \nabla_\mu \left( 2 \ri{\alpha \beta}{\sigma}{\sigma}
      \delta^{\mu}_{\lambda} \de{V}^{\nu \rho \alpha \beta} \right) = 0
      \label{eq:C2}
    \\
    & D_1 : &
    &
      \nabla_\mu \left( -2 \bt{\sigma}{\theta}{\lambda}
      \bt{\theta}{\sigma}{\alpha} \de{V}^{\mu\nu\rho\alpha} \right) =
      2 \bt{\lambda}{[\nu}{\alpha} \nabla_\beta
      \bt{\gamma}{\rho]}{\delta} \de{V}^{\alpha \beta \gamma \delta}
      + 2 \bt{\lambda}{[\nu|}{\sigma} \nabla_\beta
      \bt{\gamma}{\sigma}{\delta} \de{V}^{|\rho] \beta \gamma \delta} 
    \\
    & D_2 : &
    &
      \nabla_\mu \left( 2 \bt{\alpha}{\sigma}{\beta}
      \bt{\sigma}{\mu}{\lambda} \de{V}^{\nu \rho \alpha \beta} \right) =
      2 \bt{\lambda}{\mu}{\sigma} \nabla_\mu \bt{\alpha}{\sigma}{\beta}
      \de{V}^{\nu \rho \alpha \beta} + 2 \bt{\alpha}{[\nu}{\beta}
      \nabla_\lambda \bt{\gamma}{\rho]}{\delta} \de{V}^{ \alpha \beta \gamma \delta}
    \\
    & D_3 : &
    &
      \nabla_\mu \left( - 2 \bt{\alpha}{\nu}{\lambda}
      \bt{\beta}{\mu}{\gamma} \de{V}^{\rho \alpha \beta \gamma} \right) =
      2 \bt{\beta}{\mu}{\gamma}  \nabla_\mu \bt{\lambda}{[\rho}{\delta}
      \de{V}^{\nu] \beta \gamma \delta} + 2 \bt{\gamma}{\mu}{\sigma}
      \nabla_\lambda \bt{\mu}{\sigma}{\delta} \de{V}^{\nu \rho \gamma \delta}
    \\
    & D_4 : &
    &
      4 \bt{\alpha}{\sigma}{\beta} \nabla_{(\lambda} \Ag_{\sigma)}
      \de{V}^{\nu \rho \alpha \beta} = 0
    \\
    & D_5 : &
    &
      \nabla_\mu \left( 2 \bt{\alpha}{\mu}{\beta} \Ag_\lambda
      \de{V}^{\nu\rho\alpha\beta} \right) =
      2 \nabla_\lambda \bt{\alpha}{\sigma}{\beta} \Ag_\sigma \de{V}^{\nu\rho\alpha\beta}
    \\
    & D_6 : &
    &
      2 \Ag_{\gamma} \nabla_{\lambda} \Ag_{\delta}
      \de{V}^{\nu\rho\gamma\delta} = 0
    \\
    & D_7 : &
    &
      2 \Ag_\lambda F_{\gamma \delta} \de{V}^{\nu\rho\gamma\delta} = 0
    \\
    & E_1 : &
    &
      \nabla_\mu \left(4 \delta ^{\mu}_{\lambda} \nabla_{\sigma}
      \bt{\alpha}{\sigma}{\beta}\de{V}^{\nu\rho\alpha\beta} \right) = 0
    \\
    & E_2 : &
    &
      \nabla_\mu \left(2 \delta ^{\mu}_{\lambda} \Fg_{\alpha \beta}
      \de{V}^{\nu\rho \alpha \beta} \right) = 0
    \\
    & F_1 : &
    &
      4 \bt{\alpha}{\mu}{\beta} \bt{\mu}{\sigma}{\tau} \bt{\lambda}{\tau}{\sigma} \de{V}^{\nu \rho \alpha \beta}
      + 4 \bt{\alpha}{\mu}{\beta} \bt{\gamma}{[\nu}{\delta}
      \bt{\mu}{\rho]}{\lambda} \de{V}^{\alpha \beta \gamma \delta} = 0
    \\
    & F_2 : &
    &
      2 \bt{\alpha}{\mu}{\sigma} \bt{\beta}{\sigma}{\tau}
      \bt{\lambda}{\tau}{\mu} \de{V}^{\nu \rho \alpha \beta}
      + 2 \bt{\alpha}{\mu}{\beta} \bt{\mu}{\sigma}{\lambda} 
      \bt{\gamma}{[\rho}{\sigma} \de{V}^{\nu] \alpha \beta \gamma}
      \notag
    \\
    &       &
    &
      - 2 \bt{\alpha}{\mu}{\beta} \bt{\gamma}{\sigma}{\lambda}
      \bt{\mu}{[\rho}{\sigma} \de{V}^{\nu] \alpha \beta \gamma}
      + 2 \bt{\alpha}{[\nu}{\beta} \bt{\gamma}{\rho]}{\sigma}
      \bt{\delta}{\sigma}{\lambda} \de{V}^{\alpha\beta\gamma\delta} = 0
    \\
    & F_3 : &
    &
      2 \bt{\lambda}{[\nu}{\alpha} \bt{\beta}{\rho ]}{\gamma} \Ag_\delta
      \de{V}^{\alpha \beta \gamma \delta}
      + 2 \bt{\alpha}{\sigma}{\beta} \Ag_\gamma \bt{\lambda}{[\nu}{\sigma}
      \de{V}^{\rho] \alpha \beta \gamma}
      + 2 \bt{\sigma}{\mu}{\lambda} \bt{\mu}{\sigma}{\alpha}
      \Ag_\beta \de{V}^{\nu \rho \alpha \beta} = 0
    \\
    & F_4 : &
    &
      4 \bt{\alpha}{\mu}{\beta} \Ag_\mu \Ag_\lambda
      \de{V}^{\nu\rho\alpha\beta} = 0
  \end{align}
\end{widetext}

\subsection{Field equations for \(\Ag_\mu\)}
\label{app:field_eq_A}
In this section, we show the \emph{complete} Euler--Lagrage equations for
the \(\Ag\) field.
\begin{align}
  & B_3: & &
             \ri{\sigma\tau}{\sigma}{\alpha} \bt{\beta}{\tau}{\ga}
             \de{V}^{\alpha\beta\ga\nu}
             = 0
  \\
  & B_4: & &
             \ri{\alpha\beta}{\nu}{\sigma} \bt{\ga}{\sigma}{\delta}
             \de{V}^{\alpha\beta\ga\delta}
             = 0
  \\
  & B_5: & &
             \ri{\alpha\beta}{\sigma}{\sigma} \bt{\ga}{\nu}{\delta}
             \de{V}^{\alpha\beta\ga\delta}
             = 0
  \\
  & D_4 & &
            \nabla_\mu \Big[
            \bt{\alpha}{\mu}{\beta} \bt{\ga}{\nu}{\delta}
            \de{V}^{\alpha\beta\ga\delta} \Big]
            = 0
  \\
  & D_5: & &
             \bt{\alpha}{\sigma}{\beta} \nabla_\sigma \bt{\ga}{\nu}{\delta}
             \de{V}^{\alpha\beta\ga\delta}
             = 0
  \\
  & D_6: & &
             \nabla_\mu \Big[
             \bt{\alpha}{\mu}{\beta} \Ag_\ga \de{V}^{\alpha\beta\ga\nu}
             \Big]
             + \bt{\alpha}{\mu}{\beta} \nabla_\mu \Ag_\ga
             \de{V}^{\alpha\beta\ga\nu}
             = 0
  \\
  & D_7: & &
             \nabla_\mu \Big[
             \bt{\alpha}{\sigma}{\beta} \Ag_\sigma \de{V}^{\alpha\beta\mu\nu}
             \Big]
             + \bt{\alpha}{\nu}{\beta} \Fg_{\ga\delta}
             \de{V}^{\alpha\beta\ga\delta}
             = 0
  \\
  & E_2: & &
             \nabla_\mu \Big[ \nabla_\sigma \bt{\alpha}{\sigma}{\beta}
             \de{V}^{\alpha\beta\mu\nu} \Big]
             = 0
  \\
  & F_3: & &
             \bt{\sigma}{\tau}{\lambda} \bt{\tau}{\sigma}{\alpha}
             \bt{\beta}{\lambda}{\ga} \de{V}^{\alpha\beta\ga\nu}
             = 0
  \\
  & F_4: & &
             2 \bt{\alpha}{\sigma}{\beta} \bt{\ga}{\nu}{\delta} \Ag_\sigma
             \de{V}^{\alpha\beta\ga\delta}
             = 0
             \label{eq:last_feq}
\end{align}


\begin{thebibliography}{88}%
\makeatletter
\providecommand \@ifxundefined [1]{%
 \@ifx{#1\undefined}
}%
\providecommand \@ifnum [1]{%
 \ifnum #1\expandafter \@firstoftwo
 \else \expandafter \@secondoftwo
 \fi
}%
\providecommand \@ifx [1]{%
 \ifx #1\expandafter \@firstoftwo
 \else \expandafter \@secondoftwo
 \fi
}%
\providecommand \natexlab [1]{#1}%
\providecommand \enquote  [1]{``#1''}%
\providecommand \bibnamefont  [1]{#1}%
\providecommand \bibfnamefont [1]{#1}%
\providecommand \citenamefont [1]{#1}%
\providecommand \href@noop [0]{\@secondoftwo}%
\providecommand \href [0]{\begingroup \@sanitize@url \@href}%
\providecommand \@href[1]{\@@startlink{#1}\@@href}%
\providecommand \@@href[1]{\endgroup#1\@@endlink}%
\providecommand \@sanitize@url [0]{\catcode `\\12\catcode `\$12\catcode
  `\&12\catcode `\#12\catcode `\^12\catcode `\_12\catcode `\%12\relax}%
\providecommand \@@startlink[1]{}%
\providecommand \@@endlink[0]{}%
\providecommand \url  [0]{\begingroup\@sanitize@url \@url }%
\providecommand \@url [1]{\endgroup\@href {#1}{\urlprefix }}%
\providecommand \urlprefix  [0]{URL }%
\providecommand \Eprint [0]{\href }%
\providecommand \doibase [0]{http://dx.doi.org/}%
\providecommand \selectlanguage [0]{\@gobble}%
\providecommand \bibinfo  [0]{\@secondoftwo}%
\providecommand \bibfield  [0]{\@secondoftwo}%
\providecommand \translation [1]{[#1]}%
\providecommand \BibitemOpen [0]{}%
\providecommand \bibitemStop [0]{}%
\providecommand \bibitemNoStop [0]{.\EOS\space}%
\providecommand \EOS [0]{\spacefactor3000\relax}%
\providecommand \BibitemShut  [1]{\csname bibitem#1\endcsname}%
\let\auto@bib@innerbib\@empty
\bibitem [{\citenamefont
  {Einstein}(1915{\natexlab{a}})}]{einstein15_grund_allgem_relat_und_anwen}%
  \BibitemOpen
  \bibfield  {author} {\bibinfo {author} {\bibfnamefont {Albert}\ \bibnamefont
  {Einstein}},\ }\bibfield  {title} {\enquote {\bibinfo {title} {Grundgedanken
  der allgemeinen relativit{\"a}tstheorie und anwendung dieser theorie in der
  astronomie},}\ }\href@noop {} {\bibfield  {journal} {\bibinfo  {journal}
  {Preussische Akademie der Wissenschaften, Sitzungsberichte}\ }\textbf
  {\bibinfo {volume} {315}},\ \bibinfo {pages} {778--786} (\bibinfo {year}
  {1915}{\natexlab{a}})}\BibitemShut {NoStop}%
\bibitem [{\citenamefont
  {Einstein}(1915{\natexlab{b}})}]{einstein15_zur_allgem_relat}%
  \BibitemOpen
  \bibfield  {author} {\bibinfo {author} {\bibfnamefont {Albert}\ \bibnamefont
  {Einstein}},\ }\bibfield  {title} {\enquote {\bibinfo {title} {Zur
  allgemeinen relativit{\"a}tstheorie},}\ }\href@noop {} {\bibfield  {journal}
  {\bibinfo  {journal} {Sitzungsber. preuss. Akad. Wiss.}\ }\textbf {\bibinfo
  {volume} {1}},\ \bibinfo {pages} {778} (\bibinfo {year}
  {1915}{\natexlab{b}})}\BibitemShut {NoStop}%
\bibitem [{\citenamefont
  {Einstein}(1915{\natexlab{c}})}]{einstein15_erklar_perih_merkur_aus_allgem_relat}%
  \BibitemOpen
  \bibfield  {author} {\bibinfo {author} {\bibfnamefont {Albert}\ \bibnamefont
  {Einstein}},\ }\bibfield  {title} {\enquote {\bibinfo {title} {Erklarung der
  perihelionbewegung der merkur aus der allgemeinen relativitatstheorie},}\
  }\href@noop {} {\bibfield  {journal} {\bibinfo  {journal} {Sitzungsber.
  preuss. Akad. Wiss.}\ }\textbf {\bibinfo {volume} {47}},\ \bibinfo {pages}
  {831} (\bibinfo {year} {1915}{\natexlab{c}})}\BibitemShut {NoStop}%
\bibitem [{\citenamefont
  {Einstein}(1915{\natexlab{d}})}]{einstein15_feldg_gravit}%
  \BibitemOpen
  \bibfield  {author} {\bibinfo {author} {\bibfnamefont {Albert}\ \bibnamefont
  {Einstein}},\ }\bibfield  {title} {\enquote {\bibinfo {title} {Die
  feldgleichungen der gravitation},}\ }\href@noop {} {\bibfield  {journal}
  {\bibinfo  {journal} {Sitzungsber. preuss. Akad. Wiss.}\ }\textbf {\bibinfo
  {volume} {1}},\ \bibinfo {pages} {844} (\bibinfo {year}
  {1915}{\natexlab{d}})}\BibitemShut {NoStop}%
\bibitem [{\citenamefont {Einstein}(1916)}]{einstein16_grund_allgem_relat}%
  \BibitemOpen
  \bibfield  {author} {\bibinfo {author} {\bibfnamefont {Albert}\ \bibnamefont
  {Einstein}},\ }\bibfield  {title} {\enquote {\bibinfo {title} {Die grundlage
  der allgemeinen relativit\"atstheorie},}\ }\href@noop {} {\bibfield
  {journal} {\bibinfo  {journal} {Ann. Phys.}\ }\textbf {\bibinfo {volume}
  {49}},\ \bibinfo {pages} {284--339} (\bibinfo {year} {1916})}\BibitemShut
  {NoStop}%
\bibitem [{\citenamefont {Will}(2014)}]{will14_confr_between_geren_relat}%
  \BibitemOpen
  \bibfield  {author} {\bibinfo {author} {\bibfnamefont {Clifford~M.}\
  \bibnamefont {Will}},\ }\bibfield  {title} {\enquote {\bibinfo {title} {{The
  Confrontation Between Gerenal Relativity and experiment}},}\ }\href {\doibase
  10.12942/lrr-2014-4} {\bibfield  {journal} {\bibinfo  {journal} {Living Rev.
  Rel.}\ }\textbf {\bibinfo {volume} {17}},\ \bibinfo {pages} {4} (\bibinfo
  {year} {2014})},\ \Eprint {http://arxiv.org/abs/1403.7377} {arXiv:1403.7377
  [gr-qc]} \BibitemShut {NoStop}%
\bibitem [{\citenamefont {Abbott}\ \emph {et~al.}(2016)\citenamefont {Abbott}
  \emph {et~al.}}]{abbott16_gw151}%
  \BibitemOpen
  \bibfield  {author} {\bibinfo {author} {\bibfnamefont {B.~P.}\ \bibnamefont
  {Abbott}} \emph {et~al.} (\bibinfo {collaboration} {Virgo, LIGO
  Scientific}),\ }\bibfield  {title} {\enquote {\bibinfo {title} {{GW151226:
  Observation of Gravitational Waves From a 22-solar-Mass Binary Black Hole
  Coalescence}},}\ }\href {\doibase 10.1103/PhysRevLett.116.241103} {\bibfield
  {journal} {\bibinfo  {journal} {Phys. Rev. Lett.}\ }\textbf {\bibinfo
  {volume} {116}},\ \bibinfo {pages} {241103} (\bibinfo {year} {2016})},\
  \Eprint {http://arxiv.org/abs/1606.04855} {arXiv:1606.04855 [gr-qc]}
  \BibitemShut {NoStop}%
\bibitem [{\citenamefont {Abbott}\ \emph {et~al.}(2017)\citenamefont {Abbott}
  \emph {et~al.}}]{abbott17_gravit_waves_gamma}%
  \BibitemOpen
  \bibfield  {author} {\bibinfo {author} {\bibfnamefont {B.~P.}\ \bibnamefont
  {Abbott}} \emph {et~al.},\ }\bibfield  {title} {\enquote {\bibinfo {title}
  {Gravitational waves and gamma-rays from a binary neutron star merger:
  Gw170817 and grb170817a},}\ }\href {\doibase 10.3847/2041-8213/aa920c}
  {\bibfield  {journal} {\bibinfo  {journal} {Astrophys. J.}\ }\textbf
  {\bibinfo {volume} {848}},\ \bibinfo {pages} {L13} (\bibinfo {year}
  {2017})}\BibitemShut {NoStop}%
\bibitem [{\citenamefont
  {DeWitt}(1967{\natexlab{a}})}]{dewitt67_quant_theor_gravit_I}%
  \BibitemOpen
  \bibfield  {author} {\bibinfo {author} {\bibfnamefont {Bryce~S.}\
  \bibnamefont {DeWitt}},\ }\bibfield  {title} {\enquote {\bibinfo {title}
  {{Quantum Theory of Gravity. 1. The Canonical Theory}},}\ }\href {\doibase
  10.1103/PhysRev.160.1113} {\bibfield  {journal} {\bibinfo  {journal} {Phys.
  Rev.}\ }\textbf {\bibinfo {volume} {160}},\ \bibinfo {pages} {1113} (\bibinfo
  {year} {1967}{\natexlab{a}})}\BibitemShut {NoStop}%
\bibitem [{\citenamefont
  {DeWitt}(1967{\natexlab{b}})}]{dewitt67_quant_theor_gravit_II}%
  \BibitemOpen
  \bibfield  {author} {\bibinfo {author} {\bibfnamefont {Bryce~S.}\
  \bibnamefont {DeWitt}},\ }\bibfield  {title} {\enquote {\bibinfo {title}
  {{Quantum Theory of Gravity. 2. The Manifestly Covariant Theory}},}\ }\href
  {\doibase 10.1103/PhysRev.162.1195} {\bibfield  {journal} {\bibinfo
  {journal} {Phys. Rev.}\ }\textbf {\bibinfo {volume} {162}},\ \bibinfo {pages}
  {1195} (\bibinfo {year} {1967}{\natexlab{b}})}\BibitemShut {NoStop}%
\bibitem [{\citenamefont {Deser}\ and\ \citenamefont {van
  Nieuwenhuizen}(1974{\natexlab{a}})}]{deser74_one_loop_diver_quant_einst_maxwel_field}%
  \BibitemOpen
  \bibfield  {author} {\bibinfo {author} {\bibfnamefont {Stanley}\ \bibnamefont
  {Deser}}\ and\ \bibinfo {author} {\bibfnamefont {P.}~\bibnamefont {van
  Nieuwenhuizen}},\ }\bibfield  {title} {\enquote {\bibinfo {title} {{One Loop
  Divergences of Quantized Einstein-Maxwell Fields}},}\ }\href {\doibase
  10.1103/PhysRevD.10.401} {\bibfield  {journal} {\bibinfo  {journal} {Phys.
  Rev. D}\ }\textbf {\bibinfo {volume} {10}},\ \bibinfo {pages} {401} (\bibinfo
  {year} {1974}{\natexlab{a}})}\BibitemShut {NoStop}%
\bibitem [{\citenamefont {Deser}\ and\ \citenamefont {van
  Nieuwenhuizen}(1974{\natexlab{b}})}]{deser74_nonren_quant_dirac_einst_system}%
  \BibitemOpen
  \bibfield  {author} {\bibinfo {author} {\bibfnamefont {Stanley}\ \bibnamefont
  {Deser}}\ and\ \bibinfo {author} {\bibfnamefont {P.}~\bibnamefont {van
  Nieuwenhuizen}},\ }\bibfield  {title} {\enquote {\bibinfo {title}
  {{Nonrenormalizability of the Quantized Dirac-Einstein System}},}\ }\href
  {\doibase 10.1103/PhysRevD.10.411} {\bibfield  {journal} {\bibinfo  {journal}
  {Phys. Rev. D}\ }\textbf {\bibinfo {volume} {10}},\ \bibinfo {pages} {411}
  (\bibinfo {year} {1974}{\natexlab{b}})}\BibitemShut {NoStop}%
\bibitem [{\citenamefont {'t~Hooft}\ and\ \citenamefont
  {Veltman}(1974)}]{hooft74_one_loop_diver_theor}%
  \BibitemOpen
  \bibfield  {author} {\bibinfo {author} {\bibfnamefont {Gerard}\ \bibnamefont
  {'t~Hooft}}\ and\ \bibinfo {author} {\bibfnamefont {M.~J.~G.}\ \bibnamefont
  {Veltman}},\ }\bibfield  {title} {\enquote {\bibinfo {title} {{One Loop
  Divergencies in the Theory of gravitation}},}\ }\href@noop {} {\bibfield
  {journal} {\bibinfo  {journal} {Annales Poincare Phys. Theor. A}\ }\textbf
  {\bibinfo {volume} {20}},\ \bibinfo {pages} {69} (\bibinfo {year}
  {1974})}\BibitemShut {NoStop}%
\bibitem [{\citenamefont
  {Ashtekar}(1986)}]{ashtekar86_new_variab_class_quant_gravit}%
  \BibitemOpen
  \bibfield  {author} {\bibinfo {author} {\bibfnamefont {A.}~\bibnamefont
  {Ashtekar}},\ }\bibfield  {title} {\enquote {\bibinfo {title} {{New Variables
  for Classical and Quantum Gravity}},}\ }\href {\doibase
  10.1103/PhysRevLett.57.2244} {\bibfield  {journal} {\bibinfo  {journal}
  {Phys. Rev. Lett.}\ }\textbf {\bibinfo {volume} {57}},\ \bibinfo {pages}
  {2244--2247} (\bibinfo {year} {1986})}\BibitemShut {NoStop}%
\bibitem [{\citenamefont
  {Ashtekar}(1987)}]{ashtekar87_new_hamil_formul_gener_relat}%
  \BibitemOpen
  \bibfield  {author} {\bibinfo {author} {\bibfnamefont {A.}~\bibnamefont
  {Ashtekar}},\ }\bibfield  {title} {\enquote {\bibinfo {title} {{New
  Hamiltonian Formulation of General Relativity}},}\ }\href {\doibase
  10.1103/PhysRevD.36.1587} {\bibfield  {journal} {\bibinfo  {journal} {Phys.
  Rev. D}\ }\textbf {\bibinfo {volume} {36}},\ \bibinfo {pages} {1587}
  (\bibinfo {year} {1987})}\BibitemShut {NoStop}%
\bibitem [{\citenamefont {Zwicky}(1937)}]{zwicky37_masses_nebul_clust_nebul}%
  \BibitemOpen
  \bibfield  {author} {\bibinfo {author} {\bibfnamefont {F.}~\bibnamefont
  {Zwicky}},\ }\bibfield  {title} {\enquote {\bibinfo {title} {On the masses of
  nebulae and of clusters of nebulae},}\ }\href {\doibase 10.1086/143864}
  {\bibfield  {journal} {\bibinfo  {journal} {Astrophys. J.}\ }\textbf
  {\bibinfo {volume} {86}},\ \bibinfo {pages} {217} (\bibinfo {year}
  {1937})}\BibitemShut {NoStop}%
\bibitem [{\citenamefont {Rubin}\ and\ \citenamefont
  {W.~Kent}(1970)}]{rubin70_rotat_androm_nebul_from_spect}%
  \BibitemOpen
  \bibfield  {author} {\bibinfo {author} {\bibfnamefont {Vera~C.}\ \bibnamefont
  {Rubin}}\ and\ \bibinfo {author} {\bibfnamefont {Jr.}\ \bibnamefont
  {W.~Kent}, \bibfnamefont {Ford}},\ }\bibfield  {title} {\enquote {\bibinfo
  {title} {Rotation of the andromeda nebula from a spectroscopic survey of
  emission regions},}\ }\href {\doibase 10.1086/150317} {\bibfield  {journal}
  {\bibinfo  {journal} {Astrophys. J.}\ }\textbf {\bibinfo {volume} {159}},\
  \bibinfo {pages} {379} (\bibinfo {year} {1970})}\BibitemShut {NoStop}%
\bibitem [{\citenamefont {Sofue}\ and\ \citenamefont
  {Rubin}(2001)}]{sofue01_rotat_curves_spiral}%
  \BibitemOpen
  \bibfield  {author} {\bibinfo {author} {\bibfnamefont {Yoshiaki}\
  \bibnamefont {Sofue}}\ and\ \bibinfo {author} {\bibfnamefont {Vera}\
  \bibnamefont {Rubin}},\ }\bibfield  {title} {\enquote {\bibinfo {title}
  {{Rotation Curves of Spiral galaxies}},}\ }\href {\doibase
  10.1146/annurev.astro.39.1.137} {\bibfield  {journal} {\bibinfo  {journal}
  {Ann. Rev. Astron. Astrophys.}\ }\textbf {\bibinfo {volume} {39}},\ \bibinfo
  {pages} {137} (\bibinfo {year} {2001})},\ \Eprint
  {http://arxiv.org/abs/astro-ph/0010594} {arXiv:astro-ph/0010594 [astro-ph]}
  \BibitemShut {NoStop}%
\bibitem [{\citenamefont {Riess}\ \emph {et~al.}(1998)\citenamefont {Riess}
  \emph {et~al.}}]{riess98_obser_eviden_from_super_accel}%
  \BibitemOpen
  \bibfield  {author} {\bibinfo {author} {\bibfnamefont {Adam~G.}\ \bibnamefont
  {Riess}} \emph {et~al.} (\bibinfo {collaboration} {Supernova Search Team}),\
  }\bibfield  {title} {\enquote {\bibinfo {title} {{Observational Evidence From
  Supernovae for an Accelerating Universe and a Cosmological constant}},}\
  }\href {\doibase 10.1086/300499} {\bibfield  {journal} {\bibinfo  {journal}
  {Astron. J.}\ }\textbf {\bibinfo {volume} {116}},\ \bibinfo {pages} {1009}
  (\bibinfo {year} {1998})},\ \Eprint {http://arxiv.org/abs/astro-ph/9805201}
  {arXiv:astro-ph/9805201 [astro-ph]} \BibitemShut {NoStop}%
\bibitem [{\citenamefont {Perlmutter}\ \emph {et~al.}(1999)\citenamefont
  {Perlmutter} \emph
  {et~al.}}]{perlmutter99_measur_oemeg_lambd_from_high_redsh_super}%
  \BibitemOpen
  \bibfield  {author} {\bibinfo {author} {\bibfnamefont {S.}~\bibnamefont
  {Perlmutter}} \emph {et~al.} (\bibinfo {collaboration} {{Supernova Cosmology
  Project}}),\ }\bibfield  {title} {\enquote {\bibinfo {title} {{Measurements
  of \(\Omega\) and \(\Lambda\) From 42 High-Redshift Supernovae}},}\ }\href
  {\doibase 10.1086/307221} {\bibfield  {journal} {\bibinfo  {journal}
  {Astrophys. J.}\ }\textbf {\bibinfo {volume} {517}},\ \bibinfo {pages} {565}
  (\bibinfo {year} {1999})}\BibitemShut {NoStop}%
\bibitem [{\citenamefont {Cartan}(1922)}]{cartan22_sur_une_de_la_notion}%
  \BibitemOpen
  \bibfield  {author} {\bibinfo {author} {\bibfnamefont {Elie}\ \bibnamefont
  {Cartan}},\ }\bibfield  {title} {\enquote {\bibinfo {title} {Sur une
  g\'en\'eralisation de la notion de courbure de riemann et les espaces \`a
  torsion},}\ }\href {http://gallica.bnf.fr/ark:/12148/bpt6k3127j.image.langFR}
  {\bibfield  {journal} {\bibinfo  {journal} {C. R. Acad. Sci. Paris}\ }\textbf
  {\bibinfo {volume} {174}},\ \bibinfo {pages} {593} (\bibinfo {year}
  {1922})}\BibitemShut {NoStop}%
\bibitem [{\citenamefont {Cartan}(1923)}]{cartan23_sur_les_connex_affin_et}%
  \BibitemOpen
  \bibfield  {author} {\bibinfo {author} {\bibfnamefont {Elie}\ \bibnamefont
  {Cartan}},\ }\bibfield  {title} {\enquote {\bibinfo {title} {Sur les
  vari{\'e}t{\'e}s {\`a} connexion affine et la th{\'e}orie de la
  relativit{\'e} g{\'e}n{\'e}ralis{\'e}e (premi{\`e}re partie)},}\ }\href
  {http://archive.numdam.org/article/ASENS_1923_3_40__325_0.pdf} {\bibfield
  {journal} {\bibinfo  {journal} {Ann. Ec. Norm. Super.}\ }\textbf {\bibinfo
  {volume} {40}},\ \bibinfo {pages} {325} (\bibinfo {year} {1923})}\BibitemShut
  {NoStop}%
\bibitem [{\citenamefont {Cartan}(1924)}]{cartan24_sur_les_connex_affin_et}%
  \BibitemOpen
  \bibfield  {author} {\bibinfo {author} {\bibfnamefont {Elie}\ \bibnamefont
  {Cartan}},\ }\bibfield  {title} {\enquote {\bibinfo {title} {Sur les
  vari\'et\'es \`a connexion affine, et la th\'eorie de la relativit\'e
  g\'en\'eralis\'ee (premi\`ere partie) (suite)},}\ }\href
  {http://www.numdam.org/numdam-bin/item?id=ASENS_1924_3_41__1_0} {\bibfield
  {journal} {\bibinfo  {journal} {Ann. Ec. Norm. Super.}\ }\textbf {\bibinfo
  {volume} {41}},\ \bibinfo {pages} {1} (\bibinfo {year} {1924})}\BibitemShut
  {NoStop}%
\bibitem [{\citenamefont {Cartan}(1925)}]{cartan25_sur_les_connex_affin_et}%
  \BibitemOpen
  \bibfield  {author} {\bibinfo {author} {\bibfnamefont {Elie}\ \bibnamefont
  {Cartan}},\ }\bibfield  {title} {\enquote {\bibinfo {title} {Sur les
  vari\'et\'es \`a connexion affine et la th\'eorie de la relativit\'e
  g\'en\'eralis\'ee, (deuxi\`eme partie)},}\ }\href
  {http://www.numdam.org/numdam-bin/item?id=ASENS_1925_3_42__17_0} {\bibfield
  {journal} {\bibinfo  {journal} {Ann. Ec. Norm. Super.}\ }\textbf {\bibinfo
  {volume} {42}},\ \bibinfo {pages} {17} (\bibinfo {year} {1925})}\BibitemShut
  {NoStop}%
\bibitem [{\citenamefont {Kaluza}(1921)}]{kaluza21_probl_unity_physic}%
  \BibitemOpen
  \bibfield  {author} {\bibinfo {author} {\bibfnamefont {Theodor}\ \bibnamefont
  {Kaluza}},\ }\bibfield  {title} {\enquote {\bibinfo {title} {{On the Problem
  of Unity in Physics}},}\ }\href@noop {} {\bibfield  {journal} {\bibinfo
  {journal} {Sitzungsber. Preuss. Akad. Wiss. Berlin (Math. Phys.)}\ }\textbf
  {\bibinfo {volume} {1921}},\ \bibinfo {pages} {966} (\bibinfo {year}
  {1921})}\BibitemShut {NoStop}%
\bibitem [{\citenamefont {Klein}(1926)}]{klein26_quant_theor_five_dimen_theor}%
  \BibitemOpen
  \bibfield  {author} {\bibinfo {author} {\bibfnamefont {O.}~\bibnamefont
  {Klein}},\ }\bibfield  {title} {\enquote {\bibinfo {title} {{Quantum Theory
  and Five-Dimensional Theory of relativity}},}\ }\href {\doibase
  10.1007/BF01397481} {\bibfield  {journal} {\bibinfo  {journal} {Z. Phys.}\
  }\textbf {\bibinfo {volume} {37}},\ \bibinfo {pages} {895} (\bibinfo {year}
  {1926})}\BibitemShut {NoStop}%
\bibitem [{\citenamefont {Lovelock}(1971)}]{lovelock71_einst_tensor_its}%
  \BibitemOpen
  \bibfield  {author} {\bibinfo {author} {\bibfnamefont {David}\ \bibnamefont
  {Lovelock}},\ }\bibfield  {title} {\enquote {\bibinfo {title} {{The Einstein
  Tensor and Its generalizations}},}\ }\href {\doibase 10.1063/1.1665613}
  {\bibfield  {journal} {\bibinfo  {journal} {J. Math. Phys.}\ }\textbf
  {\bibinfo {volume} {12}},\ \bibinfo {pages} {498--501} (\bibinfo {year}
  {1971})}\BibitemShut {NoStop}%
\bibitem [{\citenamefont {Hehl}\ \emph {et~al.}(1995)\citenamefont {Hehl},
  \citenamefont {McCrea}, \citenamefont {Mielke},\ and\ \citenamefont
  {Ne'eman}}]{hehl95_metric_affin_gauge_theor_gravit}%
  \BibitemOpen
  \bibfield  {author} {\bibinfo {author} {\bibfnamefont {Friedrich~W.}\
  \bibnamefont {Hehl}}, \bibinfo {author} {\bibfnamefont {J.~Dermott}\
  \bibnamefont {McCrea}}, \bibinfo {author} {\bibfnamefont {Eckehard~W.}\
  \bibnamefont {Mielke}}, \ and\ \bibinfo {author} {\bibfnamefont {Yuval}\
  \bibnamefont {Ne'eman}},\ }\bibfield  {title} {\enquote {\bibinfo {title}
  {{Metric Affine Gauge Theory of Gravity: Field Equations, Noether Identities,
  World Spinors, and Breaking of Dilation invariance}},}\ }\href {\doibase
  10.1016/0370-1573(94)00111-F} {\bibfield  {journal} {\bibinfo  {journal}
  {Phys. Rep.}\ }\textbf {\bibinfo {volume} {258}},\ \bibinfo {pages} {1--171}
  (\bibinfo {year} {1995})},\ \Eprint {http://arxiv.org/abs/gr-qc/9402012}
  {arXiv:gr-qc/9402012 [gr-qc]} \BibitemShut {NoStop}%
\bibitem [{\citenamefont {Mardones}\ and\ \citenamefont
  {Zanelli}(1991)}]{mardones91_lovel}%
  \BibitemOpen
  \bibfield  {author} {\bibinfo {author} {\bibfnamefont {Alejandro}\
  \bibnamefont {Mardones}}\ and\ \bibinfo {author} {\bibfnamefont {Jorge}\
  \bibnamefont {Zanelli}},\ }\bibfield  {title} {\enquote {\bibinfo {title}
  {{Lovelock--Cartan Theory of gravity}},}\ }\href {\doibase
  10.1088/0264-9381/8/8/018} {\bibfield  {journal} {\bibinfo  {journal} {Class.
  Quant. Grav.}\ }\textbf {\bibinfo {volume} {8}},\ \bibinfo {pages} {1545}
  (\bibinfo {year} {1991})}\BibitemShut {NoStop}%
\bibitem [{\citenamefont {Eddington}(1923)}]{eddington23}%
  \BibitemOpen
  \bibfield  {author} {\bibinfo {author} {\bibfnamefont {Arthur~S.}\
  \bibnamefont {Eddington}},\ }\href@noop {} {\emph {\bibinfo {title} {The
  mathematical theory of relativity}}}\ (\bibinfo  {publisher} {Cambridge
  University Press},\ \bibinfo {year} {1923})\BibitemShut {NoStop}%
\bibitem [{\citenamefont {Schr{\"o}dinger}(1950)}]{schroedinger50_space}%
  \BibitemOpen
  \bibfield  {author} {\bibinfo {author} {\bibfnamefont {Erwin}\ \bibnamefont
  {Schr{\"o}dinger}},\ }\href@noop {} {\emph {\bibinfo {title} {Space-time
  structure}}}\ (\bibinfo  {publisher} {Cambridge University Press},\ \bibinfo
  {year} {1950})\BibitemShut {NoStop}%
\bibitem [{\citenamefont
  {Kijowski}(1978)}]{kijowski78_new_variat_princ_gener_relat}%
  \BibitemOpen
  \bibfield  {author} {\bibinfo {author} {\bibfnamefont {Jerzy}\ \bibnamefont
  {Kijowski}},\ }\bibfield  {title} {\enquote {\bibinfo {title} {On a new
  variational principle in general relativity and the energy of the
  gravitational field},}\ }\href {\doibase 10.1007/bf00759646} {\bibfield
  {journal} {\bibinfo  {journal} {Gen. Rel. Grav.}\ }\textbf {\bibinfo {volume}
  {9}},\ \bibinfo {pages} {857} (\bibinfo {year} {1978})}\BibitemShut {NoStop}%
\bibitem [{\citenamefont
  {Krasnov}(2006)}]{krasnov06_renor_non_metric_quant_gravit}%
  \BibitemOpen
  \bibfield  {author} {\bibinfo {author} {\bibfnamefont {Kirill}\ \bibnamefont
  {Krasnov}},\ }\bibfield  {title} {\enquote {\bibinfo {title} {{Renormalizable
  Non-Metric Quantum Gravity?}}}\ }\href@noop {} {\  (\bibinfo {year}
  {2006})},\ \Eprint {http://arxiv.org/abs/hep-th/0611182}
  {arXiv:hep-th/0611182 [hep-th]} \BibitemShut {NoStop}%
\bibitem [{\citenamefont {Krasnov}(2007)}]{krasnov07_non_metric_gravit}%
  \BibitemOpen
  \bibfield  {author} {\bibinfo {author} {\bibfnamefont {Kirill}\ \bibnamefont
  {Krasnov}},\ }\bibfield  {title} {\enquote {\bibinfo {title} {{Non-Metric
  Gravity: A Status report}},}\ }\href {\doibase 10.1142/S021773230702590X}
  {\bibfield  {journal} {\bibinfo  {journal} {Mod. Phys. Lett. A}\ }\textbf
  {\bibinfo {volume} {22}},\ \bibinfo {pages} {3013--3026} (\bibinfo {year}
  {2007})},\ \Eprint {http://arxiv.org/abs/0711.0697} {arXiv:0711.0697 [gr-qc]}
  \BibitemShut {NoStop}%
\bibitem [{\citenamefont {Krasnov}(2008)}]{krasnov08_non_metric_gravit_I}%
  \BibitemOpen
  \bibfield  {author} {\bibinfo {author} {\bibfnamefont {Kirill}\ \bibnamefont
  {Krasnov}},\ }\bibfield  {title} {\enquote {\bibinfo {title} {{Non-Metric
  Gravity. I. Field Equations}},}\ }\href {\doibase
  10.1088/0264-9381/25/2/025001} {\bibfield  {journal} {\bibinfo  {journal}
  {Class. Quant. Grav.}\ }\textbf {\bibinfo {volume} {25}},\ \bibinfo {pages}
  {025001} (\bibinfo {year} {2008})},\ \Eprint
  {http://arxiv.org/abs/gr-qc/0703002} {arXiv:gr-qc/0703002 [gr-qc]}
  \BibitemShut {NoStop}%
\bibitem [{\citenamefont {Krasnov}\ and\ \citenamefont
  {Shtanov}(2008)}]{krasnov08_non_metric_gravit_II}%
  \BibitemOpen
  \bibfield  {author} {\bibinfo {author} {\bibfnamefont {Kirill}\ \bibnamefont
  {Krasnov}}\ and\ \bibinfo {author} {\bibfnamefont {Yuri}\ \bibnamefont
  {Shtanov}},\ }\bibfield  {title} {\enquote {\bibinfo {title} {{Non-Metric
  Gravity. II. Spherically Symmetric Solution, Missing Mass and Redshifts of
  Quasars}},}\ }\href {\doibase 10.1088/0264-9381/25/2/025002} {\bibfield
  {journal} {\bibinfo  {journal} {Class. Quant. Grav.}\ }\textbf {\bibinfo
  {volume} {25}},\ \bibinfo {pages} {025002} (\bibinfo {year} {2008})},\
  \Eprint {http://arxiv.org/abs/0705.2047} {arXiv:0705.2047 [gr-qc]}
  \BibitemShut {NoStop}%
\bibitem [{\citenamefont
  {Krasnov}(2011)}]{krasnov11_pure_connec_action_princ_gener_relat}%
  \BibitemOpen
  \bibfield  {author} {\bibinfo {author} {\bibfnamefont {Kirill}\ \bibnamefont
  {Krasnov}},\ }\bibfield  {title} {\enquote {\bibinfo {title} {{Pure
  Connection Action Principle for General Relativity}},}\ }\href {\doibase
  10.1103/PhysRevLett.106.251103} {\bibfield  {journal} {\bibinfo  {journal}
  {Phys. Rev. Lett.}\ }\textbf {\bibinfo {volume} {106}},\ \bibinfo {pages}
  {251103} (\bibinfo {year} {2011})},\ \Eprint {http://arxiv.org/abs/1103.4498}
  {arXiv:1103.4498 [gr-qc]} \BibitemShut {NoStop}%
\bibitem [{\citenamefont
  {Pop{\l}awski}(2007{\natexlab{a}})}]{poplawski07_unified_purel_affin_theor_gravit_elect}%
  \BibitemOpen
  \bibfield  {author} {\bibinfo {author} {\bibfnamefont {Nikodem~J.}\
  \bibnamefont {Pop{\l}awski}},\ }\bibfield  {title} {\enquote {\bibinfo
  {title} {A unified, purely affine theory of gravitation and
  electromagnetism},}\ }\href@noop {} {\  (\bibinfo {year}
  {2007}{\natexlab{a}})},\ \Eprint {http://arxiv.org/abs/0705.0351}
  {arXiv:0705.0351 [gr-qc]} \BibitemShut {NoStop}%
\bibitem [{\citenamefont
  {Pop{\l}awski}(2007{\natexlab{b}})}]{poplawski07_nonsy_purel_affin}%
  \BibitemOpen
  \bibfield  {author} {\bibinfo {author} {\bibfnamefont {Nikodem~J.}\
  \bibnamefont {Pop{\l}awski}},\ }\bibfield  {title} {\enquote {\bibinfo
  {title} {{On the Nonsymmetric Purely Affine gravity}},}\ }\href {\doibase
  10.1142/s0217732307025662} {\bibfield  {journal} {\bibinfo  {journal} {Mod.
  Phys. Lett. A}\ }\textbf {\bibinfo {volume} {22}},\ \bibinfo {pages} {2701}
  (\bibinfo {year} {2007}{\natexlab{b}})},\ \Eprint
  {http://arxiv.org/abs/gr-qc/0610132} {arXiv:gr-qc/0610132 [gr-qc]}
  \BibitemShut {NoStop}%
\bibitem [{\citenamefont {Pop{\l}awski}(2014)}]{poplawski14_affin_theor}%
  \BibitemOpen
  \bibfield  {author} {\bibinfo {author} {\bibfnamefont {Nikodem~J.}\
  \bibnamefont {Pop{\l}awski}},\ }\bibfield  {title} {\enquote {\bibinfo
  {title} {{Affine Theory of gravitation}},}\ }\href {\doibase
  10.1007/s10714-013-1625-7} {\bibfield  {journal} {\bibinfo  {journal} {Gen.
  Rel. Grav.}\ }\textbf {\bibinfo {volume} {46}},\ \bibinfo {pages} {1625}
  (\bibinfo {year} {2014})},\ \Eprint {http://arxiv.org/abs/1203.0294}
  {arXiv:1203.0294 [gr-qc]} \BibitemShut {NoStop}%
\bibitem [{\citenamefont {Castillo-Felisola}\ and\ \citenamefont
  {Skirzewski}(2015)}]{castillo-felisola15_polyn_model_purel_affin_gravit}%
  \BibitemOpen
  \bibfield  {author} {\bibinfo {author} {\bibfnamefont {Oscar}\ \bibnamefont
  {Castillo-Felisola}}\ and\ \bibinfo {author} {\bibfnamefont {Aureliano}\
  \bibnamefont {Skirzewski}},\ }\bibfield  {title} {\enquote {\bibinfo {title}
  {{A Polynomial Model of Purely Affine Gravity}},}\ }\href@noop {} {\bibfield
  {journal} {\bibinfo  {journal} {Rev. Mex. Fis.}\ }\textbf {\bibinfo {volume}
  {61}},\ \bibinfo {pages} {421} (\bibinfo {year} {2015})},\ \Eprint
  {http://arxiv.org/abs/1410.6183} {arXiv:1410.6183 [gr-qc]} \BibitemShut
  {NoStop}%
\bibitem [{\citenamefont {Castillo-Felisola}\ and\ \citenamefont
  {Skirzewski}(2018)}]{castillo-felisola18_einst_gravit_from_polyn_affin_model}%
  \BibitemOpen
  \bibfield  {author} {\bibinfo {author} {\bibfnamefont {Oscar}\ \bibnamefont
  {Castillo-Felisola}}\ and\ \bibinfo {author} {\bibfnamefont {Aureliano}\
  \bibnamefont {Skirzewski}},\ }\bibfield  {title} {\enquote {\bibinfo {title}
  {Einstein's gravity from a polynomial affine model},}\ }\href {\doibase
  10.1088/1361-6382/aaa699} {\bibfield  {journal} {\bibinfo  {journal} {Class.
  Quant. Grav.}\ }\textbf {\bibinfo {volume} {35}},\ \bibinfo {pages} {055012}
  (\bibinfo {year} {2018})},\ \Eprint {http://arxiv.org/abs/1505.04634}
  {arXiv:1505.04634 [gr-qc]} \BibitemShut {NoStop}%
\bibitem [{\citenamefont
  {Castillo-Felisola}(2018)}]{castillo-felisola18_beyond_einstein}%
  \BibitemOpen
  \bibfield  {author} {\bibinfo {author} {\bibfnamefont {Oscar}\ \bibnamefont
  {Castillo-Felisola}},\ }\enquote {\bibinfo {title} {Gravity},}\ \ (\bibinfo
  {publisher} {IntechOpen},\ \bibinfo {year} {2018})\ Chap.\ \bibinfo {chapter}
  {Beyond Einstein: A Polynomial Affine Model of Gravity}, pp.\ \bibinfo
  {pages} {183--201},\ \Eprint {http://arxiv.org/abs/1902.09131}
  {arXiv:1902.09131 [gr-qc]} \BibitemShut {NoStop}%
\bibitem [{\citenamefont {Castillo-Felisola}\ \emph {et~al.}(2019)\citenamefont
  {Castillo-Felisola}, \citenamefont {Perdiguero},\ and\ \citenamefont
  {Orellana}}]{castillo-felisola18_cosmol}%
  \BibitemOpen
  \bibfield  {author} {\bibinfo {author} {\bibfnamefont {Oscar}\ \bibnamefont
  {Castillo-Felisola}}, \bibinfo {author} {\bibfnamefont {Jos\'e}\ \bibnamefont
  {Perdiguero}}, \ and\ \bibinfo {author} {\bibfnamefont {Oscar}\ \bibnamefont
  {Orellana}},\ }\enquote {\bibinfo {title} {Redifining standard model
  cosmology},}\ \ (\bibinfo  {publisher} {IntechOpen},\ \bibinfo {year}
  {2019})\ Chap.\ \bibinfo {chapter} {Cosmological Solutions to Polynomial
  Affine Gravity in the Torsion-Free Sector}, p.~\bibinfo {pages} {NA},\
  \Eprint {http://arxiv.org/abs/1808.05970} {arXiv:1808.05970 [gr-qc]}
  \BibitemShut {NoStop}%
\bibitem [{\citenamefont {McGady}\ and\ \citenamefont
  {Rodina}(2014)}]{mcgady14_higher_spin_massl_s_dimen}%
  \BibitemOpen
  \bibfield  {author} {\bibinfo {author} {\bibfnamefont {David~A.}\
  \bibnamefont {McGady}}\ and\ \bibinfo {author} {\bibfnamefont {Laurentiu}\
  \bibnamefont {Rodina}},\ }\bibfield  {title} {\enquote {\bibinfo {title}
  {{Higher-Spin Massless $S$-matrices in four-Dimensions}},}\ }\href {\doibase
  10.1103/PhysRevD.90.084048} {\bibfield  {journal} {\bibinfo  {journal} {Phys.
  Rev. D}\ }\textbf {\bibinfo {volume} {90}},\ \bibinfo {pages} {084048}
  (\bibinfo {year} {2014})},\ \Eprint {http://arxiv.org/abs/1311.2938}
  {arXiv:1311.2938 [hep-th]} \BibitemShut {NoStop}%
\bibitem [{\citenamefont {Camanho}\ \emph {et~al.}(2016)\citenamefont
  {Camanho}, \citenamefont {Edelstein}, \citenamefont {Maldacena},\ and\
  \citenamefont {Zhiboedov}}]{camanho16_causal_const_correc_to_gravit}%
  \BibitemOpen
  \bibfield  {author} {\bibinfo {author} {\bibfnamefont {Xian~O.}\ \bibnamefont
  {Camanho}}, \bibinfo {author} {\bibfnamefont {Jose~D.}\ \bibnamefont
  {Edelstein}}, \bibinfo {author} {\bibfnamefont {Juan}\ \bibnamefont
  {Maldacena}}, \ and\ \bibinfo {author} {\bibfnamefont {Alexander}\
  \bibnamefont {Zhiboedov}},\ }\bibfield  {title} {\enquote {\bibinfo {title}
  {{Causality Constraints on Corrections To the Graviton Three-Point
  Coupling}},}\ }\href {\doibase 10.1007/JHEP02(2016)020} {\bibfield  {journal}
  {\bibinfo  {journal} {J. High Energy Phys.}\ }\textbf {\bibinfo {volume}
  {02}},\ \bibinfo {pages} {020} (\bibinfo {year} {2016})},\ \Eprint
  {http://arxiv.org/abs/1407.5597} {arXiv:1407.5597 [hep-th]} \BibitemShut
  {NoStop}%
\bibitem [{\citenamefont
  {Zerwekh}(2013)}]{zerwekh13_quant_chrom_massiv_vector_field_adjoin_repres}%
  \BibitemOpen
  \bibfield  {author} {\bibinfo {author} {\bibfnamefont {Alfonso~R.}\
  \bibnamefont {Zerwekh}},\ }\bibfield  {title} {\enquote {\bibinfo {title} {On
  the quantum chromodynamics of a massive vector field in the adjoint
  representation},}\ }\href {\doibase 10.1142/s0217751x13500541} {\bibfield
  {journal} {\bibinfo  {journal} {Int. J. Mod. Phys. A}\ }\textbf {\bibinfo
  {volume} {28}},\ \bibinfo {pages} {1350054} (\bibinfo {year}
  {2013})}\BibitemShut {NoStop}%
\bibitem [{\citenamefont
  {Iosifidis}(2019{\natexlab{a}})}]{iosifidis19_metric_affin_gravit_cosmol_torsion}%
  \BibitemOpen
  \bibfield  {author} {\bibinfo {author} {\bibfnamefont {Damianos}\
  \bibnamefont {Iosifidis}},\ }\emph {\bibinfo {title} {Metric-Affine Gravity
  and Cosmology: aspects of Torsion and Non-Metricity in Gravity Theories}},\
  \href {http://arxiv.org/abs/1902.09643v1} {Ph.D. thesis},\ \bibinfo  {school}
  {Institute of Theoretical Physics - Physics Department of Aristotle
  University of Thessaloniki} (\bibinfo {year} {2019}{\natexlab{a}}),\ \Eprint
  {http://arxiv.org/abs/1902.09643} {arXiv:1902.09643 [gr-qc]} \BibitemShut
  {NoStop}%
\bibitem [{\citenamefont
  {Iosifidis}(2019{\natexlab{b}})}]{iosifidis19_exact_solvab_connec_metric_affin_gravit}%
  \BibitemOpen
  \bibfield  {author} {\bibinfo {author} {\bibfnamefont {Damianos}\
  \bibnamefont {Iosifidis}},\ }\bibfield  {title} {\enquote {\bibinfo {title}
  {Exactly solvable connections in metric-affine gravity},}\ }\href {\doibase
  10.1088/1361-6382/ab0be2} {\bibfield  {journal} {\bibinfo  {journal} {Class.
  Quant. Grav.}\ }\textbf {\bibinfo {volume} {36}},\ \bibinfo {pages} {085001}
  (\bibinfo {year} {2019}{\natexlab{b}})},\ \Eprint
  {http://arxiv.org/abs/1812.04031} {arXiv:1812.04031 [gr-qc]} \BibitemShut
  {NoStop}%
\bibitem [{\citenamefont {Parattu}\ \emph {et~al.}(2016)\citenamefont
  {Parattu}, \citenamefont {Chakraborty}, \citenamefont {Majhi},\ and\
  \citenamefont
  {Padmanabhan}}]{parattu16_bound_term_gravit_action_with_null_bound}%
  \BibitemOpen
  \bibfield  {author} {\bibinfo {author} {\bibfnamefont {Krishnamohan}\
  \bibnamefont {Parattu}}, \bibinfo {author} {\bibfnamefont {Sumanta}\
  \bibnamefont {Chakraborty}}, \bibinfo {author} {\bibfnamefont {{Bibhas
  Ranjan}}\ \bibnamefont {Majhi}}, \ and\ \bibinfo {author} {\bibfnamefont
  {T.}~\bibnamefont {Padmanabhan}},\ }\bibfield  {title} {\enquote {\bibinfo
  {title} {A boundary term for the gravitational action with null
  boundaries},}\ }\href {\doibase 10.1007/s10714-016-2093-7} {\bibfield
  {journal} {\bibinfo  {journal} {Gen. Rel. Grav.}\ }\textbf {\bibinfo {volume}
  {48}},\ \bibinfo {pages} {94} (\bibinfo {year} {2016})}\BibitemShut {NoStop}%
\bibitem [{\citenamefont {Krishnan}\ \emph {et~al.}(2017)\citenamefont
  {Krishnan}, \citenamefont {Maheshwari},\ and\ \citenamefont
  {Subramanian}}]{krishnan17_robin_gravit}%
  \BibitemOpen
  \bibfield  {author} {\bibinfo {author} {\bibfnamefont {Chethan}\ \bibnamefont
  {Krishnan}}, \bibinfo {author} {\bibfnamefont {Shubham}\ \bibnamefont
  {Maheshwari}}, \ and\ \bibinfo {author} {\bibfnamefont {{P. N. Bala}}\
  \bibnamefont {Subramanian}},\ }\bibfield  {title} {\enquote {\bibinfo {title}
  {Robin gravity},}\ }\href {\doibase 10.1088/1742-6596/883/1/012011}
  {\bibfield  {journal} {\bibinfo  {journal} {Journal of Physics: Conference
  Series}\ }\textbf {\bibinfo {volume} {883}},\ \bibinfo {pages} {012011}
  (\bibinfo {year} {2017})}\BibitemShut {NoStop}%
\bibitem [{\citenamefont {Krishnan}\ and\ \citenamefont
  {Raju}(2017)}]{krishnan17_neuman_bound_term_gravit}%
  \BibitemOpen
  \bibfield  {author} {\bibinfo {author} {\bibfnamefont {Chethan}\ \bibnamefont
  {Krishnan}}\ and\ \bibinfo {author} {\bibfnamefont {Avinash}\ \bibnamefont
  {Raju}},\ }\bibfield  {title} {\enquote {\bibinfo {title} {A neumann boundary
  term for gravity},}\ }\href {\doibase 10.1142/s0217732317500778} {\bibfield
  {journal} {\bibinfo  {journal} {Mod. Phys. Lett. A}\ }\textbf {\bibinfo
  {volume} {32}},\ \bibinfo {pages} {1750077} (\bibinfo {year}
  {2017})}\BibitemShut {NoStop}%
\bibitem [{\citenamefont {Lehner}\ \emph {et~al.}(2016)\citenamefont {Lehner},
  \citenamefont {Myers}, \citenamefont {Poisson},\ and\ \citenamefont
  {Sorkin}}]{lehner16_gravit_action_with_null_bound}%
  \BibitemOpen
  \bibfield  {author} {\bibinfo {author} {\bibfnamefont {Luis}\ \bibnamefont
  {Lehner}}, \bibinfo {author} {\bibfnamefont {{Robert C.}}\ \bibnamefont
  {Myers}}, \bibinfo {author} {\bibfnamefont {Eric}\ \bibnamefont {Poisson}}, \
  and\ \bibinfo {author} {\bibfnamefont {{Rafael D.}}\ \bibnamefont {Sorkin}},\
  }\bibfield  {title} {\enquote {\bibinfo {title} {Gravitational action with
  null boundaries},}\ }\href {\doibase 10.1103/physrevd.94.084046} {\bibfield
  {journal} {\bibinfo  {journal} {Phys. Rev. D}\ }\textbf {\bibinfo {volume}
  {94}},\ \bibinfo {pages} {084046} (\bibinfo {year} {2016})}\BibitemShut
  {NoStop}%
\bibitem [{\citenamefont {Hopfm{\"u}ller}\ and\ \citenamefont
  {Freidel}(2017)}]{hopfmueller17_gravit_degrees_freed_null_surfac}%
  \BibitemOpen
  \bibfield  {author} {\bibinfo {author} {\bibfnamefont {Florian}\ \bibnamefont
  {Hopfm{\"u}ller}}\ and\ \bibinfo {author} {\bibfnamefont {Laurent}\
  \bibnamefont {Freidel}},\ }\bibfield  {title} {\enquote {\bibinfo {title}
  {Gravity degrees of freedom on a null surface},}\ }\href {\doibase
  10.1103/physrevd.95.104006} {\bibfield  {journal} {\bibinfo  {journal} {Phys.
  Rev. D}\ }\textbf {\bibinfo {volume} {95}},\ \bibinfo {pages} {104006}
  (\bibinfo {year} {2017})}\BibitemShut {NoStop}%
\bibitem [{\citenamefont {Jubb}\ \emph {et~al.}(2017)\citenamefont {Jubb},
  \citenamefont {Samuel}, \citenamefont {Sorkin},\ and\ \citenamefont
  {Surya}}]{jubb17_bound_corner_terms_action_gener_relat}%
  \BibitemOpen
  \bibfield  {author} {\bibinfo {author} {\bibfnamefont {Ian}\ \bibnamefont
  {Jubb}}, \bibinfo {author} {\bibfnamefont {Joseph}\ \bibnamefont {Samuel}},
  \bibinfo {author} {\bibfnamefont {{Rafael D.}}\ \bibnamefont {Sorkin}}, \
  and\ \bibinfo {author} {\bibfnamefont {Sumati}\ \bibnamefont {Surya}},\
  }\bibfield  {title} {\enquote {\bibinfo {title} {Boundary and corner terms in
  the action for general relativity},}\ }\href {\doibase
  10.1088/1361-6382/aa6014} {\bibfield  {journal} {\bibinfo  {journal} {Class.
  Quant. Grav.}\ }\textbf {\bibinfo {volume} {34}},\ \bibinfo {pages} {065006}
  (\bibinfo {year} {2017})}\BibitemShut {NoStop}%
\bibitem [{\citenamefont {Nomizu}\ and\ \citenamefont
  {Sasaki}(1994)}]{nomizu94_affin}%
  \BibitemOpen
  \bibfield  {author} {\bibinfo {author} {\bibfnamefont {Katsumi}\ \bibnamefont
  {Nomizu}}\ and\ \bibinfo {author} {\bibfnamefont {Takeshi}\ \bibnamefont
  {Sasaki}},\ }\href@noop {} {\emph {\bibinfo {title} {Affine differential
  geometry}}}\ (\bibinfo  {publisher} {Cambridge University Press},\ \bibinfo
  {year} {1994})\BibitemShut {NoStop}%
\bibitem [{\citenamefont {Besse}(2007)}]{besse07_einst}%
  \BibitemOpen
  \bibfield  {author} {\bibinfo {author} {\bibfnamefont {Arthur~L.}\
  \bibnamefont {Besse}},\ }\href@noop {} {\emph {\bibinfo {title} {Einstein
  manifolds}}}\ (\bibinfo  {publisher} {Springer},\ \bibinfo {year}
  {2007})\BibitemShut {NoStop}%
\bibitem [{\citenamefont
  {Stephenson}(1958)}]{stephenson58_quadr_lagran_gener_relat}%
  \BibitemOpen
  \bibfield  {author} {\bibinfo {author} {\bibfnamefont {G.}~\bibnamefont
  {Stephenson}},\ }\bibfield  {title} {\enquote {\bibinfo {title} {Quadratic
  lagrangians and general relativity},}\ }\href {\doibase 10.1007/bf02724929}
  {\bibfield  {journal} {\bibinfo  {journal} {Nuovo Cimento}\ }\textbf
  {\bibinfo {volume} {9}},\ \bibinfo {pages} {263--269} (\bibinfo {year}
  {1958})}\BibitemShut {NoStop}%
\bibitem [{\citenamefont {Kilmister}\ and\ \citenamefont
  {Newman}(1961)}]{kilmister61_use_alg_struct_phys}%
  \BibitemOpen
  \bibfield  {author} {\bibinfo {author} {\bibfnamefont {C.~W.}\ \bibnamefont
  {Kilmister}}\ and\ \bibinfo {author} {\bibfnamefont {D.~J.}\ \bibnamefont
  {Newman}},\ }\bibfield  {title} {\enquote {\bibinfo {title} {The use of
  algebraic structures in physics},}\ }in\ \href {\doibase
  10.1017/S0305004100036008} {\emph {\bibinfo {booktitle} {Mathematical
  Proceedings of the Cambridge Philosophical Society}}},\ Vol.~\bibinfo
  {volume} {57}\ (\bibinfo  {publisher} {Cambridge University Press},\ \bibinfo
  {year} {1961})\ p.\ \bibinfo {pages} {851}\BibitemShut {NoStop}%
\bibitem [{\citenamefont {Yang}(1974)}]{yang74_integ_formal_gauge_field}%
  \BibitemOpen
  \bibfield  {author} {\bibinfo {author} {\bibfnamefont {C.~N.}\ \bibnamefont
  {Yang}},\ }\bibfield  {title} {\enquote {\bibinfo {title} {Integral formalism
  for gauge fields},}\ }\href {\doibase 10.1103/PhysRevLett.33.445} {\bibfield
  {journal} {\bibinfo  {journal} {Phys. Rev. Lett.}\ }\textbf {\bibinfo
  {volume} {33}},\ \bibinfo {pages} {445} (\bibinfo {year} {1974})}\BibitemShut
  {NoStop}%
\bibitem [{\citenamefont
  {Pavelle}(1975)}]{pavelle75_unphy_solut_yangs_gravit_field_equat}%
  \BibitemOpen
  \bibfield  {author} {\bibinfo {author} {\bibfnamefont {Richard}\ \bibnamefont
  {Pavelle}},\ }\bibfield  {title} {\enquote {\bibinfo {title} {Unphysical
  solutions of yang's gravitational-field equations},}\ }\href {\doibase
  10.1103/physrevlett.34.1114} {\bibfield  {journal} {\bibinfo  {journal}
  {Phys. Rev. Lett.}\ }\textbf {\bibinfo {volume} {34}},\ \bibinfo {pages}
  {1114--1114} (\bibinfo {year} {1975})}\BibitemShut {NoStop}%
\bibitem [{\citenamefont
  {Thompson}(1975)}]{thompson75_geomet_degen_solut_kilmis_yang_equat}%
  \BibitemOpen
  \bibfield  {author} {\bibinfo {author} {\bibfnamefont {A.~H.}\ \bibnamefont
  {Thompson}},\ }\bibfield  {title} {\enquote {\bibinfo {title} {Geometrically
  degenerate solutions of the kilmister-yang equations},}\ }\href {\doibase
  10.1103/PhysRevLett.35.320} {\bibfield  {journal} {\bibinfo  {journal} {Phys.
  Rev. Lett.}\ }\textbf {\bibinfo {volume} {35}},\ \bibinfo {pages} {320}
  (\bibinfo {year} {1975})}\BibitemShut {NoStop}%
\bibitem [{\citenamefont
  {Einstein}(1919)}]{einstein19_spiel_gravit_im_auf_ber}%
  \BibitemOpen
  \bibfield  {author} {\bibinfo {author} {\bibfnamefont {Albert}\ \bibnamefont
  {Einstein}},\ }\bibfield  {title} {\enquote {\bibinfo {title} {Spielen
  gravitationsfelder im auf ber der materiellen elementar-teilchen eine
  wesentliche rolle?}}\ }\href@noop {} {\bibfield  {journal} {\bibinfo
  {journal} {Sitzungsber. Preuss. Akad. Wiss. (Berlin)}\ ,\ \bibinfo {pages}
  {349--356}} (\bibinfo {year} {1919})}\BibitemShut {NoStop}%
\bibitem [{\citenamefont {Garfinkle}\ \emph {et~al.}(2018)\citenamefont
  {Garfinkle}, \citenamefont {Mead},\ and\ \citenamefont
  {Ringermacher}}]{garfinkle18_shape_orbit_flrw_spacet}%
  \BibitemOpen
  \bibfield  {author} {\bibinfo {author} {\bibfnamefont {David}\ \bibnamefont
  {Garfinkle}}, \bibinfo {author} {\bibfnamefont {Lawrence~R.}\ \bibnamefont
  {Mead}}, \ and\ \bibinfo {author} {\bibfnamefont {H.~I.}\ \bibnamefont
  {Ringermacher}},\ }\href {http://arxiv.org/abs/1808.06683v1} {\enquote
  {\bibinfo {title} {The shape of the orbit in flrw spacetimes},}\ } (\bibinfo
  {year} {2018}),\ \Eprint {http://arxiv.org/abs/1808.06683} {arXiv:1808.06683
  [gr-qc]} \BibitemShut {NoStop}%
\bibitem [{\citenamefont {Mielke}(2017)}]{mielke17_geomet_gauge_field}%
  \BibitemOpen
  \bibfield  {author} {\bibinfo {author} {\bibfnamefont {Eckehard~W.}\
  \bibnamefont {Mielke}},\ }\href {\doibase 10.1007/978-3-319-29734-7} {\emph
  {\bibinfo {title} {Geometrodynamics of Gauge Fields}}},\ Mathematical Physics
  Studies\ (\bibinfo  {publisher} {Springer International Publishing},\
  \bibinfo {year} {2017})\ p.\ \bibinfo {pages} {373}\BibitemShut {NoStop}%
\bibitem [{\citenamefont {Weinberg}(1989)}]{weinberg89_cosmol_const_probl}%
  \BibitemOpen
  \bibfield  {author} {\bibinfo {author} {\bibfnamefont {Steven}\ \bibnamefont
  {Weinberg}},\ }\bibfield  {title} {\enquote {\bibinfo {title} {The
  cosmological constant problem},}\ }\href {\doibase 10.1103/revmodphys.61.1}
  {\bibfield  {journal} {\bibinfo  {journal} {Rev. Mod. Phys.}\ }\textbf
  {\bibinfo {volume} {61}},\ \bibinfo {pages} {1--23} (\bibinfo {year}
  {1989})}\BibitemShut {NoStop}%
\bibitem [{\citenamefont {Ng}\ and\ \citenamefont {van
  Dam}(1991)}]{ng91_unimod_theor_gravit_cosmol_const}%
  \BibitemOpen
  \bibfield  {author} {\bibinfo {author} {\bibfnamefont {Y.~Jack}\ \bibnamefont
  {Ng}}\ and\ \bibinfo {author} {\bibfnamefont {H.}~\bibnamefont {van Dam}},\
  }\bibfield  {title} {\enquote {\bibinfo {title} {Unimodular theory of gravity
  and the cosmological constant},}\ }\href {\doibase 10.1063/1.529283}
  {\bibfield  {journal} {\bibinfo  {journal} {J. Math. Phys.}\ }\textbf
  {\bibinfo {volume} {32}},\ \bibinfo {pages} {1337--1340} (\bibinfo {year}
  {1991})}\BibitemShut {NoStop}%
\bibitem [{\citenamefont
  {Smolin}(2009)}]{smolin09_quant_unimod_gravit_cosmol_const_probl}%
  \BibitemOpen
  \bibfield  {author} {\bibinfo {author} {\bibfnamefont {Lee}\ \bibnamefont
  {Smolin}},\ }\bibfield  {title} {\enquote {\bibinfo {title} {Quantization of
  unimodular gravity and the cosmological constant problems},}\ }\href
  {\doibase 10.1103/physrevd.80.084003} {\bibfield  {journal} {\bibinfo
  {journal} {Phys. Rev. D}\ }\textbf {\bibinfo {volume} {80}} (\bibinfo {year}
  {2009}),\ 10.1103/physrevd.80.084003},\ \Eprint
  {http://arxiv.org/abs/0904.4841} {arXiv:0904.4841 [hep-th]} \BibitemShut
  {NoStop}%
\bibitem [{\citenamefont {Ellis}\ \emph {et~al.}(2011)\citenamefont {Ellis},
  \citenamefont {van Elst}, \citenamefont {Murugan},\ and\ \citenamefont
  {Uzan}}]{ellis11_trace_free_einst_equat_as}%
  \BibitemOpen
  \bibfield  {author} {\bibinfo {author} {\bibfnamefont {George~F.~R.}\
  \bibnamefont {Ellis}}, \bibinfo {author} {\bibfnamefont {Henk}\ \bibnamefont
  {van Elst}}, \bibinfo {author} {\bibfnamefont {Jeff}\ \bibnamefont
  {Murugan}}, \ and\ \bibinfo {author} {\bibfnamefont {Jean-Philippe}\
  \bibnamefont {Uzan}},\ }\bibfield  {title} {\enquote {\bibinfo {title} {On
  the trace-free einstein equations as a viable alternative to general
  relativity},}\ }\href {\doibase 10.1088/0264-9381/28/22/225007} {\bibfield
  {journal} {\bibinfo  {journal} {Class. Quant. Grav.}\ }\textbf {\bibinfo
  {volume} {28}},\ \bibinfo {pages} {225007} (\bibinfo {year} {2011})},\
  \Eprint {http://arxiv.org/abs/1008.1196} {arXiv:1008.1196 [gr-qc]}
  \BibitemShut {NoStop}%
\bibitem [{\citenamefont {Boskoff}\ and\ \citenamefont
  {Capozziello}(2019)}]{boskoff19_recov_cosmol_const_from_affin_geomet}%
  \BibitemOpen
  \bibfield  {author} {\bibinfo {author} {\bibfnamefont {Wladimir-Georges}\
  \bibnamefont {Boskoff}}\ and\ \bibinfo {author} {\bibfnamefont {Salvatore}\
  \bibnamefont {Capozziello}},\ }\bibfield  {title} {\enquote {\bibinfo {title}
  {Recovering the cosmological constant from affine geometry},}\ }\href
  {http://arxiv.org/abs/1908.02340v1} {\  (\bibinfo {year} {2019})},\ \Eprint
  {http://arxiv.org/abs/1908.02340} {arXiv:1908.02340 [gr-qc]} \BibitemShut
  {NoStop}%
\bibitem [{\citenamefont {Ivey}\ and\ \citenamefont
  {Landsberg}(2003)}]{ivey03_cartan_begin}%
  \BibitemOpen
  \bibfield  {author} {\bibinfo {author} {\bibfnamefont {Thomas~A.}\
  \bibnamefont {Ivey}}\ and\ \bibinfo {author} {\bibfnamefont {J.~M.}\
  \bibnamefont {Landsberg}},\ }\href@noop {} {\emph {\bibinfo {title} {Cartan
  for Beginners}}},\ \bibinfo {series} {Graduate Studies in Mathematics},
  Vol.~\bibinfo {volume} {61}\ (\bibinfo  {publisher} {AMS},\ \bibinfo {year}
  {2003})\BibitemShut {NoStop}%
\bibitem [{\citenamefont {Baez}(1994)}]{baez94_gauge}%
  \BibitemOpen
  \bibfield  {author} {\bibinfo {author} {\bibfnamefont {John}\ \bibnamefont
  {Baez}},\ }\href@noop {} {\emph {\bibinfo {title} {Gauge fields, knots, and
  gravity}}}\ (\bibinfo  {publisher} {World Scientific},\ \bibinfo {address}
  {Singapore River Edge, N.J},\ \bibinfo {year} {1994})\BibitemShut {NoStop}%
\bibitem [{\citenamefont
  {L\'evy-Leblond}(1965)}]{levy-leblond65_une_nouvel_limit_non_relat}%
  \BibitemOpen
  \bibfield  {author} {\bibinfo {author} {\bibfnamefont {Jean-Marc}\
  \bibnamefont {L\'evy-Leblond}},\ }\bibfield  {title} {\enquote {\bibinfo
  {title} {Une nouvelle limite non-relativiste du groupe de poincar\'e},}\
  }\href {http://www.numdam.org/item/AIHPA_1965__3_1_1_0} {\bibfield  {journal}
  {\bibinfo  {journal} {Annales de l'I.H.P. Physique th\'eorique}\ }\textbf
  {\bibinfo {volume} {3}},\ \bibinfo {pages} {1--12} (\bibinfo {year}
  {1965})}\BibitemShut {NoStop}%
\bibitem [{\citenamefont {Josset}\ \emph {et~al.}(2017)\citenamefont {Josset},
  \citenamefont {Perez},\ and\ \citenamefont
  {Sudarsky}}]{josset17_dark_energ_from_violat_energ_conser}%
  \BibitemOpen
  \bibfield  {author} {\bibinfo {author} {\bibfnamefont {Thibaut}\ \bibnamefont
  {Josset}}, \bibinfo {author} {\bibfnamefont {Alejandro}\ \bibnamefont
  {Perez}}, \ and\ \bibinfo {author} {\bibfnamefont {Daniel}\ \bibnamefont
  {Sudarsky}},\ }\bibfield  {title} {\enquote {\bibinfo {title} {Dark energy
  from violation of energy conservation},}\ }\href {\doibase
  10.1103/physrevlett.118.021102} {\bibfield  {journal} {\bibinfo  {journal}
  {Phys. Rev. Lett.}\ }\textbf {\bibinfo {volume} {118}},\ \bibinfo {pages}
  {021102} (\bibinfo {year} {2017})}\BibitemShut {NoStop}%
\bibitem [{\citenamefont {Brensinger}\ \emph {et~al.}(2019)\citenamefont
  {Brensinger}, \citenamefont {Heitritter}, \citenamefont {Rodgers},
  \citenamefont {Stiffler},\ and\ \citenamefont
  {Whiting}}]{brensinger19_dark_energ_from_dynam_projec_connec}%
  \BibitemOpen
  \bibfield  {author} {\bibinfo {author} {\bibfnamefont {Samuel}\ \bibnamefont
  {Brensinger}}, \bibinfo {author} {\bibfnamefont {Kenneth}\ \bibnamefont
  {Heitritter}}, \bibinfo {author} {\bibfnamefont {Vincent G.~J.}\ \bibnamefont
  {Rodgers}}, \bibinfo {author} {\bibfnamefont {Kory}\ \bibnamefont
  {Stiffler}}, \ and\ \bibinfo {author} {\bibfnamefont {Catherine~A.}\
  \bibnamefont {Whiting}},\ }\bibfield  {title} {\enquote {\bibinfo {title}
  {Dark energy from dynamical projective connections},}\ }\href
  {http://arxiv.org/abs/1907.05334v1} {\  (\bibinfo {year} {2019})},\ \Eprint
  {http://arxiv.org/abs/1907.05334} {arXiv:1907.05334 [hep-th]} \BibitemShut
  {NoStop}%
\bibitem [{\citenamefont {Cervantes-Cota}\ and\ \citenamefont
  {Liebscher}(2016)}]{cervantes-cota16_const_purel_affin_theor_with_matter}%
  \BibitemOpen
  \bibfield  {author} {\bibinfo {author} {\bibfnamefont {Jorge~L.}\
  \bibnamefont {Cervantes-Cota}}\ and\ \bibinfo {author} {\bibfnamefont
  {Dierck-Ekkehard}\ \bibnamefont {Liebscher}},\ }\bibfield  {title} {\enquote
  {\bibinfo {title} {On constructing purely affine theories with matter},}\
  }\href {\doibase 10.1007/s10714-016-2103-9} {\bibfield  {journal} {\bibinfo
  {journal} {Gen. Rel. Grav.}\ }\textbf {\bibinfo {volume} {48}},\ \bibinfo
  {pages} {108} (\bibinfo {year} {2016})}\BibitemShut {NoStop}%
\bibitem [{\citenamefont {Barrow}\ \emph {et~al.}(2019)\citenamefont {Barrow},
  \citenamefont {Tsagas},\ and\ \citenamefont
  {Fanaras}}]{barrow19_fried_like_univer_with_torsion}%
  \BibitemOpen
  \bibfield  {author} {\bibinfo {author} {\bibfnamefont {J.~D.}\ \bibnamefont
  {Barrow}}, \bibinfo {author} {\bibfnamefont {C.~G.}\ \bibnamefont {Tsagas}},
  \ and\ \bibinfo {author} {\bibfnamefont {G.}~\bibnamefont {Fanaras}},\
  }\bibfield  {title} {\enquote {\bibinfo {title} {Friedmann-like universes
  with torsion: a dynamical system approach},}\ }\href
  {http://arxiv.org/abs/1907.07586v1} {\  (\bibinfo {year} {2019})},\ \Eprint
  {http://arxiv.org/abs/1907.07586} {arXiv:1907.07586 [gr-qc]} \BibitemShut
  {NoStop}%
\bibitem [{\citenamefont {Kilic}(2019)}]{kilic19_diffeom_field_revis}%
  \BibitemOpen
  \bibfield  {author} {\bibinfo {author} {\bibfnamefont {Delalcan}\
  \bibnamefont {Kilic}},\ }\bibfield  {title} {\enquote {\bibinfo {title} {The
  diffeomorphism field revisited},}\ }\href {http://arxiv.org/abs/1907.08850v1}
  {\  (\bibinfo {year} {2019})},\ \Eprint {http://arxiv.org/abs/1907.08850}
  {arXiv:1907.08850 [hep-th]} \BibitemShut {NoStop}%
\bibitem [{\citenamefont {Roberts}(1995)}]{roberts95_projec_connec_t}%
  \BibitemOpen
  \bibfield  {author} {\bibinfo {author} {\bibfnamefont {Craig~W.}\
  \bibnamefont {Roberts}},\ }\bibfield  {title} {\enquote {\bibinfo {title}
  {The projective connections of t.y. thomas and j.h.c. whitehead applied to
  invariant connections},}\ }\href {\doibase 10.1016/0926-2245(95)92848-y}
  {\bibfield  {journal} {\bibinfo  {journal} {Differential Geometry and its
  Applications}\ }\textbf {\bibinfo {volume} {5}},\ \bibinfo {pages} {237--255}
  (\bibinfo {year} {1995})}\BibitemShut {NoStop}%
\bibitem [{\citenamefont {Chothe}\ \emph {et~al.}(2019)\citenamefont {Chothe},
  \citenamefont {Dutta},\ and\ \citenamefont
  {Sur}}]{chothe19_cosmol_dark_sector_from_mimet}%
  \BibitemOpen
  \bibfield  {author} {\bibinfo {author} {\bibfnamefont {Hiyang~Ramo}\
  \bibnamefont {Chothe}}, \bibinfo {author} {\bibfnamefont {Ashim}\
  \bibnamefont {Dutta}}, \ and\ \bibinfo {author} {\bibfnamefont {Sourav}\
  \bibnamefont {Sur}},\ }\bibfield  {title} {\enquote {\bibinfo {title}
  {Cosmological dark sector from a mimetic-metric-torsion perspective},}\
  }\href {http://arxiv.org/abs/1907.12429v1} {\  (\bibinfo {year} {2019})},\
  \Eprint {http://arxiv.org/abs/1907.12429} {arXiv:1907.12429 [gr-qc]}
  \BibitemShut {NoStop}%
\bibitem [{\citenamefont {Stein}\ \emph {et~al.}(2019)\citenamefont {Stein}
  \emph {et~al.}}]{stein18_sage_mathem_softw_version}%
  \BibitemOpen
  \bibfield  {author} {\bibinfo {author} {\bibfnamefont {W.~A.}\ \bibnamefont
  {Stein}} \emph {et~al.},\ }\href {http://www.sagemath.org} {\emph {\bibinfo
  {title} {{S}age {M}athematics {S}oftware ({V}ersion 8.6)}}},\ \bibinfo
  {organization} {The Sage Development Team} (\bibinfo {year}
  {2019})\BibitemShut {NoStop}%
\bibitem [{\citenamefont {Gourgoulhon}\ \emph {et~al.}(2019)\citenamefont
  {Gourgoulhon}, \citenamefont {Bejger} \emph
  {et~al.}}]{gourgoulhon18_sagem_version}%
  \BibitemOpen
  \bibfield  {author} {\bibinfo {author} {\bibfnamefont {Eric}\ \bibnamefont
  {Gourgoulhon}}, \bibinfo {author} {\bibfnamefont {Michal}\ \bibnamefont
  {Bejger}},  \emph {et~al.},\ }\href
  {http://sagemanifolds.obspm.fr/index.html} {\emph {\bibinfo {title}
  {{S}age{M}anifolds ({V}ersion 8.6)}}},\ \bibinfo {organization}
  {SageManifolds Development Team} (\bibinfo {year} {2019})\BibitemShut
  {NoStop}%
\bibitem [{\citenamefont {Gourgoulhon}\ \emph {et~al.}(2015)\citenamefont
  {Gourgoulhon}, \citenamefont {Bejger},\ and\ \citenamefont
  {Mancini}}]{gourgoulhon15_tensor_calcul_with_open_sourc_softw}%
  \BibitemOpen
  \bibfield  {author} {\bibinfo {author} {\bibfnamefont {Eric}\ \bibnamefont
  {Gourgoulhon}}, \bibinfo {author} {\bibfnamefont {Michal}\ \bibnamefont
  {Bejger}}, \ and\ \bibinfo {author} {\bibfnamefont {Marco}\ \bibnamefont
  {Mancini}},\ }\bibfield  {title} {\enquote {\bibinfo {title} {Tensor calculus
  with open-source software: the sagemanifolds project},}\ }\href {\doibase
  10.1088/1742-6596/600/1/012002} {\bibfield  {journal} {\bibinfo  {journal}
  {Journal of Physics: Conference Series}\ }\textbf {\bibinfo {volume} {600}},\
  \bibinfo {pages} {012002} (\bibinfo {year} {2015})},\ \Eprint
  {http://arxiv.org/abs/1412.4765} {arXiv:1412.4765 [gr-qc]} \BibitemShut
  {NoStop}%
\bibitem [{\citenamefont {Gourgoulhon}\ and\ \citenamefont
  {Mancini}(2018)}]{gourgoulhon18_symbol_tensor_calcul_manif}%
  \BibitemOpen
  \bibfield  {author} {\bibinfo {author} {\bibfnamefont {{\'E}ric}\
  \bibnamefont {Gourgoulhon}}\ and\ \bibinfo {author} {\bibfnamefont {Marco}\
  \bibnamefont {Mancini}},\ }\bibfield  {title} {\enquote {\bibinfo {title}
  {Symbolic tensor calculus on manifolds: a sagemath implementation},}\ }\href
  {\doibase 10.5802/ccirm.26} {\bibfield  {journal} {\bibinfo  {journal} {Les
  cours du CIRM}\ }\textbf {\bibinfo {volume} {6}},\ \bibinfo {pages} {1--54}
  (\bibinfo {year} {2018})},\ \Eprint {http://arxiv.org/abs/1804.07346}
  {arXiv:1804.07346 [gr-qc]} \BibitemShut {NoStop}%
\bibitem [{\citenamefont {Tulczyjew}(1974)}]{tulczyjew74_hamil_lagr_syst}%
  \BibitemOpen
  \bibfield  {author} {\bibinfo {author} {\bibfnamefont {Wlodzimierz~M.}\
  \bibnamefont {Tulczyjew}},\ }\bibfield  {title} {\enquote {\bibinfo {title}
  {Hamiltonian systems, lagrangian systems and the legendre tranformation},}\ \
  }(\bibinfo  {publisher} {Academic Press},\ \bibinfo {year} {1974})\ pp.\
  \bibinfo {pages} {247--258}\BibitemShut {NoStop}%
\bibitem [{\citenamefont
  {Tulczyjew}(1975{\natexlab{a}})}]{tulczyjew75_symp_form_part_dyna}%
  \BibitemOpen
  \bibfield  {author} {\bibinfo {author} {\bibfnamefont {Wlodzimierz~M.}\
  \bibnamefont {Tulczyjew}},\ }\bibfield  {title} {\enquote {\bibinfo {title}
  {{A Symplectic Formulation of Particle Dynamics}},}\ }in\ \href@noop {}
  {\emph {\bibinfo {booktitle} {{Conference on Differential Geometrical Methods
  in Mathematical Physics Bonn, Germany, July 1-4, 1975}}}}\ (\bibinfo {year}
  {1975})\ pp.\ \bibinfo {pages} {457--463}\BibitemShut {NoStop}%
\bibitem [{\citenamefont
  {Tulczyjew}(1975{\natexlab{b}})}]{tulczyjew75_symp_form_field_dyna}%
  \BibitemOpen
  \bibfield  {author} {\bibinfo {author} {\bibfnamefont {Wlodzimierz~M.}\
  \bibnamefont {Tulczyjew}},\ }\bibfield  {title} {\enquote {\bibinfo {title}
  {{A Symplectic Formulation of Field Dynamics}},}\ }in\ \href@noop {} {\emph
  {\bibinfo {booktitle} {{Conference on Differential Geometrical Methods in
  Mathematical Physics Bonn, Germany, July 1-4, 1975}}}}\ (\bibinfo {year}
  {1975})\ pp.\ \bibinfo {pages} {464--468}\BibitemShut {NoStop}%
\bibitem [{\citenamefont {Kijowski}\ and\ \citenamefont
  {Tulczyjew}(1979)}]{kijowski79}%
  \BibitemOpen
  \bibfield  {author} {\bibinfo {author} {\bibfnamefont {Jerzy}\ \bibnamefont
  {Kijowski}}\ and\ \bibinfo {author} {\bibfnamefont {Wlodzimierz~M.}\
  \bibnamefont {Tulczyjew}},\ }\href@noop {} {\emph {\bibinfo {title} {A
  symplectic framework for field theories}}}\ (\bibinfo  {publisher}
  {Springer-Verlag},\ \bibinfo {year} {1979})\BibitemShut {NoStop}%
\end{thebibliography}
%

\end{document}